\documentclass[prl,twocolumn,english,amsmath,amssymb,floatfix,superscriptaddress,longbibliography]{revtex4-2}
\usepackage{graphicx}
\usepackage[usenames,dvipsnames]{xcolor}
\definecolor{oceanboatblue}{rgb}{0.0, 0.47, 0.75}
\definecolor{orange}{rgb}{1,0.5,0}
\definecolor{goodgreen}{rgb}{0.1,0.5,0}
\definecolor{goodred}{rgb}{0.7,0,0}
\usepackage[colorlinks,urlcolor=goodgreen,citecolor=blue,linkcolor=goodred]{hyperref}
\usepackage{bm,amsmath}

\usepackage[normalem]{ulem}

\begin{document}
\title{Wannier center spectroscopy to identify boundary-obstructed topological insulators}
\author{R.A.M. Ligthart}
\affiliation{Debye Institute for Nanomaterials Science, Utrecht University, 3584 CC Utrecht, The Netherlands}
\author{M.A.J. Herrera}
\affiliation{Donostia International Physics Center (DIPC), Manuel de Lardizbal 4, 20018 San Sebasti\'an, Spain}
\author{A.C.H. Visser}
\affiliation{Debye Institute for Nanomaterials Science, Utrecht University, 3584 CC Utrecht, The Netherlands}
\author{A. Vlasblom}
\affiliation{Debye Institute for Nanomaterials Science, Utrecht University, 3584 CC Utrecht, The Netherlands}
\author{D. Bercioux}
\affiliation{Donostia International Physics Center (DIPC), Manuel de Lardizbal 4, 20018 San Sebasti\'an, Spain}
\affiliation{IKERBASQUE, Basque Foundation for Science, Plaza Euskadi 5
48009 Bilbao, Spain}
\author{I. Swart}
\email{Corresponding author: i.swart@uu.nl}
\affiliation{Debye Institute for Nanomaterials Science, Utrecht University, 3584 CC Utrecht, The Netherlands}

\begin{abstract}
The hallmark of topological crystalline insulators is the emergence of a robust electronic state in a bandgap localized at the boundary of the material. However, end, edge, and surface states can also have a non-topological origin. Unfortunately, topological invariants such as the winding number and Zak-phase are often not directly experimentally accessible for solids. In addition to topological invariants, the position of the Wannier centers provides a fingerprint for the topological character of a material. Here, we demonstrate a method to experimentally determine the location of Wannier centers in artificial lattices made of Cs/InAs(111)A by integrating the density of states. We determine the locations of the Wannier centers for various 1D chains, topological and trivial, and corroborate our findings with tight-binding simulations.
\end{abstract}

\maketitle

Topological insulators are renowned for their robust edge modes, which offer the prospect of dissipationless transport of electrons~\cite{Zhang2005a, St-Jean2017, Konig2007, Kane2005, Breunig2022, Hasan2010}. Topological boundary modes emerge in systems with certain (combinations of) symmetries: time-reversal, crystalline symmetry, and particle-hole. The boundary state manifests in one dimension lower than the material itself, e.g., 0D end states for 1D chains. These topological systems are characterized by Wannier states that cannot be associated with atomic orbitals~\cite{Soluyanov2011, Bradlyn_2017, Kruthoff_2017, BlancodePaz2022}.In addition, there are boundary-obstructed topological~(BOT) phases~\cite{Khalaf_2021}. The topological obstruction in these systems arises from the structure of the boundary rather than the bulk (i.e., the boundary modes do not arise from the bulk-boundary correspondence, and these systems allow for a description using localized Wannier functions). Examples include the Su-Schrieffer-Heeger (SSH) chain~\cite{Su1979}, the Kekul\'e lattice~\cite{Freeney2020}, and some higher-order topological insulators~\cite{Khalaf_2021}. However, a boundary mode does not necessarily have a topological origin. For example, the well-known surface state of crystals of noble metals with (111) termination is topologically trivial. Hence, an experimental probe to distinguish between topologically trivial and non-trivial materials boundary modes is, therefore, highly desirable. To establish the nature of a boundary mode, topological invariants such as the Chern number, the winding number, or the Zak phase can be computed directly from the phase of the wavefunction~\cite{Grushin2020, Asboth2015, Lee2022, Chiu2016}. For certain photonic and cold atom systems, topological invariants and wavefunction phase are experimentally accessible~\cite{Xu2018, Schine2019, Ozawa2019, DePaz2019, Atala2013}. For solid-state materials, the phase of the electronic wave function is difficult to measure. Instead, topological invariants can be extracted from transport experiments via the quantization of conductance at values proportional to the Chern number~\cite{Thouless_1982, Zhao2020}. However, transport measurements do not provide local information, and it is difficult to contact the edge (end) mode of 2D (1D) materials~\cite{Leis2022}. Local probe experiments have revealed robust boundary modes in low-dimensional materials, but the topological nature is always inferred by comparison to theoretical calculations~\cite{Nadj-Perge2014, Schneider2022, Kezilebieke2020, Moes2023}. Therefore, there is a compelling need for a local probe-based experimental tool to establish the nature of a boundary mode. 
For BOT insulators, the spatial position of Wannier centers is a fingerprint of the topological phase of the material.
In the simplest case of the SSH chain~\cite{Su1979}, it has been shown that the Wannier centers are localized respectively at the center or the edge of the unit cell for the trivial/non-trivial case and that this property affects the polarization of the system~\cite{Vanderbilt_1993, Resta_1994, Neupert_2018,Vanderbilt2018}.

Here, we demonstrate a method to experimentally determine the position of Wannier centers using a Scanning Tunnelling Microscope (STM), thus providing a tool to unambiguously distinguish if a potential BOT insulator is topologically trivial or non-trivial. The technique is based on spatially resolved integration of the local density of states (vide infra). Measurements were performed on various artificial electronic lattices, topological and trivial, and corroborated with tight-binding-based calculations. In particular, we made and characterized the atomic, Rice-Mele~\cite{Rice_1982, Allen_2022}, SSH, and trimer chains~\cite{Su1981, MartinezAlvarez2019}. 1D topological chains such as the SSH model are extensively studied for their topological phase transition~\cite{Su1979, Meier2016, Drost2017a, Pham2022}. They are well-suited candidates to demonstrate the capability to determine the position of Wannier centres experimentally. 
%
%
\begin{figure*}[!t]
 \centering
 \includegraphics[width=\textwidth]{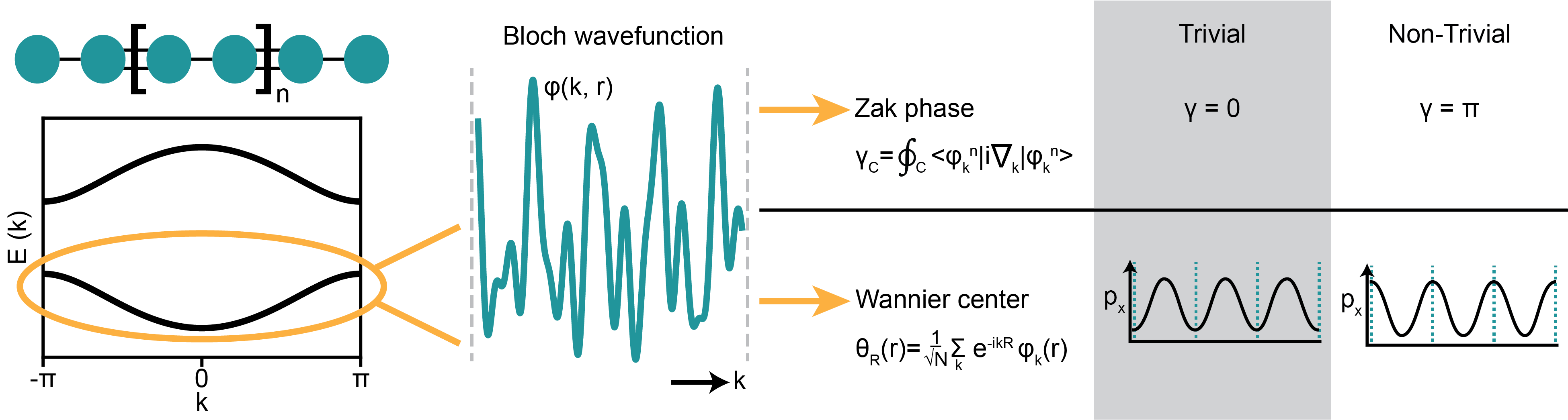}
 \caption{Scheme to derive information on topology. A given band --- in this case of the SSH chain --- is described by a Bloch wavefunction. From there, either the Zak phase or the Wannier representation can be obtained as a fingerprint for the topological character.}
 \label{TheorySchematic}
\end{figure*}
%
%
The chains were created by manipulating Cs atoms on InAs(111)A~\cite{Ligthart2024}. A detailed discussion on constructing artificial electronic lattices can be found elsewhere~\cite{Gomes2012a, Kempkes2018a, Slot2017a}. In brief, the positively charged adatoms confine electrons, forming particle-in-a-box-like states that can be coupled to form electronic lattices~\cite{Folsch2014, Sierda2023}. The on-site energy and hopping strength can be controlled via the number of Cs atoms in the artificial atom and the distance between the artificial atoms, respectively. The states have energies located in the bandgap of the substrate, allowing a higher energy resolution than metal-based platforms~\cite{Folsch2014, Sierda2023}. This is beneficial for the procedure described below.

The InAs(111)A sample was prepared by decapping the amorphous As layer inside the Scienta Omicron LT-STM. The measurements were performed at 4.3 K and a pressure below $10^{-10}$~mbar. The Cs atoms were evaporated (SAES Getters) on a cold InAs(111)A surface. Vertical and lateral manipulations were done to position the Cs atoms to the desired locations~\cite{Ligthart2024}. Further details on the sample growth and measurement settings are described in the Supplemental Material (SM)~\cite{SM}. 

The Wannier function is the Fourier transform of the Bloch wavefunction of a specific band. Hence, the Wannier representation is the real space localization of the electron density of a given band in reciprocal space; see schematic Fig.~\ref{TheorySchematic}. Note that this function is necessarily periodic (following the crystal structure). The Wannier center corresponds to the unit cell position where the Wannier function/electron density is maximum. Scanning Tunneling Spectroscopy (STS) provides access to the Local Density of States (LDOS) $\rho(x,E)$ via differential conductance measurements. By integrating the LDOS in the energy range of a specific band, one obtains the charge density at that particular position of the chosen band. We normalize the charge density by the total LDOS at the specific site. This procedure sets the total amount of states in the system to one and acts as an internal standard. We define the normalized charge $p_x$ as:
%
%
\begin{equation}\label{WannierSTM}
    p_x = \frac{ \int_{E_n^\text{min}}^{E_n^\text{max}} \rho(x, E) dE}{\int_\text{All bands} \rho(x, E) dE}
\end{equation}
%
%
where, $E_n^\text{min}(E_n^\text{max})$ is the minimum (maximum) energy of the $n$-th band. In the STM, the LDOS is acquired at a given $x,y$ coordinate | the position of the tip | allowing the spatial electron density mapping. In the following, we limit ourselves to the 1D case, for which it is sufficient to take spectra along one line. The maxima in $p_x$  correspond to the Wannier centers' positions. The Wannier representation can be obtained for each separated band by altering the integration values of the numerator to include the specific band. To apply this method, well-separated energy bands are required, \emph{i.e.} the energy resolution should be high enough to resolve band gaps. When bands partially overlap in energy, it becomes difficult to correctly identify the limits to be used in the integration. Therefore, we focus on artificial lattices constructed using Cs on InAs(111)A, which provide the required energy resolution. In addition, both trivial and topological states of matter can be engineered using this platform.

\begin{figure}[b!]
    \centering
    \includegraphics{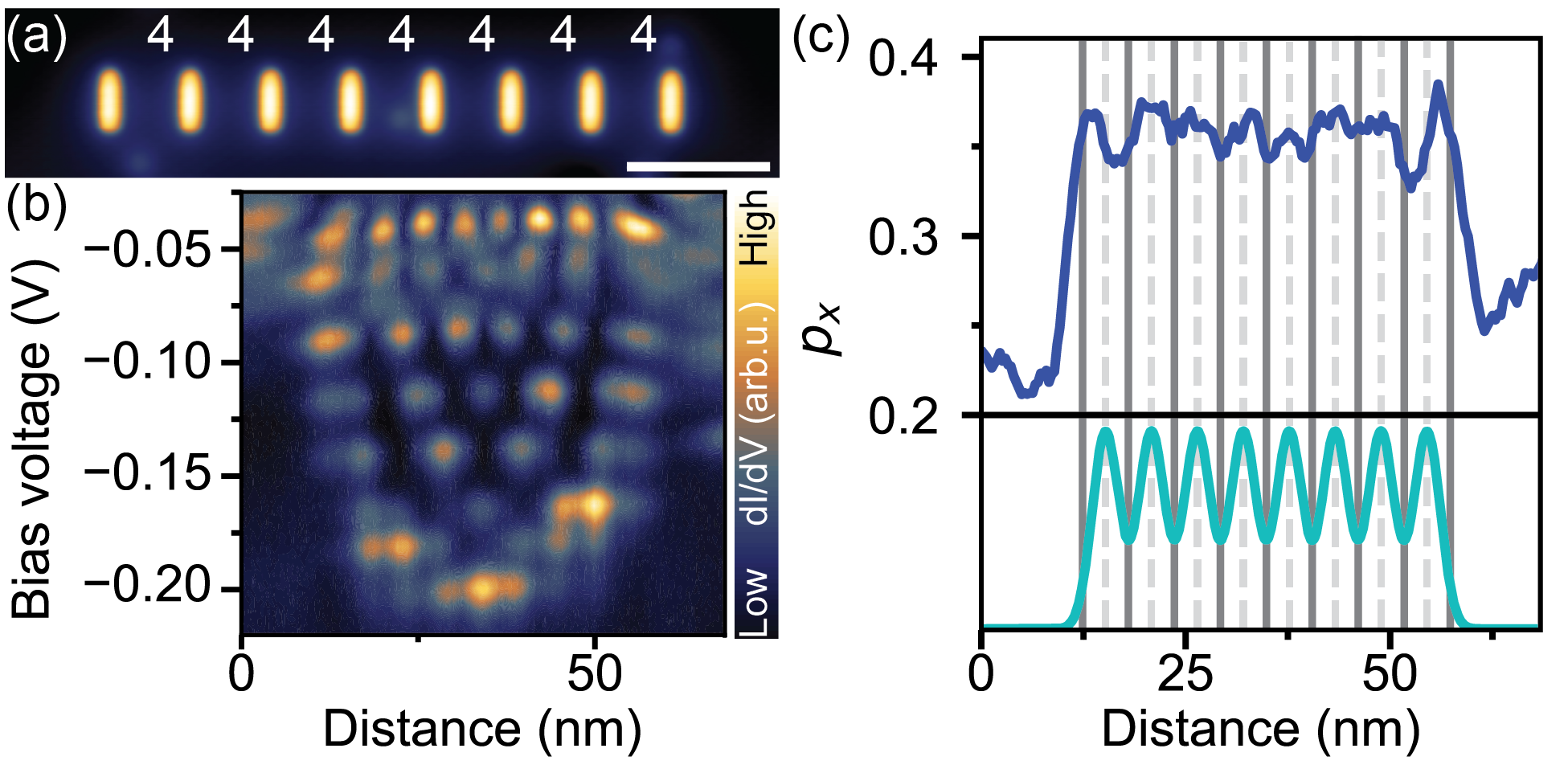}
    \caption{(a) STM topography image of the atomic chain with 8 unit cells. The distance between the chains is $4a\sqrt{3}$ nm, which corresponds to a hopping of 38 meV. Image taken at 0.5 V and 30 pA, scale bar is 10 nm. (b) Contour plot of $dI/dV$ spectra taken above the atomic chain in (a). (c) The experimental (dark blue) and theoretical (cyan) spatial dependence of the charge density. Positions of the atoms and the boundary of the unit cell are indicated by light dashed and dark gray lines, respectively. A 3-point running average was applied to the experimental data.}
    \label{AtomicChain}
\end{figure}

We now apply this method to various atomically well-defined chains. We start with the simplest chain imaginable: the atomic chain. Figure~\ref{AtomicChain}a shows a chain consisting of 8 artificial atoms, with an equidistant spacing of $4a\sqrt{3}$, with $a = 0.857$~nm, the lattice constant of the InAs(111)A (2$\times$2) surface reconstruction. The on-site energy of the artificial atoms is controlled by the number of Cs atoms in the chain, here 5~Cs, see~\cite{SM}. In this configuration, the coupling strength between sites is $38\pm 2$~meV, determined by measuring the peak-spacing as a function of the distance between two artificial atoms, see~\cite{SM, FN1}. 
STS is performed above the chain with 209 points and is depicted as a contour plot in Fig.~\ref{AtomicChain}b. For the computation of the charge density/Wannier representation of the lowest energy band, the integration values were set to include all states with energies lower than the bandgap. A background correction is performed to remove the onset of the conduction band of the InAs(111)A and potential tip states, see~\cite{SM}. The atomic chain has only one band, so no normalization was performed. The resulting charge density is plotted in Fig.~\ref{AtomicChain}c (solid blue line). These results are qualitatively robust with respect to changes in the integration limits and background correction procedures used, see\cite{SM}. The maxima, \emph{i.e.} the Wannier centers positions, are located around the position of the atoms, as expected for the atomic chain.
Furthermore, the same features are observed for every unit cell qualitatively. Tight binding calculations for a similar system (cyan line) reproduce the experimental findings. When we apply the procedure to an atomic Rice-Mele chain, the Wannier centers are found to coincide with the tight-binding calculations as well, see~\cite{SM}. 

\begin{figure}
 \centering
 \includegraphics{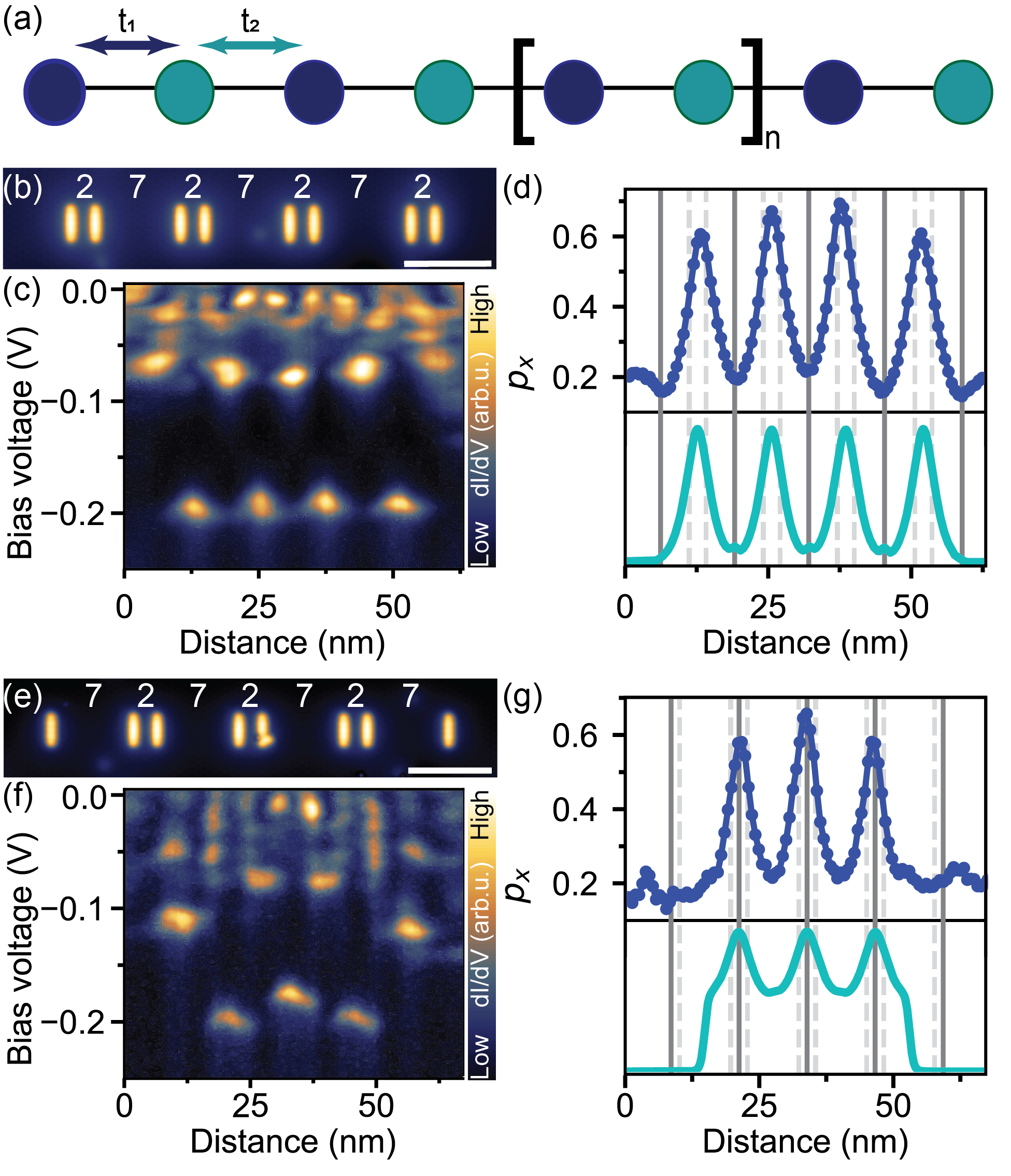}
 \caption{(a) Schematic of the SSH chain with two atoms per unit cell. (b, e) STM topography image of the trivial (b) and non-trivial (e) SSH chain with 4 unit cells and a distance of $2a\sqrt{3}$ and $7a\sqrt{3}$ nm between the artificial atoms. Images taken at 0.8 V (b) and 0.5 V (e), and 30 pA, scale bar is 10 nm. (c, f) Contour plot of $dI/dV$ spectra taken along a line above the trivial and non-trivial SSH chains in (b) and (e), respectively. (d, g) Charge density plot experiment (dark blue) and theory (cyan) with the positions of the atoms (light grey) and the boundary of the unit cell (dark grey).}
 \label{SSH}
\end{figure}

Next, we focus on the simplest well-established BOT insulator: the SSH chain displayed in Fig.~\ref{SSH}a. A topograph of the trivial SSH chain is shown in Fig.~\ref{SSH}b. The intra-cell distance between artificial atoms --- $d_1=2a\sqrt{3}$~[nm] --- is smaller than the inter-cell distance --- $d_2=7a\sqrt{3}$~[nm]. This corresponds to two different hopping terms, $t_1$ and $t_2$ for the intra- and inter-cell processes, respectively. This configuration leads to a hopping ratio of $ t_1/t_2=5.6 \gg 1$, and the chain is, therefore, trivial~\cite{Asboth2015}. The contour plot in Fig.~\ref{SSH}c of the $dI/dV$ spectroscopy obtained above the chain shows a clear bandgap of $111$~mV, as expected for the trivial case, and no edge modes are present in the gap. The value of the gap is consistent with the value of $t_1/t_2 =5.6$. The charge density, blue curve in Fig.~\ref{SSH}d, shows clear peaks in the centres of the unit cell, in agreement with earlier theoretical work and our calculations (cyan line)~\cite{Benalcazar2019}.

The non-trivial SSH chain with hopping terms $t_1$ and $t_2$ switched is shown in Fig.~\ref{SSH}e, with the corresponding contour plot in Fig.~\ref{SSH}f. For this configuration, we have $t_1/t_2=0.18\ll1$. A defect in the substrate prevented us from positioning all Cs atoms in the fifth artificial atom from the left in a straight line. One of the five Cs atoms had to be positioned one site further away, leading to a lower on-site energy of that artificial atom. This, in turn, leads to a slightly higher energy of the bonding band around -200 meV, which can indeed be seen in the contour plot in Fig.~\ref{SSH}f. The non-trivial end-localized states are observed on the left and right ends of the chain at $-110$~mV. 
The end-states are not in the centre of the bandgap since the electronic surrounding of the edge sites is different from those in the bulk | they miss a neighbour.
This changes the onsite energy, pushing the end modes up in energy, as confirmed by tight-binding calculations.
The charge density of the valence band of the non-trivial SSH chain is shown in Fig.~\ref{SSH}g. The integration values are chosen to include the valence band of the chain, from $-300$~mV to $-144$~mV. The charge density now peaks at the unit cell boundaries (dark gray solid lines). Again, the experimental data is corroborated by our tight-binding calculations (cyan line). Hence, the experimentally determined peak positions in $p_x$ coincide with the position of the Wannier centers for both the trivial and non-trivial SSH chains. As mentioned above, the atomic scale defect modifies the on-site energy of one particular site. Note that the type of defect does not break the chiral symmetry protecting the topological character of the chain and should, therefore, not affect the position of the Wannier center. Indeed, the maximum charge density of the valence band is also located at the boundary of the unit cell containing the defect. This highlights the robustness of the method. 

\begin{figure}
 \centering
 \includegraphics{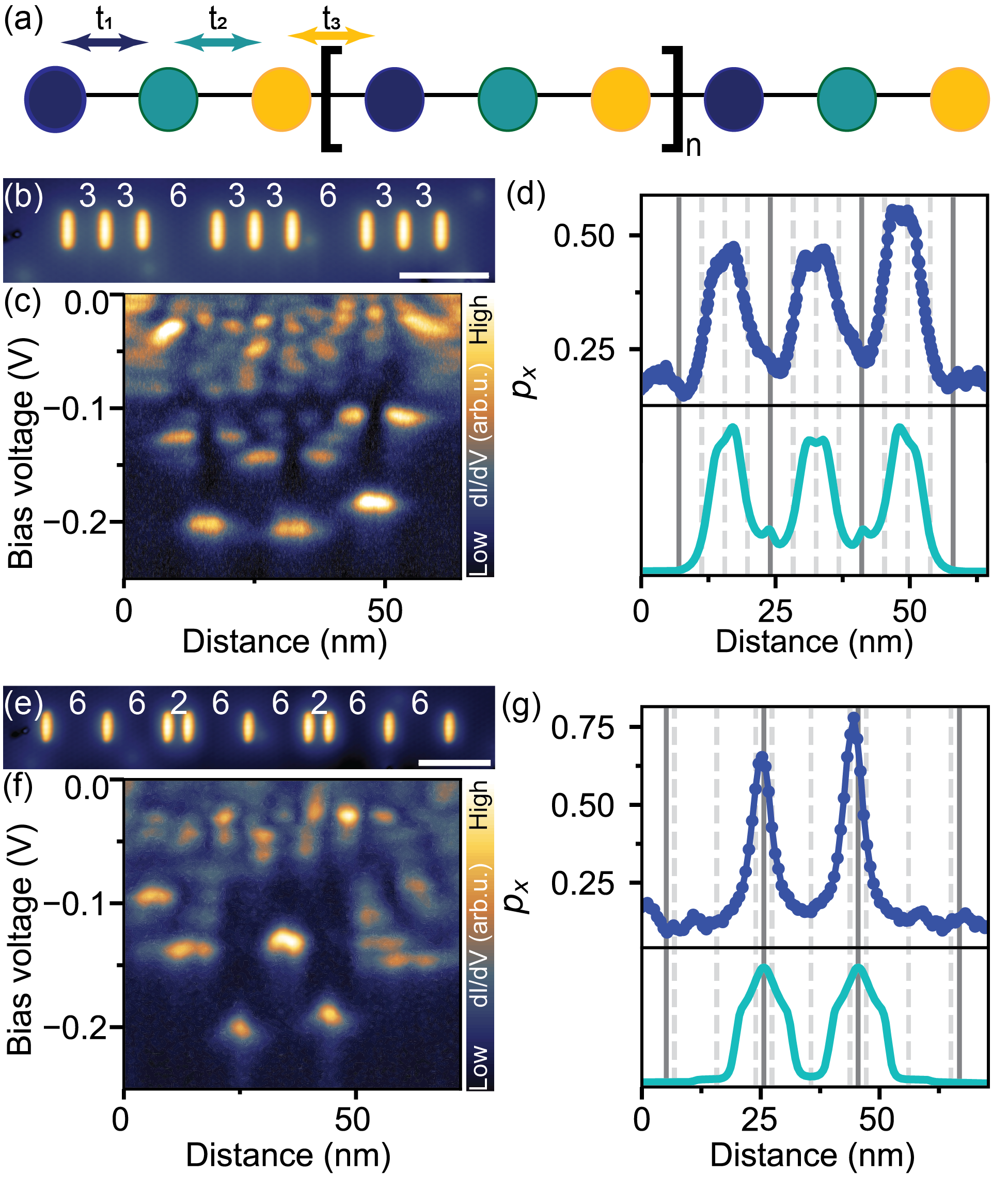}
 \caption{(a) Schematic of the trimer chain with three atoms per unit cell. (b, e) STM topography image of the trivial (b) and non-trivial (e) symmetric trimer chain with three unit cells and a distance of respectively $3a\sqrt{3}$, $3a\sqrt{3}$ and $6a\sqrt{3}$ nm and $6a\sqrt{3}$, $6a\sqrt{3}$ and $2a\sqrt{3}$ nm between the artificial atoms. Images taken at 0.8~V (b) and 0.1~V (e), and 30~pA, scale bar is 10~nm. (c, f) Contour plot of $dI/dV$ spectra taken along a line above the trivial and non-trivial trimer chains in (b) and (e), respectively. (d, g) Charge density plot experiment (dark blue) and theory (cyan) with the positions of the atoms (light grey) and the boundary of the unit cell (dark grey). }
 \label{Trimer}
\end{figure}

To demonstrate the generality of the method, we also applied it to the trimer chain~\cite{Su1981, MartinezAlvarez2019}, shown in Fig.~\ref{Trimer}a. Differently from the SSH model, the hopping parameters trimer chain can be chosen so that the chain possesses the inversion symmetry or not~\cite{MartinezAlvarez2019}. End modes are present when the intercell hopping is larger than the intracell hopping. Since the chain has three atoms per unit cell, there are three bands separated by two band gaps. End-localized states can emerge in both gaps~\cite{Su1981}. 

An inversion symmetric trivial chain is shown in Fig.~\ref{Trimer}b: The distances between artificial atoms are $3a\sqrt{3}$~[nm],  $3a\sqrt{3}$~[nm] and $6a\sqrt{3}$~[nm]. We could not position three rows of five Cs atoms with a spacing of $2a\sqrt{3}$. In this configuration, the density of positively charged Cs atoms is so high that it causes surface instabilities~\cite{Yang2011b}. However, since the intracell hopping is still larger than the intercell hopping, the chain is topologically trivial. The smaller hopping ratio results in a smaller bandgap --- see Fig.~\ref{Trimer}c. Two bandgaps can still be distinguished at $-180$~mV and $-100$~mV. For the charge density, Fig.~\ref{Trimer}d, the integration values were chosen to include the lowest band. The Wannier centres are localised at the centre of the unit cell. The third peak is slightly off from the centre of the unit cell, likely caused by drift during the spectroscopy measurement.

The experimentally realised symmetric non-trivial trimer chain is shown in Fig.~\ref{Trimer}e. The distances were chosen to be $6a\sqrt{3}$, $6a\sqrt{3}$ and $2a\sqrt{3}$~[nm], corresponding to hopping parameters $t_1=19$~meV, $t_2=19$~meV, and $t_3=85$~meV, respectively. Figure~\ref{Trimer}f shows the contour plot of the differential conductance line spectra. The two topological edge modes on both sides of the chain are found at $-137$~mV and $-93$~mV, respectively. The edge modes are close in energy with the middle non-bonding band due to deviation from the trimerized limit of the chain. Figure~\ref{Trimer}g shows the spatially resolved charge density of the lowest energy band (blue for experimental data, cyan for tight-binding results). We observe again that the Wannier centres are located at the boundaries of the unit cells, as expected for the non-trivial chain. 

We also analyzed the non-inversion symmetric trimer chain and found that the edge states present in this system are still localized at the boundary of the chain~\cite{SM}. The corresponding charge density $p_x$ peaks at the unit cell's boundary in the topological configuration and its center in the trivial configuration~\cite{SM}. This appears to contradict the prediction of the Zak phase, which is not quantized in this configuration~\cite{SM}. However, this result can be related to latent symmetries that we have not accounted for~\cite{Roentgen_2021, Roentgen_2023}.

In summary, we demonstrate a method to experimentally determine the location of Wannier centers in solid-state artificial lattices. This method is based on integrating and normalising $dI/dV$ spectra, which are proportional to the charge density, as a function of position. The peaks of the charge densities, $p_x$, correspond to the location of the Wannier centres in the system. We applied the method to an atomic chain, the Rice-Mele chain, the SSH chain, and the trimer chain. For the trivial systems, the Wannier centers are located in the middle of the unit cells, whereas they are located at the boundaries for topological chains, exactly as theoretically expected. The experimental results are corroborated by tight-binding simulations.
Being able to determine the positions of Wannier centers in a scanning tunnelling microscopy experiment provides an unambiguous method that can distinguish between the different phases of BOT insulators. Our method can be extended to 2D-BOT insulators, where it is possible to correlate, in specific cases, the position of the Wannier centers to the expectation value of the position operator over the Wannier functions~\cite{Ni_2018,Kempkes_2019,Herrera_2022,Tassi_2024}. This mitigates the reliance of local probe methods on the observation of a boundary state, which can have many origins. In principle, the method described here can be applied to any solid-state material measurable by STM.   

\emph{Note added in proof.} Recently, we became aware of 
two preprints reporting results for atomically obstructed insulators in 2D systems one on WSe$_2$~\cite{Holbrook2024} and one on NbSe$_2$~\cite{Calugaru2025}.

\begin{acknowledgments}
R.A.M.L. and I.S. acknowledge funding from the European Research Council (Horizon 2020 “FRACTAL,” 865570). A.V. and I.S. acknowledge the research program “Materials for the Quantum Age” (QuMat) for financial support. This program (Registration No. 024.005.006) is part of the Gravitation program financed by the Dutch Ministry of Education, Culture and Science (OCW). D.B. and M.A.J.H. acknowledge support from the Spanish Ministerio de Ciencia, Innovación y Universidades -- Agencia Estatal de Investigación through Project No. PID2020-120614GB-I00 (ENACT), the Transnational Common Laboratory \emph{Quantum-ChemPhys}, the Hezkuntza, Hizkuntza Politika Eta Kultura Saila, Eusko Jaurlaritza through the project PIBA\_2023\_1\_0007 (STRAINER), and financial support received from the IKUR Strategy under the collaboration agreement between the Ikerbasque Foundation and DIPC on behalf of the Department of Education of the Basque Government and the Diputación Foral de Gipuzkoa within the QUAN-000021-01 project. 
\end{acknowledgments}

\bibliography{Caesium-Trimer} 

\begin{thebibliography}{58}%
\makeatletter
\providecommand \@ifxundefined [1]{%
 \@ifx{#1\undefined}
}%
\providecommand \@ifnum [1]{%
 \ifnum #1\expandafter \@firstoftwo
 \else \expandafter \@secondoftwo
 \fi
}%
\providecommand \@ifx [1]{%
 \ifx #1\expandafter \@firstoftwo
 \else \expandafter \@secondoftwo
 \fi
}%
\providecommand \natexlab [1]{#1}%
\providecommand \enquote  [1]{``#1''}%
\providecommand \bibnamefont  [1]{#1}%
\providecommand \bibfnamefont [1]{#1}%
\providecommand \citenamefont [1]{#1}%
\providecommand \href@noop [0]{\@secondoftwo}%
\providecommand \href [0]{\begingroup \@sanitize@url \@href}%
\providecommand \@href[1]{\@@startlink{#1}\@@href}%
\providecommand \@@href[1]{\endgroup#1\@@endlink}%
\providecommand \@sanitize@url [0]{\catcode `\\12\catcode `\$12\catcode
  `\&12\catcode `\#12\catcode `\^12\catcode `\_12\catcode `\%12\relax}%
\providecommand \@@startlink[1]{}%
\providecommand \@@endlink[0]{}%
\providecommand \url  [0]{\begingroup\@sanitize@url \@url }%
\providecommand \@url [1]{\endgroup\@href {#1}{\urlprefix }}%
\providecommand \urlprefix  [0]{URL }%
\providecommand \Eprint [0]{\href }%
\providecommand \doibase [0]{https://doi.org/}%
\providecommand \selectlanguage [0]{\@gobble}%
\providecommand \bibinfo  [0]{\@secondoftwo}%
\providecommand \bibfield  [0]{\@secondoftwo}%
\providecommand \translation [1]{[#1]}%
\providecommand \BibitemOpen [0]{}%
\providecommand \bibitemStop [0]{}%
\providecommand \bibitemNoStop [0]{.\EOS\space}%
\providecommand \EOS [0]{\spacefactor3000\relax}%
\providecommand \BibitemShut  [1]{\csname bibitem#1\endcsname}%
\let\auto@bib@innerbib\@empty
\bibitem [{\citenamefont {Zhang}\ \emph {et~al.}(2005)\citenamefont {Zhang},
  \citenamefont {Tan}, \citenamefont {Stormer},\ and\ \citenamefont
  {Kim}}]{Zhang2005a}%
  \BibitemOpen
  \bibfield  {author} {\bibinfo {author} {\bibfnamefont {Y.}~\bibnamefont
  {Zhang}}, \bibinfo {author} {\bibfnamefont {Y.-W.}\ \bibnamefont {Tan}},
  \bibinfo {author} {\bibfnamefont {H.~L.}\ \bibnamefont {Stormer}},\ and\
  \bibinfo {author} {\bibfnamefont {P.}~\bibnamefont {Kim}},\ }\bibfield
  {title} {\bibinfo {title} {{Experimental observation of the quantum Hall
  effect and Berry's phase in graphene}},\ }\href
  {https://doi.org/10.1038/nature04235} {\bibfield  {journal} {\bibinfo
  {journal} {Nature}\ }\textbf {\bibinfo {volume} {438}},\ \bibinfo {pages}
  {201} (\bibinfo {year} {2005})}\BibitemShut {NoStop}%
\bibitem [{\citenamefont {St-Jean}\ \emph {et~al.}(2017)\citenamefont
  {St-Jean}, \citenamefont {Goblot}, \citenamefont {Galopin}, \citenamefont
  {Lema{\^{i}}tre}, \citenamefont {Ozawa}, \citenamefont {{Le Gratiet}},
  \citenamefont {Sagnes}, \citenamefont {Bloch},\ and\ \citenamefont
  {Amo}}]{St-Jean2017}%
  \BibitemOpen
  \bibfield  {author} {\bibinfo {author} {\bibfnamefont {P.}~\bibnamefont
  {St-Jean}}, \bibinfo {author} {\bibfnamefont {V.}~\bibnamefont {Goblot}},
  \bibinfo {author} {\bibfnamefont {E.}~\bibnamefont {Galopin}}, \bibinfo
  {author} {\bibfnamefont {A.}~\bibnamefont {Lema{\^{i}}tre}}, \bibinfo
  {author} {\bibfnamefont {T.}~\bibnamefont {Ozawa}}, \bibinfo {author}
  {\bibfnamefont {L.}~\bibnamefont {{Le Gratiet}}}, \bibinfo {author}
  {\bibfnamefont {I.}~\bibnamefont {Sagnes}}, \bibinfo {author} {\bibfnamefont
  {J.}~\bibnamefont {Bloch}},\ and\ \bibinfo {author} {\bibfnamefont
  {A.}~\bibnamefont {Amo}},\ }\bibfield  {title} {\bibinfo {title} {{Lasing in
  topological edge states of a one-dimensional lattice}},\ }\href
  {https://doi.org/10.1038/s41566-017-0006-2} {\bibfield  {journal} {\bibinfo
  {journal} {Nat. Photonics}\ }\textbf {\bibinfo {volume} {11}},\ \bibinfo
  {pages} {651} (\bibinfo {year} {2017})}\BibitemShut {NoStop}%
\bibitem [{\citenamefont {K{\"{o}}nig}\ \emph {et~al.}(2007)\citenamefont
  {K{\"{o}}nig}, \citenamefont {Wiedmann}, \citenamefont {Br{\"{u}}ne},
  \citenamefont {Roth}, \citenamefont {Buhmann}, \citenamefont {Molenkamp},
  \citenamefont {Qi},\ and\ \citenamefont {Zhang}}]{Konig2007}%
  \BibitemOpen
  \bibfield  {author} {\bibinfo {author} {\bibfnamefont {M.}~\bibnamefont
  {K{\"{o}}nig}}, \bibinfo {author} {\bibfnamefont {S.}~\bibnamefont
  {Wiedmann}}, \bibinfo {author} {\bibfnamefont {C.}~\bibnamefont
  {Br{\"{u}}ne}}, \bibinfo {author} {\bibfnamefont {A.}~\bibnamefont {Roth}},
  \bibinfo {author} {\bibfnamefont {H.}~\bibnamefont {Buhmann}}, \bibinfo
  {author} {\bibfnamefont {L.~W.}\ \bibnamefont {Molenkamp}}, \bibinfo {author}
  {\bibfnamefont {X.~L.}\ \bibnamefont {Qi}},\ and\ \bibinfo {author}
  {\bibfnamefont {S.~C.}\ \bibnamefont {Zhang}},\ }\bibfield  {title} {\bibinfo
  {title} {{Quantum spin Hall insulator state in HgTe quantum wells}},\ }\href
  {https://doi.org/10.1126/science.1148047} {\bibfield  {journal} {\bibinfo
  {journal} {Science}\ }\textbf {\bibinfo {volume} {318}},\ \bibinfo {pages}
  {766} (\bibinfo {year} {2007})}\BibitemShut {NoStop}%
\bibitem [{\citenamefont {Kane}\ and\ \citenamefont {Mele}(2005)}]{Kane2005}%
  \BibitemOpen
  \bibfield  {author} {\bibinfo {author} {\bibfnamefont {C.~L.}\ \bibnamefont
  {Kane}}\ and\ \bibinfo {author} {\bibfnamefont {E.~J.}\ \bibnamefont
  {Mele}},\ }\bibfield  {title} {\bibinfo {title} {{Quantum Spin Hall effect in
  graphene}},\ }\href {https://doi.org/10.1103/PhysRevLett.95.226801}
  {\bibfield  {journal} {\bibinfo  {journal} {Phys. Rev. Lett.}\ }\textbf
  {\bibinfo {volume} {95}},\ \bibinfo {pages} {226801} (\bibinfo {year}
  {2005})}\BibitemShut {NoStop}%
\bibitem [{\citenamefont {Breunig}\ and\ \citenamefont
  {Ando}(2021)}]{Breunig2022}%
  \BibitemOpen
  \bibfield  {author} {\bibinfo {author} {\bibfnamefont {O.}~\bibnamefont
  {Breunig}}\ and\ \bibinfo {author} {\bibfnamefont {Y.}~\bibnamefont {Ando}},\
  }\bibfield  {title} {\bibinfo {title} {Opportunities in topological insulator
  devices},\ }\href {https://doi.org/10.1038/s42254-021-00402-6} {\bibfield
  {journal} {\bibinfo  {journal} {Nat. Rev. Phys.}\ }\textbf {\bibinfo {volume}
  {4}},\ \bibinfo {pages} {184} (\bibinfo {year} {2021})}\BibitemShut {NoStop}%
\bibitem [{\citenamefont {Hasan}\ and\ \citenamefont {Kane}(2010)}]{Hasan2010}%
  \BibitemOpen
  \bibfield  {author} {\bibinfo {author} {\bibfnamefont {M.~Z.}\ \bibnamefont
  {Hasan}}\ and\ \bibinfo {author} {\bibfnamefont {C.~L.}\ \bibnamefont
  {Kane}},\ }\bibfield  {title} {\bibinfo {title} {{Colloquium: Topological
  insulators}},\ }\href
  {https://doi.org/10.1103/REVMODPHYS.82.3045/FIGURES/20/MEDIUM} {\bibfield
  {journal} {\bibinfo  {journal} {Rev. Mod. Phys.}\ }\textbf {\bibinfo {volume}
  {82}},\ \bibinfo {pages} {3045} (\bibinfo {year} {2010})}\BibitemShut
  {NoStop}%
\bibitem [{\citenamefont {Soluyanov}\ and\ \citenamefont
  {Vanderbilt}(2011)}]{Soluyanov2011}%
  \BibitemOpen
  \bibfield  {author} {\bibinfo {author} {\bibfnamefont {A.~A.}\ \bibnamefont
  {Soluyanov}}\ and\ \bibinfo {author} {\bibfnamefont {D.}~\bibnamefont
  {Vanderbilt}},\ }\bibfield  {title} {\bibinfo {title} {{Wannier
  representation of ${\mathbb{Z}}_{2}$ topological insulators}},\ }\href
  {https://doi.org/10.1103/PhysRevB.83.035108} {\bibfield  {journal} {\bibinfo
  {journal} {Phys. Rev. B}\ }\textbf {\bibinfo {volume} {83}},\ \bibinfo
  {pages} {035108} (\bibinfo {year} {2011})}\BibitemShut {NoStop}%
\bibitem [{\citenamefont {Bradlyn}\ \emph {et~al.}(2017)\citenamefont
  {Bradlyn}, \citenamefont {Elcoro}, \citenamefont {Cano}, \citenamefont
  {Vergniory}, \citenamefont {Wang}, \citenamefont {Felser}, \citenamefont
  {Aroyo},\ and\ \citenamefont {Bernevig}}]{Bradlyn_2017}%
  \BibitemOpen
  \bibfield  {author} {\bibinfo {author} {\bibfnamefont {B.}~\bibnamefont
  {Bradlyn}}, \bibinfo {author} {\bibfnamefont {L.}~\bibnamefont {Elcoro}},
  \bibinfo {author} {\bibfnamefont {J.}~\bibnamefont {Cano}}, \bibinfo {author}
  {\bibfnamefont {M.~G.}\ \bibnamefont {Vergniory}}, \bibinfo {author}
  {\bibfnamefont {Z.}~\bibnamefont {Wang}}, \bibinfo {author} {\bibfnamefont
  {C.}~\bibnamefont {Felser}}, \bibinfo {author} {\bibfnamefont {M.~I.}\
  \bibnamefont {Aroyo}},\ and\ \bibinfo {author} {\bibfnamefont {B.~A.}\
  \bibnamefont {Bernevig}},\ }\bibfield  {title} {\bibinfo {title}
  {{Topological quantum chemistry}},\ }\href
  {https://doi.org/10.1038/nature23268} {\bibfield  {journal} {\bibinfo
  {journal} {Nature}\ }\textbf {\bibinfo {volume} {547}},\ \bibinfo {pages}
  {298} (\bibinfo {year} {2017})}\BibitemShut {NoStop}%
\bibitem [{\citenamefont {Kruthoff}\ \emph {et~al.}(2017)\citenamefont
  {Kruthoff}, \citenamefont {de~Boer}, \citenamefont {van Wezel}, \citenamefont
  {Kane},\ and\ \citenamefont {Slager}}]{Kruthoff_2017}%
  \BibitemOpen
  \bibfield  {author} {\bibinfo {author} {\bibfnamefont {J.}~\bibnamefont
  {Kruthoff}}, \bibinfo {author} {\bibfnamefont {J.}~\bibnamefont {de~Boer}},
  \bibinfo {author} {\bibfnamefont {J.}~\bibnamefont {van Wezel}}, \bibinfo
  {author} {\bibfnamefont {C.~L.}\ \bibnamefont {Kane}},\ and\ \bibinfo
  {author} {\bibfnamefont {R.-J.}\ \bibnamefont {Slager}},\ }\bibfield  {title}
  {\bibinfo {title} {Topological classification of crystalline insulators
  through band structure combinatorics},\ }\href
  {https://doi.org/10.1103/PhysRevX.7.041069} {\bibfield  {journal} {\bibinfo
  {journal} {Phys. Rev. X}\ }\textbf {\bibinfo {volume} {7}},\ \bibinfo {pages}
  {041069} (\bibinfo {year} {2017})}\BibitemShut {NoStop}%
\bibitem [{\citenamefont {Blanco~de Paz}\ \emph {et~al.}(2022)\citenamefont
  {Blanco~de Paz}, \citenamefont {Herrera}, \citenamefont {Arroyo~Huidobro},
  \citenamefont {Alaeian}, \citenamefont {Vergniory}, \citenamefont {Bradlyn},
  \citenamefont {Giedke}, \citenamefont {García-Etxarri},\ and\ \citenamefont
  {Bercioux}}]{BlancodePaz2022}%
  \BibitemOpen
  \bibfield  {author} {\bibinfo {author} {\bibfnamefont {M.}~\bibnamefont
  {Blanco~de Paz}}, \bibinfo {author} {\bibfnamefont {M.~A.~J.}\ \bibnamefont
  {Herrera}}, \bibinfo {author} {\bibfnamefont {P.}~\bibnamefont
  {Arroyo~Huidobro}}, \bibinfo {author} {\bibfnamefont {H.}~\bibnamefont
  {Alaeian}}, \bibinfo {author} {\bibfnamefont {M.~G.}\ \bibnamefont
  {Vergniory}}, \bibinfo {author} {\bibfnamefont {B.}~\bibnamefont {Bradlyn}},
  \bibinfo {author} {\bibfnamefont {G.}~\bibnamefont {Giedke}}, \bibinfo
  {author} {\bibfnamefont {A.}~\bibnamefont {García-Etxarri}},\ and\ \bibinfo
  {author} {\bibfnamefont {D.}~\bibnamefont {Bercioux}},\ }\bibfield  {title}
  {\bibinfo {title} {Energy density as a probe of band representations in
  photonic crystals},\ }\href {https://doi.org/10.1088/1361-648x/ac73cf}
  {\bibfield  {journal} {\bibinfo  {journal} {J. Phys. Condens. Matter}\
  }\textbf {\bibinfo {volume} {34}},\ \bibinfo {pages} {314002} (\bibinfo
  {year} {2022})}\BibitemShut {NoStop}%
\bibitem [{\citenamefont {Khalaf}\ \emph {et~al.}(2021)\citenamefont {Khalaf},
  \citenamefont {Benalcazar}, \citenamefont {Hughes},\ and\ \citenamefont
  {Queiroz}}]{Khalaf_2021}%
  \BibitemOpen
  \bibfield  {author} {\bibinfo {author} {\bibfnamefont {E.}~\bibnamefont
  {Khalaf}}, \bibinfo {author} {\bibfnamefont {W.~A.}\ \bibnamefont
  {Benalcazar}}, \bibinfo {author} {\bibfnamefont {T.~L.}\ \bibnamefont
  {Hughes}},\ and\ \bibinfo {author} {\bibfnamefont {R.}~\bibnamefont
  {Queiroz}},\ }\bibfield  {title} {\bibinfo {title} {Boundary-obstructed
  topological phases},\ }\href
  {https://doi.org/10.1103/PhysRevResearch.3.013239} {\bibfield  {journal}
  {\bibinfo  {journal} {Phys. Rev. Res.}\ }\textbf {\bibinfo {volume} {3}},\
  \bibinfo {pages} {013239} (\bibinfo {year} {2021})}\BibitemShut {NoStop}%
\bibitem [{\citenamefont {Su}\ \emph {et~al.}(1979)\citenamefont {Su},
  \citenamefont {Schrieffer},\ and\ \citenamefont {Heeger}}]{Su1979}%
  \BibitemOpen
  \bibfield  {author} {\bibinfo {author} {\bibfnamefont {W.~P.}\ \bibnamefont
  {Su}}, \bibinfo {author} {\bibfnamefont {J.~R.}\ \bibnamefont {Schrieffer}},\
  and\ \bibinfo {author} {\bibfnamefont {A.~J.}\ \bibnamefont {Heeger}},\
  }\bibfield  {title} {\bibinfo {title} {{Solitons in polyacetylene}},\ }\href
  {https://doi.org/10.1103/PhysRevLett.42.1698} {\bibfield  {journal} {\bibinfo
   {journal} {Phys. Rev. Lett.}\ }\textbf {\bibinfo {volume} {42}},\ \bibinfo
  {pages} {1698} (\bibinfo {year} {1979})}\BibitemShut {NoStop}%
\bibitem [{\citenamefont {Freeney}\ \emph {et~al.}(2020)\citenamefont
  {Freeney}, \citenamefont {van~den Broeke}, \citenamefont {{Harsveld van der
  Veen}}, \citenamefont {Swart},\ and\ \citenamefont {{Morais
  Smith}}}]{Freeney2020}%
  \BibitemOpen
  \bibfield  {author} {\bibinfo {author} {\bibfnamefont {S.~E.}\ \bibnamefont
  {Freeney}}, \bibinfo {author} {\bibfnamefont {J.~J.}\ \bibnamefont {van~den
  Broeke}}, \bibinfo {author} {\bibfnamefont {A.~J.~J.}\ \bibnamefont
  {{Harsveld van der Veen}}}, \bibinfo {author} {\bibfnamefont
  {I.}~\bibnamefont {Swart}},\ and\ \bibinfo {author} {\bibfnamefont
  {C.}~\bibnamefont {{Morais Smith}}},\ }\bibfield  {title} {\bibinfo {title}
  {{Edge-Dependent Topology in Kekul{\'{e}} Lattices}},\ }\href
  {https://doi.org/10.1103/PhysRevLett.124.236404} {\bibfield  {journal}
  {\bibinfo  {journal} {Phys. Rev. Lett.}\ }\textbf {\bibinfo {volume} {124}},\
  \bibinfo {pages} {236404} (\bibinfo {year} {2020})}\BibitemShut {NoStop}%
\bibitem [{\citenamefont {Grushin}(2020)}]{Grushin2020}%
  \BibitemOpen
  \bibfield  {author} {\bibinfo {author} {\bibfnamefont {A.~G.}\ \bibnamefont
  {Grushin}},\ }\href
  {https://grushingroup.cnrs.fr/wp-content/uploads/2024/10/intrototopo_v2.pdf}
  {\emph {\bibinfo {title} {{Introduction to topological Phases in Condensed
  Matter}}}}\ (\bibinfo {year} {2020})\ p.~\bibinfo {pages} {9}\BibitemShut
  {NoStop}%
\bibitem [{\citenamefont {Asb{\'{o}}th}\ \emph {et~al.}(2016)\citenamefont
  {Asb{\'{o}}th}, \citenamefont {Oroszl{\'{a}}ny},\ and\ \citenamefont
  {P{\'{a}}lyi}}]{Asboth2015}%
  \BibitemOpen
  \bibfield  {author} {\bibinfo {author} {\bibfnamefont {J.~K.}\ \bibnamefont
  {Asb{\'{o}}th}}, \bibinfo {author} {\bibfnamefont {L.}~\bibnamefont
  {Oroszl{\'{a}}ny}},\ and\ \bibinfo {author} {\bibfnamefont {A.}~\bibnamefont
  {P{\'{a}}lyi}},\ }\href {https://doi.org/10.1007/978-3-319-25607-8} {\emph
  {\bibinfo {title} {{A Short Course on Topological Insulators}}}},\ \bibinfo
  {series} {Lecture Notes in Physics}, Vol.\ \bibinfo {volume} {919}\ (\bibinfo
   {publisher} {Springer International Publishing},\ \bibinfo {address}
  {Cham},\ \bibinfo {year} {2016})\BibitemShut {NoStop}%
\bibitem [{\citenamefont {Lee}\ \emph {et~al.}(2022)\citenamefont {Lee},
  \citenamefont {Io},\ and\ \citenamefont {chung Kao}}]{Lee2022}%
  \BibitemOpen
  \bibfield  {author} {\bibinfo {author} {\bibfnamefont {C.~S.}\ \bibnamefont
  {Lee}}, \bibinfo {author} {\bibfnamefont {I.~F.}\ \bibnamefont {Io}},\ and\
  \bibinfo {author} {\bibfnamefont {H.}~\bibnamefont {chung Kao}},\ }\bibfield
  {title} {\bibinfo {title} {{Winding number and Zak phase in multi-band SSH
  models}},\ }\href {https://doi.org/10.1016/j.cjph.2022.05.007} {\bibfield
  {journal} {\bibinfo  {journal} {Chin. J. Phys.}\ }\textbf {\bibinfo {volume}
  {78}},\ \bibinfo {pages} {96} (\bibinfo {year} {2022})}\BibitemShut {NoStop}%
\bibitem [{\citenamefont {Chiu}\ \emph {et~al.}(2016)\citenamefont {Chiu},
  \citenamefont {Teo}, \citenamefont {Schnyder},\ and\ \citenamefont
  {Ryu}}]{Chiu2016}%
  \BibitemOpen
  \bibfield  {author} {\bibinfo {author} {\bibfnamefont {C.-K.}\ \bibnamefont
  {Chiu}}, \bibinfo {author} {\bibfnamefont {J.~C.~Y.}\ \bibnamefont {Teo}},
  \bibinfo {author} {\bibfnamefont {A.~P.}\ \bibnamefont {Schnyder}},\ and\
  \bibinfo {author} {\bibfnamefont {S.}~\bibnamefont {Ryu}},\ }\bibfield
  {title} {\bibinfo {title} {Classification of topological quantum matter with
  symmetries},\ }\href {https://doi.org/10.1103/RevModPhys.88.035005}
  {\bibfield  {journal} {\bibinfo  {journal} {Rev. Mod. Phys.}\ }\textbf
  {\bibinfo {volume} {88}},\ \bibinfo {pages} {035005} (\bibinfo {year}
  {2016})}\BibitemShut {NoStop}%
\bibitem [{\citenamefont {Xu}\ \emph {et~al.}(2018)\citenamefont {Xu},
  \citenamefont {Wang}, \citenamefont {Pan}, \citenamefont {Sun}, \citenamefont
  {Xu}, \citenamefont {Chen}, \citenamefont {Tang}, \citenamefont {Gong},
  \citenamefont {Han}, \citenamefont {Li},\ and\ \citenamefont {Guo}}]{Xu2018}%
  \BibitemOpen
  \bibfield  {author} {\bibinfo {author} {\bibfnamefont {X.~Y.}\ \bibnamefont
  {Xu}}, \bibinfo {author} {\bibfnamefont {Q.~Q.}\ \bibnamefont {Wang}},
  \bibinfo {author} {\bibfnamefont {W.~W.}\ \bibnamefont {Pan}}, \bibinfo
  {author} {\bibfnamefont {K.}~\bibnamefont {Sun}}, \bibinfo {author}
  {\bibfnamefont {J.~S.}\ \bibnamefont {Xu}}, \bibinfo {author} {\bibfnamefont
  {G.}~\bibnamefont {Chen}}, \bibinfo {author} {\bibfnamefont {J.~S.}\
  \bibnamefont {Tang}}, \bibinfo {author} {\bibfnamefont {M.}~\bibnamefont
  {Gong}}, \bibinfo {author} {\bibfnamefont {Y.~J.}\ \bibnamefont {Han}},
  \bibinfo {author} {\bibfnamefont {C.~F.}\ \bibnamefont {Li}},\ and\ \bibinfo
  {author} {\bibfnamefont {G.~C.}\ \bibnamefont {Guo}},\ }\bibfield  {title}
  {\bibinfo {title} {{Measuring the Winding Number in a Large-Scale Chiral
  Quantum Walk}},\ }\href {https://doi.org/10.1103/PhysRevLett.120.260501}
  {\bibfield  {journal} {\bibinfo  {journal} {Phys. Rev. Lett.}\ }\textbf
  {\bibinfo {volume} {120}},\ \bibinfo {pages} {260501} (\bibinfo {year}
  {2018})}\BibitemShut {NoStop}%
\bibitem [{\citenamefont {Schine}\ \emph {et~al.}(2019)\citenamefont {Schine},
  \citenamefont {Chalupnik}, \citenamefont {Can}, \citenamefont {Gromov},\ and\
  \citenamefont {Simon}}]{Schine2019}%
  \BibitemOpen
  \bibfield  {author} {\bibinfo {author} {\bibfnamefont {N.}~\bibnamefont
  {Schine}}, \bibinfo {author} {\bibfnamefont {M.}~\bibnamefont {Chalupnik}},
  \bibinfo {author} {\bibfnamefont {T.}~\bibnamefont {Can}}, \bibinfo {author}
  {\bibfnamefont {A.}~\bibnamefont {Gromov}},\ and\ \bibinfo {author}
  {\bibfnamefont {J.}~\bibnamefont {Simon}},\ }\bibfield  {title} {\bibinfo
  {title} {{Electromagnetic and gravitational responses of photonic Landau
  levels}},\ }\href {https://doi.org/10.1038/s41586-018-0817-4} {\bibfield
  {journal} {\bibinfo  {journal} {Nature}\ }\textbf {\bibinfo {volume} {565}},\
  \bibinfo {pages} {173} (\bibinfo {year} {2019})}\BibitemShut {NoStop}%
\bibitem [{\citenamefont {Ozawa}\ \emph {et~al.}(2019)\citenamefont {Ozawa},
  \citenamefont {Price}, \citenamefont {Amo}, \citenamefont {Goldman},
  \citenamefont {Hafezi}, \citenamefont {Lu}, \citenamefont {Rechtsman},
  \citenamefont {Schuster}, \citenamefont {Simon}, \citenamefont {Zilberberg},\
  and\ \citenamefont {Carusotto}}]{Ozawa2019}%
  \BibitemOpen
  \bibfield  {author} {\bibinfo {author} {\bibfnamefont {T.}~\bibnamefont
  {Ozawa}}, \bibinfo {author} {\bibfnamefont {H.~M.}\ \bibnamefont {Price}},
  \bibinfo {author} {\bibfnamefont {A.}~\bibnamefont {Amo}}, \bibinfo {author}
  {\bibfnamefont {N.}~\bibnamefont {Goldman}}, \bibinfo {author} {\bibfnamefont
  {M.}~\bibnamefont {Hafezi}}, \bibinfo {author} {\bibfnamefont
  {L.}~\bibnamefont {Lu}}, \bibinfo {author} {\bibfnamefont {M.~C.}\
  \bibnamefont {Rechtsman}}, \bibinfo {author} {\bibfnamefont {D.}~\bibnamefont
  {Schuster}}, \bibinfo {author} {\bibfnamefont {J.}~\bibnamefont {Simon}},
  \bibinfo {author} {\bibfnamefont {O.}~\bibnamefont {Zilberberg}},\ and\
  \bibinfo {author} {\bibfnamefont {I.}~\bibnamefont {Carusotto}},\ }\bibfield
  {title} {\bibinfo {title} {{Topological photonics}},\ }\href
  {https://doi.org/10.1103/RevModPhys.91.015006} {\bibfield  {journal}
  {\bibinfo  {journal} {Rev. Mod. Phys.}\ }\textbf {\bibinfo {volume} {91}},\
  \bibinfo {pages} {015006} (\bibinfo {year} {2019})}\BibitemShut {NoStop}%
\bibitem [{\citenamefont {{Blanco De Paz}}\ \emph {et~al.}(2019)\citenamefont
  {{Blanco De Paz}}, \citenamefont {Vergniory}, \citenamefont {Dario},
  \citenamefont {Garci{\'{a}}-Etxarri},\ and\ \citenamefont
  {Bradlyn}}]{DePaz2019}%
  \BibitemOpen
  \bibfield  {author} {\bibinfo {author} {\bibfnamefont {M.}~\bibnamefont
  {{Blanco De Paz}}}, \bibinfo {author} {\bibfnamefont {M.~G.}\ \bibnamefont
  {Vergniory}}, \bibinfo {author} {\bibnamefont {Dario}}, \bibinfo {author}
  {\bibfnamefont {A.}~\bibnamefont {Garci{\'{a}}-Etxarri}},\ and\ \bibinfo
  {author} {\bibfnamefont {B.}~\bibnamefont {Bradlyn}},\ }\bibfield  {title}
  {\bibinfo {title} {{Engineering fragile topology in photonic crystals:
  Topological quantum chemistry of light}},\ }\href
  {https://doi.org/10.1103/PhysRevResearch.1.032005} {\bibfield  {journal}
  {\bibinfo  {journal} {Phys. Rev. Res.}\ }\textbf {\bibinfo {volume} {1}},\
  \bibinfo {pages} {1} (\bibinfo {year} {2019})}\BibitemShut {NoStop}%
\bibitem [{\citenamefont {Atala}\ \emph {et~al.}(2013)\citenamefont {Atala},
  \citenamefont {Aidelsburger}, \citenamefont {Barreiro}, \citenamefont
  {Abanin}, \citenamefont {Kitagawa}, \citenamefont {Demler},\ and\
  \citenamefont {Bloch}}]{Atala2013}%
  \BibitemOpen
  \bibfield  {author} {\bibinfo {author} {\bibfnamefont {M.}~\bibnamefont
  {Atala}}, \bibinfo {author} {\bibfnamefont {M.}~\bibnamefont {Aidelsburger}},
  \bibinfo {author} {\bibfnamefont {J.~T.}\ \bibnamefont {Barreiro}}, \bibinfo
  {author} {\bibfnamefont {D.}~\bibnamefont {Abanin}}, \bibinfo {author}
  {\bibfnamefont {T.}~\bibnamefont {Kitagawa}}, \bibinfo {author}
  {\bibfnamefont {E.}~\bibnamefont {Demler}},\ and\ \bibinfo {author}
  {\bibfnamefont {I.}~\bibnamefont {Bloch}},\ }\bibfield  {title} {\bibinfo
  {title} {{Direct measurement of the Zak phase in topological Bloch bands}},\
  }\href {https://doi.org/10.1038/nphys2790} {\bibfield  {journal} {\bibinfo
  {journal} {Nat. Phys.}\ }\textbf {\bibinfo {volume} {9}},\ \bibinfo {pages}
  {795 } (\bibinfo {year} {2013})}\BibitemShut {NoStop}%
\bibitem [{\citenamefont {Thouless}\ \emph {et~al.}(1982)\citenamefont
  {Thouless}, \citenamefont {Kohmoto}, \citenamefont {Nightingale},\ and\
  \citenamefont {den Nijs}}]{Thouless_1982}%
  \BibitemOpen
  \bibfield  {author} {\bibinfo {author} {\bibfnamefont {D.~J.}\ \bibnamefont
  {Thouless}}, \bibinfo {author} {\bibfnamefont {M.}~\bibnamefont {Kohmoto}},
  \bibinfo {author} {\bibfnamefont {M.~P.}\ \bibnamefont {Nightingale}},\ and\
  \bibinfo {author} {\bibfnamefont {M.}~\bibnamefont {den Nijs}},\ }\bibfield
  {title} {\bibinfo {title} {{Quantized Hall Conductance in a Two-Dimensional
  Periodic Potential}},\ }\href {https://doi.org/10.1103/PhysRevLett.49.405}
  {\bibfield  {journal} {\bibinfo  {journal} {Phys. Rev. Lett.}\ }\textbf
  {\bibinfo {volume} {49}},\ \bibinfo {pages} {405} (\bibinfo {year}
  {1982})}\BibitemShut {NoStop}%
\bibitem [{\citenamefont {Zhao}\ \emph {et~al.}(2020)\citenamefont {Zhao},
  \citenamefont {Zhang}, \citenamefont {Mei}, \citenamefont {Zhou},
  \citenamefont {Yi}, \citenamefont {Zhang}, \citenamefont {Yu}, \citenamefont
  {Xiao}, \citenamefont {Wang}, \citenamefont {Samarth}, \citenamefont {Chan},
  \citenamefont {Liu},\ and\ \citenamefont {Chang}}]{Zhao2020}%
  \BibitemOpen
  \bibfield  {author} {\bibinfo {author} {\bibfnamefont {Y.~F.}\ \bibnamefont
  {Zhao}}, \bibinfo {author} {\bibfnamefont {R.}~\bibnamefont {Zhang}},
  \bibinfo {author} {\bibfnamefont {R.}~\bibnamefont {Mei}}, \bibinfo {author}
  {\bibfnamefont {L.~J.}\ \bibnamefont {Zhou}}, \bibinfo {author}
  {\bibfnamefont {H.}~\bibnamefont {Yi}}, \bibinfo {author} {\bibfnamefont
  {Y.~Q.}\ \bibnamefont {Zhang}}, \bibinfo {author} {\bibfnamefont
  {J.}~\bibnamefont {Yu}}, \bibinfo {author} {\bibfnamefont {R.}~\bibnamefont
  {Xiao}}, \bibinfo {author} {\bibfnamefont {K.}~\bibnamefont {Wang}}, \bibinfo
  {author} {\bibfnamefont {N.}~\bibnamefont {Samarth}}, \bibinfo {author}
  {\bibfnamefont {M.~H.}\ \bibnamefont {Chan}}, \bibinfo {author}
  {\bibfnamefont {C.~X.}\ \bibnamefont {Liu}},\ and\ \bibinfo {author}
  {\bibfnamefont {C.~Z.}\ \bibnamefont {Chang}},\ }\bibfield  {title} {\bibinfo
  {title} {{Tuning the Chern number in quantum anomalous Hall insulators}},\
  }\href {https://doi.org/10.1038/s41586-020-3020-3} {\bibfield  {journal}
  {\bibinfo  {journal} {Nature}\ }\textbf {\bibinfo {volume} {588}},\ \bibinfo
  {pages} {419} (\bibinfo {year} {2020})}\BibitemShut {NoStop}%
\bibitem [{\citenamefont {Leis}\ \emph {et~al.}(2022)\citenamefont {Leis},
  \citenamefont {Schleenvoigt}, \citenamefont {Moors}, \citenamefont {Soltner},
  \citenamefont {Cherepanov}, \citenamefont {Sch{\"{u}}ffelgen}, \citenamefont
  {Mussler}, \citenamefont {Gr{\"{u}}tzmacher}, \citenamefont
  {Voigtl{\"{a}}nder}, \citenamefont {L{\"{u}}pke},\ and\ \citenamefont
  {Tautz}}]{Leis2022}%
  \BibitemOpen
  \bibfield  {author} {\bibinfo {author} {\bibfnamefont {A.}~\bibnamefont
  {Leis}}, \bibinfo {author} {\bibfnamefont {M.}~\bibnamefont {Schleenvoigt}},
  \bibinfo {author} {\bibfnamefont {K.}~\bibnamefont {Moors}}, \bibinfo
  {author} {\bibfnamefont {H.}~\bibnamefont {Soltner}}, \bibinfo {author}
  {\bibfnamefont {V.}~\bibnamefont {Cherepanov}}, \bibinfo {author}
  {\bibfnamefont {P.}~\bibnamefont {Sch{\"{u}}ffelgen}}, \bibinfo {author}
  {\bibfnamefont {G.}~\bibnamefont {Mussler}}, \bibinfo {author} {\bibfnamefont
  {D.}~\bibnamefont {Gr{\"{u}}tzmacher}}, \bibinfo {author} {\bibfnamefont
  {B.}~\bibnamefont {Voigtl{\"{a}}nder}}, \bibinfo {author} {\bibfnamefont
  {F.}~\bibnamefont {L{\"{u}}pke}},\ and\ \bibinfo {author} {\bibfnamefont
  {F.~S.}\ \bibnamefont {Tautz}},\ }\bibfield  {title} {\bibinfo {title}
  {{Probing Edge State Conductance in Ultra-Thin Topological Insulator
  Films}},\ }\href {https://doi.org/10.1002/qute.202200043} {\bibfield
  {journal} {\bibinfo  {journal} {Adv. Quantum Technol.}\ }\textbf {\bibinfo
  {volume} {5}},\ \bibinfo {pages} {2200043} (\bibinfo {year}
  {2022})}\BibitemShut {NoStop}%
\bibitem [{\citenamefont {Nadj-Perge}\ \emph {et~al.}(2014)\citenamefont
  {Nadj-Perge}, \citenamefont {Drozdov}, \citenamefont {Li}, \citenamefont
  {Chen}, \citenamefont {Jeon}, \citenamefont {Seo}, \citenamefont {MacDonald},
  \citenamefont {Bernevig},\ and\ \citenamefont {Yazdani}}]{Nadj-Perge2014}%
  \BibitemOpen
  \bibfield  {author} {\bibinfo {author} {\bibfnamefont {S.}~\bibnamefont
  {Nadj-Perge}}, \bibinfo {author} {\bibfnamefont {I.~K.}\ \bibnamefont
  {Drozdov}}, \bibinfo {author} {\bibfnamefont {J.}~\bibnamefont {Li}},
  \bibinfo {author} {\bibfnamefont {H.}~\bibnamefont {Chen}}, \bibinfo {author}
  {\bibfnamefont {S.}~\bibnamefont {Jeon}}, \bibinfo {author} {\bibfnamefont
  {J.}~\bibnamefont {Seo}}, \bibinfo {author} {\bibfnamefont {A.~H.}\
  \bibnamefont {MacDonald}}, \bibinfo {author} {\bibfnamefont {B.~A.}\
  \bibnamefont {Bernevig}},\ and\ \bibinfo {author} {\bibfnamefont
  {A.}~\bibnamefont {Yazdani}},\ }\bibfield  {title} {\bibinfo {title}
  {{Observation of Majorana fermions in ferromagnetic atomic chains on a
  superconductor}},\ }\href {https://doi.org/10.1126/science.1259327}
  {\bibfield  {journal} {\bibinfo  {journal} {Science}\ }\textbf {\bibinfo
  {volume} {346}},\ \bibinfo {pages} {602} (\bibinfo {year}
  {2014})}\BibitemShut {NoStop}%
\bibitem [{\citenamefont {Schneider}\ \emph {et~al.}(2022)\citenamefont
  {Schneider}, \citenamefont {Beck}, \citenamefont {Neuhaus-Steinmetz},
  \citenamefont {R{\'{o}}zsa}, \citenamefont {Posske}, \citenamefont {Wiebe},\
  and\ \citenamefont {Wiesendanger}}]{Schneider2022}%
  \BibitemOpen
  \bibfield  {author} {\bibinfo {author} {\bibfnamefont {L.}~\bibnamefont
  {Schneider}}, \bibinfo {author} {\bibfnamefont {P.}~\bibnamefont {Beck}},
  \bibinfo {author} {\bibfnamefont {J.}~\bibnamefont {Neuhaus-Steinmetz}},
  \bibinfo {author} {\bibfnamefont {L.}~\bibnamefont {R{\'{o}}zsa}}, \bibinfo
  {author} {\bibfnamefont {T.}~\bibnamefont {Posske}}, \bibinfo {author}
  {\bibfnamefont {J.}~\bibnamefont {Wiebe}},\ and\ \bibinfo {author}
  {\bibfnamefont {R.}~\bibnamefont {Wiesendanger}},\ }\bibfield  {title}
  {\bibinfo {title} {{Precursors of Majorana modes and their length-dependent
  energy oscillations probed at both ends of atomic Shiba chains}},\ }\href
  {https://doi.org/10.1038/s41565-022-01078-4} {\bibfield  {journal} {\bibinfo
  {journal} {Nat. Nanotechnol.}\ }\textbf {\bibinfo {volume} {17}},\ \bibinfo
  {pages} {384} (\bibinfo {year} {2022})}\BibitemShut {NoStop}%
\bibitem [{\citenamefont {Kezilebieke}\ \emph {et~al.}(2020)\citenamefont
  {Kezilebieke}, \citenamefont {Huda}, \citenamefont {Vano}, \citenamefont
  {Aapro}, \citenamefont {Ganguli}, \citenamefont {Silveira}, \citenamefont
  {G{\l}odzik}, \citenamefont {Foster}, \citenamefont {Ojanen},\ and\
  \citenamefont {Liljeroth}}]{Kezilebieke2020}%
  \BibitemOpen
  \bibfield  {author} {\bibinfo {author} {\bibfnamefont {S.}~\bibnamefont
  {Kezilebieke}}, \bibinfo {author} {\bibfnamefont {M.~N.}\ \bibnamefont
  {Huda}}, \bibinfo {author} {\bibfnamefont {V.}~\bibnamefont {Vano}}, \bibinfo
  {author} {\bibfnamefont {M.}~\bibnamefont {Aapro}}, \bibinfo {author}
  {\bibfnamefont {S.~C.}\ \bibnamefont {Ganguli}}, \bibinfo {author}
  {\bibfnamefont {O.~J.}\ \bibnamefont {Silveira}}, \bibinfo {author}
  {\bibfnamefont {S.}~\bibnamefont {G{\l}odzik}}, \bibinfo {author}
  {\bibfnamefont {A.~S.}\ \bibnamefont {Foster}}, \bibinfo {author}
  {\bibfnamefont {T.}~\bibnamefont {Ojanen}},\ and\ \bibinfo {author}
  {\bibfnamefont {P.}~\bibnamefont {Liljeroth}},\ }\bibfield  {title} {\bibinfo
  {title} {{Topological superconductivity in a van der Waals
  heterostructure}},\ }\href {https://doi.org/10.1038/s41586-020-2989-y}
  {\bibfield  {journal} {\bibinfo  {journal} {Nature}\ }\textbf {\bibinfo
  {volume} {588}},\ \bibinfo {pages} {424} (\bibinfo {year}
  {2020})}\BibitemShut {NoStop}%
\bibitem [{\citenamefont {Moes}\ \emph {et~al.}(2023)\citenamefont {Moes},
  \citenamefont {Vliem}, \citenamefont {de~Melo}, \citenamefont {Wigmans},
  \citenamefont {Botello-M{\'{e}}ndez}, \citenamefont {Mendes}, \citenamefont
  {van Brenk}, \citenamefont {Swart}, \citenamefont {{Maisel Licer{\'{a}}n}},
  \citenamefont {Stoof}, \citenamefont {Delerue}, \citenamefont {Zanolli},\
  and\ \citenamefont {Vanmaekelbergh}}]{Moes2023}%
  \BibitemOpen
  \bibfield  {author} {\bibinfo {author} {\bibfnamefont {J.~R.}\ \bibnamefont
  {Moes}}, \bibinfo {author} {\bibfnamefont {J.~F.}\ \bibnamefont {Vliem}},
  \bibinfo {author} {\bibfnamefont {P.~M.}\ \bibnamefont {de~Melo}}, \bibinfo
  {author} {\bibfnamefont {T.~C.}\ \bibnamefont {Wigmans}}, \bibinfo {author}
  {\bibfnamefont {A.~R.}\ \bibnamefont {Botello-M{\'{e}}ndez}}, \bibinfo
  {author} {\bibfnamefont {R.~G.}\ \bibnamefont {Mendes}}, \bibinfo {author}
  {\bibfnamefont {E.~F.}\ \bibnamefont {van Brenk}}, \bibinfo {author}
  {\bibfnamefont {I.}~\bibnamefont {Swart}}, \bibinfo {author} {\bibfnamefont
  {L.}~\bibnamefont {{Maisel Licer{\'{a}}n}}}, \bibinfo {author} {\bibfnamefont
  {H.~T.}\ \bibnamefont {Stoof}}, \bibinfo {author} {\bibfnamefont
  {C.}~\bibnamefont {Delerue}}, \bibinfo {author} {\bibfnamefont
  {Z.}~\bibnamefont {Zanolli}},\ and\ \bibinfo {author} {\bibfnamefont
  {D.}~\bibnamefont {Vanmaekelbergh}},\ }\bibfield  {title} {\bibinfo {title}
  {{Characterization of the Edge States in Colloidal Bi$_2$Se$_3$ Platelets}},\
  }\href {https://doi.org/10.1021/acs.nanolett.3c04460} {\bibfield  {journal}
  {\bibinfo  {journal} {Nano Lett.}\ }\textbf {\bibinfo {volume} {45}},\
  \bibinfo {pages} {43} (\bibinfo {year} {2023})}\BibitemShut {NoStop}%
\bibitem [{\citenamefont {Vanderbilt}\ and\ \citenamefont
  {King-Smith}(1993)}]{Vanderbilt_1993}%
  \BibitemOpen
  \bibfield  {author} {\bibinfo {author} {\bibfnamefont {D.}~\bibnamefont
  {Vanderbilt}}\ and\ \bibinfo {author} {\bibfnamefont {R.~D.}\ \bibnamefont
  {King-Smith}},\ }\bibfield  {title} {\bibinfo {title} {Electric polarization
  as a bulk quantity and its relation to surface charge},\ }\href
  {https://doi.org/10.1103/PhysRevB.48.4442} {\bibfield  {journal} {\bibinfo
  {journal} {Phys. Rev. B}\ }\textbf {\bibinfo {volume} {48}},\ \bibinfo
  {pages} {4442} (\bibinfo {year} {1993})}\BibitemShut {NoStop}%
\bibitem [{\citenamefont {Resta}(1994)}]{Resta_1994}%
  \BibitemOpen
  \bibfield  {author} {\bibinfo {author} {\bibfnamefont {R.}~\bibnamefont
  {Resta}},\ }\bibfield  {title} {\bibinfo {title} {Macroscopic polarization in
  crystalline dielectrics: the geometric phase approach},\ }\href
  {https://doi.org/10.1103/RevModPhys.66.899} {\bibfield  {journal} {\bibinfo
  {journal} {Rev. Mod. Phys.}\ }\textbf {\bibinfo {volume} {66}},\ \bibinfo
  {pages} {899} (\bibinfo {year} {1994})}\BibitemShut {NoStop}%
\bibitem [{\citenamefont {Neupert}\ and\ \citenamefont
  {Schindler}(2018)}]{Neupert_2018}%
  \BibitemOpen
  \bibfield  {author} {\bibinfo {author} {\bibfnamefont {T.}~\bibnamefont
  {Neupert}}\ and\ \bibinfo {author} {\bibfnamefont {F.}~\bibnamefont
  {Schindler}},\ }\bibinfo {title} {Topological crystalline insulators},\ in\
  \href {https://doi.org/10.1007/978-3-319-76388-0_2} {\emph {\bibinfo
  {booktitle} {Topological Matter: Lectures from the Topological Matter School
  2017}}},\ \bibinfo {editor} {edited by\ \bibinfo {editor} {\bibfnamefont
  {D.}~\bibnamefont {Bercioux}}, \bibinfo {editor} {\bibfnamefont
  {J.}~\bibnamefont {Cayssol}}, \bibinfo {editor} {\bibfnamefont {M.~G.}\
  \bibnamefont {Vergniory}},\ and\ \bibinfo {editor} {\bibfnamefont
  {M.}~\bibnamefont {Reyes~Calvo}}}\ (\bibinfo  {publisher} {Springer
  International Publishing},\ \bibinfo {address} {Cham},\ \bibinfo {year}
  {2018})\ pp.\ \bibinfo {pages} {31--61}\BibitemShut {NoStop}%
\bibitem [{\citenamefont {Vanderbilt}(2018)}]{Vanderbilt2018}%
  \BibitemOpen
  \bibfield  {author} {\bibinfo {author} {\bibfnamefont {D.}~\bibnamefont
  {Vanderbilt}},\ }\href {https://doi.org/10.1017/9781316662205} {\emph
  {\bibinfo {title} {Berry Phases in Electronic Structure Theory: Electric
  Polarization, Orbital Magnetization and Topological Insulators}}}\ (\bibinfo
  {publisher} {Cambridge University Press},\ \bibinfo {year}
  {2018})\BibitemShut {NoStop}%
\bibitem [{\citenamefont {Rice}\ and\ \citenamefont {Mele}(1982)}]{Rice_1982}%
  \BibitemOpen
  \bibfield  {author} {\bibinfo {author} {\bibfnamefont {M.~J.}\ \bibnamefont
  {Rice}}\ and\ \bibinfo {author} {\bibfnamefont {E.~J.}\ \bibnamefont
  {Mele}},\ }\bibfield  {title} {\bibinfo {title} {Elementary excitations of a
  linearly conjugated diatomic polymer},\ }\href
  {https://doi.org/10.1103/PhysRevLett.49.1455} {\bibfield  {journal} {\bibinfo
   {journal} {Phys. Rev. Lett.}\ }\textbf {\bibinfo {volume} {49}},\ \bibinfo
  {pages} {1455} (\bibinfo {year} {1982})}\BibitemShut {NoStop}%
\bibitem [{\citenamefont {Allen}\ \emph {et~al.}(2022)\citenamefont {Allen},
  \citenamefont {Gibbons}, \citenamefont {Sherlock}, \citenamefont
  {Stanfield},\ and\ \citenamefont {McCann}}]{Allen_2022}%
  \BibitemOpen
  \bibfield  {author} {\bibinfo {author} {\bibfnamefont {R.~E.~J.}\
  \bibnamefont {Allen}}, \bibinfo {author} {\bibfnamefont {H.~V.}\ \bibnamefont
  {Gibbons}}, \bibinfo {author} {\bibfnamefont {A.~M.}\ \bibnamefont
  {Sherlock}}, \bibinfo {author} {\bibfnamefont {H.~R.~M.}\ \bibnamefont
  {Stanfield}},\ and\ \bibinfo {author} {\bibfnamefont {E.}~\bibnamefont
  {McCann}},\ }\bibfield  {title} {\bibinfo {title} {{Nonsymmorphic chiral
  symmetry and solitons in the Rice-Mele model}},\ }\href
  {https://doi.org/10.1103/PhysRevB.106.165409} {\bibfield  {journal} {\bibinfo
   {journal} {Phys. Rev. B}\ }\textbf {\bibinfo {volume} {106}},\ \bibinfo
  {pages} {165409} (\bibinfo {year} {2022})}\BibitemShut {NoStop}%
\bibitem [{\citenamefont {Su}\ and\ \citenamefont {Schrieffer}(1981)}]{Su1981}%
  \BibitemOpen
  \bibfield  {author} {\bibinfo {author} {\bibfnamefont {W.~P.}\ \bibnamefont
  {Su}}\ and\ \bibinfo {author} {\bibfnamefont {J.~R.}\ \bibnamefont
  {Schrieffer}},\ }\bibfield  {title} {\bibinfo {title} {{Fractionally charged
  excitations in charge-density-wave systems with commensurability 3}},\ }\href
  {https://doi.org/10.1103/PhysRevLett.46.738} {\bibfield  {journal} {\bibinfo
  {journal} {Phys. Rev. Lett.}\ }\textbf {\bibinfo {volume} {46}},\ \bibinfo
  {pages} {738} (\bibinfo {year} {1981})}\BibitemShut {NoStop}%
\bibitem [{\citenamefont {{Martinez Alvarez}}\ and\ \citenamefont
  {Coutinho-Filho}(2019)}]{MartinezAlvarez2019}%
  \BibitemOpen
  \bibfield  {author} {\bibinfo {author} {\bibfnamefont {V.~M.}\ \bibnamefont
  {{Martinez Alvarez}}}\ and\ \bibinfo {author} {\bibfnamefont {M.~D.}\
  \bibnamefont {Coutinho-Filho}},\ }\bibfield  {title} {\bibinfo {title} {{Edge
  states in trimer lattices}},\ }\href
  {https://doi.org/10.1103/PhysRevA.99.013833} {\bibfield  {journal} {\bibinfo
  {journal} {Phys. Rev. A}\ }\textbf {\bibinfo {volume} {99}},\ \bibinfo
  {pages} {013833} (\bibinfo {year} {2019})}\BibitemShut {NoStop}%
\bibitem [{\citenamefont {Meier}\ \emph {et~al.}(2016)\citenamefont {Meier},
  \citenamefont {An},\ and\ \citenamefont {Gadway}}]{Meier2016}%
  \BibitemOpen
  \bibfield  {author} {\bibinfo {author} {\bibfnamefont {E.~J.}\ \bibnamefont
  {Meier}}, \bibinfo {author} {\bibfnamefont {F.~A.}\ \bibnamefont {An}},\ and\
  \bibinfo {author} {\bibfnamefont {B.}~\bibnamefont {Gadway}},\ }\bibfield
  {title} {\bibinfo {title} {{Observation of the topological soliton state in
  the Su-Schrieffer-Heeger model}},\ }\href
  {https://doi.org/10.1038/ncomms13986} {\bibfield  {journal} {\bibinfo
  {journal} {Nat. Commun.}\ }\textbf {\bibinfo {volume} {7}},\ \bibinfo {pages}
  {1} (\bibinfo {year} {2016})}\BibitemShut {NoStop}%
\bibitem [{\citenamefont {Drost}\ \emph {et~al.}(2017)\citenamefont {Drost},
  \citenamefont {Ojanen}, \citenamefont {Harju},\ and\ \citenamefont
  {Liljeroth}}]{Drost2017a}%
  \BibitemOpen
  \bibfield  {author} {\bibinfo {author} {\bibfnamefont {R.}~\bibnamefont
  {Drost}}, \bibinfo {author} {\bibfnamefont {T.}~\bibnamefont {Ojanen}},
  \bibinfo {author} {\bibfnamefont {A.}~\bibnamefont {Harju}},\ and\ \bibinfo
  {author} {\bibfnamefont {P.}~\bibnamefont {Liljeroth}},\ }\bibfield  {title}
  {\bibinfo {title} {{Topological states in engineered atomic lattices}},\
  }\href {https://doi.org/10.1038/nphys4080} {\bibfield  {journal} {\bibinfo
  {journal} {Nat. Phys.}\ }\textbf {\bibinfo {volume} {13}},\ \bibinfo {pages}
  {668} (\bibinfo {year} {2017})}\BibitemShut {NoStop}%
\bibitem [{\citenamefont {Pham}\ \emph {et~al.}(2022)\citenamefont {Pham},
  \citenamefont {Pan}, \citenamefont {Erwin}, \citenamefont {{Von Oppen}},
  \citenamefont {Kanisawa},\ and\ \citenamefont {F{\"{o}}lsch}}]{Pham2022}%
  \BibitemOpen
  \bibfield  {author} {\bibinfo {author} {\bibfnamefont {V.~D.}\ \bibnamefont
  {Pham}}, \bibinfo {author} {\bibfnamefont {Y.}~\bibnamefont {Pan}}, \bibinfo
  {author} {\bibfnamefont {S.~C.}\ \bibnamefont {Erwin}}, \bibinfo {author}
  {\bibfnamefont {F.}~\bibnamefont {{Von Oppen}}}, \bibinfo {author}
  {\bibfnamefont {K.}~\bibnamefont {Kanisawa}},\ and\ \bibinfo {author}
  {\bibfnamefont {S.}~\bibnamefont {F{\"{o}}lsch}},\ }\bibfield  {title}
  {\bibinfo {title} {{Topological states in dimerized quantum-dot chains
  created by atom manipulation}},\ }\href
  {https://doi.org/10.1103/PhysRevB.105.125418} {\bibfield  {journal} {\bibinfo
   {journal} {Phys. Rev. B}\ }\textbf {\bibinfo {volume} {105}},\ \bibinfo
  {pages} {1} (\bibinfo {year} {2022})}\BibitemShut {NoStop}%
\bibitem [{\citenamefont {Ligthart}\ \emph {et~al.}(2024)\citenamefont
  {Ligthart}, \citenamefont {Coinon}, \citenamefont {Desplanque}, \citenamefont
  {Wallart},\ and\ \citenamefont {Swart}}]{Ligthart2024}%
  \BibitemOpen
  \bibfield  {author} {\bibinfo {author} {\bibfnamefont {R.}~\bibnamefont
  {Ligthart}}, \bibinfo {author} {\bibfnamefont {C.}~\bibnamefont {Coinon}},
  \bibinfo {author} {\bibfnamefont {L.}~\bibnamefont {Desplanque}}, \bibinfo
  {author} {\bibfnamefont {X.}~\bibnamefont {Wallart}},\ and\ \bibinfo {author}
  {\bibfnamefont {I.}~\bibnamefont {Swart}},\ }\bibfield  {title} {\bibinfo
  {title} {{Vertical and lateral manipulation of single Cs atoms on the
  semiconductor InAs(111)A}},\ }\href
  {https://doi.org/10.21468/scipostphys.16.4.096} {\bibfield  {journal}
  {\bibinfo  {journal} {SciPost Physics}\ }\textbf {\bibinfo {volume} {16}},\
  \bibinfo {pages} {096} (\bibinfo {year} {2024})}\BibitemShut {NoStop}%
\bibitem [{\citenamefont {Gomes}\ \emph {et~al.}(2012)\citenamefont {Gomes},
  \citenamefont {Mar}, \citenamefont {Ko}, \citenamefont {Guinea},\ and\
  \citenamefont {Manoharan}}]{Gomes2012a}%
  \BibitemOpen
  \bibfield  {author} {\bibinfo {author} {\bibfnamefont {K.~K.}\ \bibnamefont
  {Gomes}}, \bibinfo {author} {\bibfnamefont {W.}~\bibnamefont {Mar}}, \bibinfo
  {author} {\bibfnamefont {W.}~\bibnamefont {Ko}}, \bibinfo {author}
  {\bibfnamefont {F.}~\bibnamefont {Guinea}},\ and\ \bibinfo {author}
  {\bibfnamefont {H.~C.}\ \bibnamefont {Manoharan}},\ }\bibfield  {title}
  {\bibinfo {title} {{Designer Dirac fermions and topological phases in
  molecular graphene}},\ }\href {https://doi.org/10.1038/nature10941}
  {\bibfield  {journal} {\bibinfo  {journal} {Nature}\ }\textbf {\bibinfo
  {volume} {483}},\ \bibinfo {pages} {306} (\bibinfo {year}
  {2012})}\BibitemShut {NoStop}%
\bibitem [{\citenamefont {Kempkes}\ \emph
  {et~al.}(2019{\natexlab{a}})\citenamefont {Kempkes}, \citenamefont {Slot},
  \citenamefont {Freeney}, \citenamefont {Zevenhuizen}, \citenamefont
  {Vanmaekelbergh}, \citenamefont {Swart},\ and\ \citenamefont {{Morais
  Smith}}}]{Kempkes2018a}%
  \BibitemOpen
  \bibfield  {author} {\bibinfo {author} {\bibfnamefont {S.~N.}\ \bibnamefont
  {Kempkes}}, \bibinfo {author} {\bibfnamefont {M.~R.}\ \bibnamefont {Slot}},
  \bibinfo {author} {\bibfnamefont {S.~E.}\ \bibnamefont {Freeney}}, \bibinfo
  {author} {\bibfnamefont {S.~J.~M.}\ \bibnamefont {Zevenhuizen}}, \bibinfo
  {author} {\bibfnamefont {D.}~\bibnamefont {Vanmaekelbergh}}, \bibinfo
  {author} {\bibfnamefont {I.}~\bibnamefont {Swart}},\ and\ \bibinfo {author}
  {\bibfnamefont {C.}~\bibnamefont {{Morais Smith}}},\ }\bibfield  {title}
  {\bibinfo {title} {{Design and characterization of electrons in a fractal
  geometry}},\ }\href {https://doi.org/10.1038/s41567-018-0328-0} {\bibfield
  {journal} {\bibinfo  {journal} {Nat. Phys.}\ }\textbf {\bibinfo {volume}
  {15}},\ \bibinfo {pages} {127} (\bibinfo {year}
  {2019}{\natexlab{a}})}\BibitemShut {NoStop}%
\bibitem [{\citenamefont {Slot}\ \emph {et~al.}(2017)\citenamefont {Slot},
  \citenamefont {Gardenier}, \citenamefont {Jacobse}, \citenamefont {van
  Miert}, \citenamefont {Kempkes}, \citenamefont {Zevenhuizen}, \citenamefont
  {{Morais Smith}}, \citenamefont {Vanmaekelbergh},\ and\ \citenamefont
  {Swart}}]{Slot2017a}%
  \BibitemOpen
  \bibfield  {author} {\bibinfo {author} {\bibfnamefont {M.~R.}\ \bibnamefont
  {Slot}}, \bibinfo {author} {\bibfnamefont {T.~S.}\ \bibnamefont {Gardenier}},
  \bibinfo {author} {\bibfnamefont {P.~H.}\ \bibnamefont {Jacobse}}, \bibinfo
  {author} {\bibfnamefont {G.~C.~P.}\ \bibnamefont {van Miert}}, \bibinfo
  {author} {\bibfnamefont {S.~N.}\ \bibnamefont {Kempkes}}, \bibinfo {author}
  {\bibfnamefont {S.~J.~M.}\ \bibnamefont {Zevenhuizen}}, \bibinfo {author}
  {\bibfnamefont {C.}~\bibnamefont {{Morais Smith}}}, \bibinfo {author}
  {\bibfnamefont {D.}~\bibnamefont {Vanmaekelbergh}},\ and\ \bibinfo {author}
  {\bibfnamefont {I.}~\bibnamefont {Swart}},\ }\bibfield  {title} {\bibinfo
  {title} {{Experimental realization and characterization of an electronic Lieb
  lattice}},\ }\href {https://doi.org/10.1038/nphys4105} {\bibfield  {journal}
  {\bibinfo  {journal} {Nat. Phys.}\ }\textbf {\bibinfo {volume} {13}},\
  \bibinfo {pages} {672} (\bibinfo {year} {2017})}\BibitemShut {NoStop}%
\bibitem [{\citenamefont {F{\"{o}}lsch}\ \emph {et~al.}(2014)\citenamefont
  {F{\"{o}}lsch}, \citenamefont {Mart{\'{i}}nez-Blanco}, \citenamefont {Yang},
  \citenamefont {Kanisawa},\ and\ \citenamefont {Erwin}}]{Folsch2014}%
  \BibitemOpen
  \bibfield  {author} {\bibinfo {author} {\bibfnamefont {S.}~\bibnamefont
  {F{\"{o}}lsch}}, \bibinfo {author} {\bibfnamefont {J.}~\bibnamefont
  {Mart{\'{i}}nez-Blanco}}, \bibinfo {author} {\bibfnamefont {J.}~\bibnamefont
  {Yang}}, \bibinfo {author} {\bibfnamefont {K.}~\bibnamefont {Kanisawa}},\
  and\ \bibinfo {author} {\bibfnamefont {S.~C.}\ \bibnamefont {Erwin}},\
  }\bibfield  {title} {\bibinfo {title} {{Quantum dots with single-atom
  precision}},\ }\href {https://doi.org/10.1038/nnano.2014.129} {\bibfield
  {journal} {\bibinfo  {journal} {Nat. Nanotechnol.}\ }\textbf {\bibinfo
  {volume} {9}},\ \bibinfo {pages} {505} (\bibinfo {year} {2014})}\BibitemShut
  {NoStop}%
\bibitem [{\citenamefont {Sierda}\ \emph {et~al.}(2023)\citenamefont {Sierda},
  \citenamefont {Huang}, \citenamefont {Badrtdinov}, \citenamefont {Kiraly},
  \citenamefont {Knol}, \citenamefont {Groenenboom}, \citenamefont
  {Katsnelson}, \citenamefont {R{\"{o}}sner}, \citenamefont {Wegner},\ and\
  \citenamefont {Khajetoorians}}]{Sierda2023}%
  \BibitemOpen
  \bibfield  {author} {\bibinfo {author} {\bibfnamefont {E.}~\bibnamefont
  {Sierda}}, \bibinfo {author} {\bibfnamefont {X.}~\bibnamefont {Huang}},
  \bibinfo {author} {\bibfnamefont {D.~I.}\ \bibnamefont {Badrtdinov}},
  \bibinfo {author} {\bibfnamefont {B.}~\bibnamefont {Kiraly}}, \bibinfo
  {author} {\bibfnamefont {E.~J.}\ \bibnamefont {Knol}}, \bibinfo {author}
  {\bibfnamefont {G.~C.}\ \bibnamefont {Groenenboom}}, \bibinfo {author}
  {\bibfnamefont {M.~I.}\ \bibnamefont {Katsnelson}}, \bibinfo {author}
  {\bibfnamefont {M.}~\bibnamefont {R{\"{o}}sner}}, \bibinfo {author}
  {\bibfnamefont {D.}~\bibnamefont {Wegner}},\ and\ \bibinfo {author}
  {\bibfnamefont {A.~A.}\ \bibnamefont {Khajetoorians}},\ }\bibfield  {title}
  {\bibinfo {title} {{Quantum simulator to emulate lower-dimensional molecular
  structure}},\ }\href {https://doi.org/10.1126/science.adf2685} {\bibfield
  {journal} {\bibinfo  {journal} {Science}\ }\textbf {\bibinfo {volume}
  {380}},\ \bibinfo {pages} {1048} (\bibinfo {year} {2023})}\BibitemShut
  {NoStop}%
\bibitem [{SM()}]{SM}%
  \BibitemOpen
  \href@noop {} {}\bibinfo {note} {{S}ee {S}upplemental {M}aterial at [URL],
  which includes details on I. Materials and methods; II. Theoretical model;
  III. Comparison experimental results and tight-binding
  calculations.}\BibitemShut {Stop}%
\bibitem [{FN1()}]{FN1}%
  \BibitemOpen
  \href@noop {} {}\bibinfo {note} {This is an estimation of the hopping, since
  the hopping also includes the experimentally inaccessible overlap integral S:
  $\Delta E =
  \frac{\epsilon+t}{\epsilon+s}-\frac{\epsilon-t}{\epsilon-s}$.}\BibitemShut
  {Stop}%
\bibitem [{\citenamefont {Benalcazar}\ \emph {et~al.}(2019)\citenamefont
  {Benalcazar}, \citenamefont {Li},\ and\ \citenamefont
  {Hughes}}]{Benalcazar2019}%
  \BibitemOpen
  \bibfield  {author} {\bibinfo {author} {\bibfnamefont {W.~A.}\ \bibnamefont
  {Benalcazar}}, \bibinfo {author} {\bibfnamefont {T.}~\bibnamefont {Li}},\
  and\ \bibinfo {author} {\bibfnamefont {T.~L.}\ \bibnamefont {Hughes}},\
  }\bibfield  {title} {\bibinfo {title} {{Quantization of fractional corner
  charge in ${C}_{n}$-symmetric higher-order topological crystalline
  insulators}},\ }\href {https://doi.org/10.1103/PhysRevB.99.245151} {\bibfield
   {journal} {\bibinfo  {journal} {Phys. Rev. B}\ }\textbf {\bibinfo {volume}
  {99}},\ \bibinfo {pages} {245151} (\bibinfo {year} {2019})}\BibitemShut
  {NoStop}%
\bibitem [{\citenamefont {Yang}\ \emph {et~al.}(2011)\citenamefont {Yang},
  \citenamefont {Erwin}, \citenamefont {Kanisawa}, \citenamefont {Nacci},\ and\
  \citenamefont {F{\"{o}}lsch}}]{Yang2011b}%
  \BibitemOpen
  \bibfield  {author} {\bibinfo {author} {\bibfnamefont {J.}~\bibnamefont
  {Yang}}, \bibinfo {author} {\bibfnamefont {S.~C.}\ \bibnamefont {Erwin}},
  \bibinfo {author} {\bibfnamefont {K.}~\bibnamefont {Kanisawa}}, \bibinfo
  {author} {\bibfnamefont {C.}~\bibnamefont {Nacci}},\ and\ \bibinfo {author}
  {\bibfnamefont {S.}~\bibnamefont {F{\"{o}}lsch}},\ }\bibfield  {title}
  {\bibinfo {title} {{Emergent multistability in assembled nanostructures}},\
  }\href {https://doi.org/10.1021/nl2009444} {\bibfield  {journal} {\bibinfo
  {journal} {Nano Lett.}\ }\textbf {\bibinfo {volume} {11}},\ \bibinfo {pages}
  {2486} (\bibinfo {year} {2011})}\BibitemShut {NoStop}%
\bibitem [{\citenamefont {R\"ontgen}\ \emph {et~al.}(2021)\citenamefont
  {R\"ontgen}, \citenamefont {Pyzh}, \citenamefont {Morfonios}, \citenamefont
  {Palaiodimopoulos}, \citenamefont {Diakonos},\ and\ \citenamefont
  {Schmelcher}}]{Roentgen_2021}%
  \BibitemOpen
  \bibfield  {author} {\bibinfo {author} {\bibfnamefont {M.}~\bibnamefont
  {R\"ontgen}}, \bibinfo {author} {\bibfnamefont {M.}~\bibnamefont {Pyzh}},
  \bibinfo {author} {\bibfnamefont {C.~V.}\ \bibnamefont {Morfonios}}, \bibinfo
  {author} {\bibfnamefont {N.~E.}\ \bibnamefont {Palaiodimopoulos}}, \bibinfo
  {author} {\bibfnamefont {F.~K.}\ \bibnamefont {Diakonos}},\ and\ \bibinfo
  {author} {\bibfnamefont {P.}~\bibnamefont {Schmelcher}},\ }\bibfield  {title}
  {\bibinfo {title} {Latent symmetry induced degeneracies},\ }\href
  {https://doi.org/10.1103/PhysRevLett.126.180601} {\bibfield  {journal}
  {\bibinfo  {journal} {Phys. Rev. Lett.}\ }\textbf {\bibinfo {volume} {126}},\
  \bibinfo {pages} {180601} (\bibinfo {year} {2021})}\BibitemShut {NoStop}%
\bibitem [{\citenamefont {Röntgen}\ \emph {et~al.}(2023)\citenamefont
  {Röntgen}, \citenamefont {Chen}, \citenamefont {Gao}, \citenamefont {Pyzh},
  \citenamefont {Schmelcher}, \citenamefont {Pagneux}, \citenamefont
  {Achilleos},\ and\ \citenamefont {Coutant}}]{Roentgen_2023}%
  \BibitemOpen
  \bibfield  {author} {\bibinfo {author} {\bibfnamefont {M.}~\bibnamefont
  {Röntgen}}, \bibinfo {author} {\bibfnamefont {X.}~\bibnamefont {Chen}},
  \bibinfo {author} {\bibfnamefont {W.}~\bibnamefont {Gao}}, \bibinfo {author}
  {\bibfnamefont {M.}~\bibnamefont {Pyzh}}, \bibinfo {author} {\bibfnamefont
  {P.}~\bibnamefont {Schmelcher}}, \bibinfo {author} {\bibfnamefont
  {V.}~\bibnamefont {Pagneux}}, \bibinfo {author} {\bibfnamefont
  {V.}~\bibnamefont {Achilleos}},\ and\ \bibinfo {author} {\bibfnamefont
  {A.}~\bibnamefont {Coutant}},\ }\href@noop {} {\bibinfo {title} {{Latent
  Su-Schrieffer-Heeger models}}} (\bibinfo {year} {2023}),\ \Eprint
  {https://arxiv.org/abs/arXiv:2310.07619} {arXiv:2310.07619} \BibitemShut
  {NoStop}%
\bibitem [{\citenamefont {Ni}\ \emph {et~al.}(2019)\citenamefont {Ni},
  \citenamefont {Weiner}, \citenamefont {Al{\`u}},\ and\ \citenamefont
  {Khanikaev}}]{Ni_2018}%
  \BibitemOpen
  \bibfield  {author} {\bibinfo {author} {\bibfnamefont {X.}~\bibnamefont
  {Ni}}, \bibinfo {author} {\bibfnamefont {M.}~\bibnamefont {Weiner}}, \bibinfo
  {author} {\bibfnamefont {A.}~\bibnamefont {Al{\`u}}},\ and\ \bibinfo {author}
  {\bibfnamefont {A.~B.}\ \bibnamefont {Khanikaev}},\ }\bibfield  {title}
  {\bibinfo {title} {Observation of higher-order topological acoustic states
  protected by generalized chiral symmetry},\ }\href
  {https://doi.org/10.1038/s41563-018-0252-9} {\bibfield  {journal} {\bibinfo
  {journal} {Nat. Mat.}\ }\textbf {\bibinfo {volume} {18}},\ \bibinfo {pages}
  {113} (\bibinfo {year} {2019})}\BibitemShut {NoStop}%
\bibitem [{\citenamefont {Kempkes}\ \emph
  {et~al.}(2019{\natexlab{b}})\citenamefont {Kempkes}, \citenamefont {Slot},
  \citenamefont {van~den Broeke}, \citenamefont {Capiod}, \citenamefont
  {Benalcazar}, \citenamefont {Vanmaekelbergh}, \citenamefont {Bercioux},
  \citenamefont {Swart},\ and\ \citenamefont {Smith}}]{Kempkes_2019}%
  \BibitemOpen
  \bibfield  {author} {\bibinfo {author} {\bibfnamefont {S.}~\bibnamefont
  {Kempkes}}, \bibinfo {author} {\bibfnamefont {M.}~\bibnamefont {Slot}},
  \bibinfo {author} {\bibfnamefont {J.}~\bibnamefont {van~den Broeke}},
  \bibinfo {author} {\bibfnamefont {P.}~\bibnamefont {Capiod}}, \bibinfo
  {author} {\bibfnamefont {W.}~\bibnamefont {Benalcazar}}, \bibinfo {author}
  {\bibfnamefont {D.}~\bibnamefont {Vanmaekelbergh}}, \bibinfo {author}
  {\bibfnamefont {D.}~\bibnamefont {Bercioux}}, \bibinfo {author}
  {\bibfnamefont {I.}~\bibnamefont {Swart}},\ and\ \bibinfo {author}
  {\bibfnamefont {C.~M.}\ \bibnamefont {Smith}},\ }\bibfield  {title} {\bibinfo
  {title} {Robust zero-energy modes in an electronic higher-order topological
  insulator},\ }\href {https://doi.org/10.1038/s41563-019-0483-4} {\bibfield
  {journal} {\bibinfo  {journal} {Nat. Mat.}\ }\textbf {\bibinfo {volume}
  {18}},\ \bibinfo {pages} {1292} (\bibinfo {year}
  {2019}{\natexlab{b}})}\BibitemShut {NoStop}%
\bibitem [{\citenamefont {Herrera}\ \emph {et~al.}(2022)\citenamefont
  {Herrera}, \citenamefont {Kempkes}, \citenamefont {de~Paz}, \citenamefont
  {Garc\'{\i}a-Etxarri}, \citenamefont {Swart}, \citenamefont {Smith},\ and\
  \citenamefont {Bercioux}}]{Herrera_2022}%
  \BibitemOpen
  \bibfield  {author} {\bibinfo {author} {\bibfnamefont {M.~A.~J.}\
  \bibnamefont {Herrera}}, \bibinfo {author} {\bibfnamefont {S.~N.}\
  \bibnamefont {Kempkes}}, \bibinfo {author} {\bibfnamefont {M.~B.}\
  \bibnamefont {de~Paz}}, \bibinfo {author} {\bibfnamefont {A.}~\bibnamefont
  {Garc\'{\i}a-Etxarri}}, \bibinfo {author} {\bibfnamefont {I.}~\bibnamefont
  {Swart}}, \bibinfo {author} {\bibfnamefont {C.~M.}\ \bibnamefont {Smith}},\
  and\ \bibinfo {author} {\bibfnamefont {D.}~\bibnamefont {Bercioux}},\
  }\bibfield  {title} {\bibinfo {title} {Corner modes of the breathing kagome
  lattice: Origin and robustness},\ }\href
  {https://doi.org/10.1103/PhysRevB.105.085411} {\bibfield  {journal} {\bibinfo
   {journal} {Phys. Rev. B}\ }\textbf {\bibinfo {volume} {105}},\ \bibinfo
  {pages} {085411} (\bibinfo {year} {2022})}\BibitemShut {NoStop}%
\bibitem [{\citenamefont {Tassi}\ and\ \citenamefont
  {Bercioux}(2024)}]{Tassi_2024}%
  \BibitemOpen
  \bibfield  {author} {\bibinfo {author} {\bibfnamefont {C.}~\bibnamefont
  {Tassi}}\ and\ \bibinfo {author} {\bibfnamefont {D.}~\bibnamefont
  {Bercioux}},\ }\bibfield  {title} {\bibinfo {title} {Implementation and
  characterization of the dice lattice in the electron quantum simulator},\
  }\href {https://doi.org/10.1002/apxr.202400038} {\bibfield  {journal}
  {\bibinfo  {journal} {Adv. Phys. Res.}\ }\textbf {\bibinfo {volume} {3}},\
  \bibinfo {pages} {2400038} (\bibinfo {year} {2024})}\BibitemShut {NoStop}%
\bibitem [{\citenamefont {Holbrook}\ \emph {et~al.}(2024)\citenamefont
  {Holbrook}, \citenamefont {Ingham}, \citenamefont {Kaplan}, \citenamefont
  {Holtzman}, \citenamefont {Bierman}, \citenamefont {Olson}, \citenamefont
  {Nashabeh}, \citenamefont {Liu}, \citenamefont {Zhu}, \citenamefont {Rhodes},
  \citenamefont {Barmak}, \citenamefont {Hone}, \citenamefont {Queiroz},\ and\
  \citenamefont {Pasupathy}}]{Holbrook2024}%
  \BibitemOpen
  \bibfield  {author} {\bibinfo {author} {\bibfnamefont {M.}~\bibnamefont
  {Holbrook}}, \bibinfo {author} {\bibfnamefont {J.}~\bibnamefont {Ingham}},
  \bibinfo {author} {\bibfnamefont {D.}~\bibnamefont {Kaplan}}, \bibinfo
  {author} {\bibfnamefont {L.}~\bibnamefont {Holtzman}}, \bibinfo {author}
  {\bibfnamefont {B.}~\bibnamefont {Bierman}}, \bibinfo {author} {\bibfnamefont
  {N.}~\bibnamefont {Olson}}, \bibinfo {author} {\bibfnamefont
  {L.}~\bibnamefont {Nashabeh}}, \bibinfo {author} {\bibfnamefont
  {S.}~\bibnamefont {Liu}}, \bibinfo {author} {\bibfnamefont {X.}~\bibnamefont
  {Zhu}}, \bibinfo {author} {\bibfnamefont {D.}~\bibnamefont {Rhodes}},
  \bibinfo {author} {\bibfnamefont {K.}~\bibnamefont {Barmak}}, \bibinfo
  {author} {\bibfnamefont {J.}~\bibnamefont {Hone}}, \bibinfo {author}
  {\bibfnamefont {R.}~\bibnamefont {Queiroz}},\ and\ \bibinfo {author}
  {\bibfnamefont {A.}~\bibnamefont {Pasupathy}},\ }\href
  {https://arxiv.org/abs/2412.02813} {\bibinfo {title} {Real-space imaging of
  the band topology of transition metal dichalcogenides}} (\bibinfo {year}
  {2024}),\ \Eprint {https://arxiv.org/abs/2412.02813} {arXiv:2412.02813
  [cond-mat.mtrl-sci]} \BibitemShut {NoStop}%
\bibitem [{\citenamefont {Călugăru}\ \emph {et~al.}(2025)\citenamefont
  {Călugăru}, \citenamefont {Jiang}, \citenamefont {Guo}, \citenamefont
  {Sajan}, \citenamefont {Wang}, \citenamefont {Hu}, \citenamefont {Yu},
  \citenamefont {Bernevig}, \citenamefont {de~Juan},\ and\ \citenamefont
  {Ugeda}}]{Calugaru2025}%
  \BibitemOpen
  \bibfield  {author} {\bibinfo {author} {\bibfnamefont {D.}~\bibnamefont
  {Călugăru}}, \bibinfo {author} {\bibfnamefont {Y.}~\bibnamefont {Jiang}},
  \bibinfo {author} {\bibfnamefont {H.}~\bibnamefont {Guo}}, \bibinfo {author}
  {\bibfnamefont {S.}~\bibnamefont {Sajan}}, \bibinfo {author} {\bibfnamefont
  {Y.}~\bibnamefont {Wang}}, \bibinfo {author} {\bibfnamefont {H.}~\bibnamefont
  {Hu}}, \bibinfo {author} {\bibfnamefont {J.}~\bibnamefont {Yu}}, \bibinfo
  {author} {\bibfnamefont {B.~A.}\ \bibnamefont {Bernevig}}, \bibinfo {author}
  {\bibfnamefont {F.}~\bibnamefont {de~Juan}},\ and\ \bibinfo {author}
  {\bibfnamefont {M.~M.}\ \bibnamefont {Ugeda}},\ }\href
  {https://arxiv.org/abs/2501.09063} {\bibinfo {title} {Probing the quantized
  berry phases in 1h-nbse$_2$ using scanning tunneling microscopy}} (\bibinfo
  {year} {2025}),\ \Eprint {https://arxiv.org/abs/2501.09063} {arXiv:2501.09063
  [cond-mat.mes-hall]} \BibitemShut {NoStop}%
\end{thebibliography}%
\onecolumngrid

\pagebreak

\renewcommand{\theequation}{SE\arabic{equation}}
\renewcommand{\thefigure}{SF\arabic{figure}}
\renewcommand{\bibnumfmt}[1]{[#1]}
\renewcommand{\citenumfont}[1]{#1}
\setcounter{equation}{0}
\setcounter{figure}{0}
\setcounter{table}{0}
\setcounter{section}{0}

\begin{center}
   \large{\textbf{SUPPLEMENTAL MATERIAL FOR:}} \\ 
    \vspace{.5cm}
   \Large{Wannier center spectroscopy to identify boundary-obstructed topological insulators}
\end{center}

\section{Materials and methods}
\subsection{Methods}
\begin{figure}[!b]
    \centering
    \includegraphics{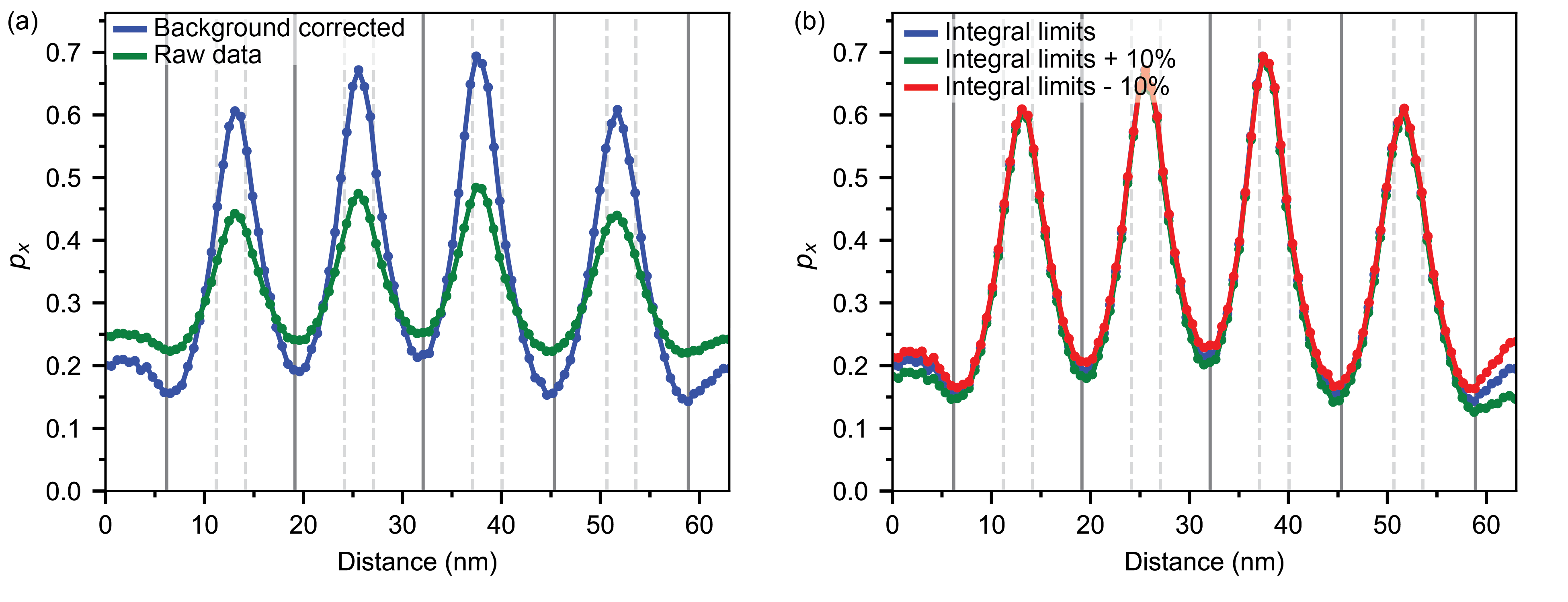}
    \caption{(a) The charge density, $p_x$, for the trivial SSH chain with green depicting the raw data and blue the background corrected. The background correction was performed by subtracting a underestimated fit of the onset of the conduction band of InAs(111)A. (b) The charge density for the trivial SSH (blue) with the upper boundary limit of the integral overestimated with 10\% (green) and underestimated with 10\% (red). }
    \label{BackgroundCorrection}
\end{figure}

The InAs(111)A sample was commercially bought from SWI, $n$-doped with a concentration between $1\times10^{17}$--$3\times10^{18}$~cm$^{-3}$. On top of the InAs(111)A wafer, a 100~nm thick InAs(111)A layer was grown to reduce the defect concentration on the surface. For the full protocol, see Ligthart \emph{et al.}~\cite{Ligthart2024} The wafer was capped with amorphous As, which was removed by heating in the preparation chamber of a Scienta Omicron LT-STM in which the measurements were performed. During measurements, the base temperature was 4.3 K, and the pressure was in the $10^{-10}$~mbar range. The Cs was evaporated (SAES Getters) on a cold InAs(111)A surface. Spectroscopy was performed with a bias ac-modulation (at 769~Hz) of 1~mV (rms amplitude) and a lock-in amplifier for detection. For the charge density graphs, an underestimated fitted background was subtracted from the experimental data to remove the onset of the conduction band of InAs(111)A at $-100$~mV and to ensure the entire area below the graph was positive. Figure \ref{BackgroundCorrection}a shows that the background correction did not alter the positions of the peaks for the trivial SSH chain. Furthermore, the sensitivity of integration limit wast tested by increasing/decreasing the upper integral limit with 10\%. Figure \ref{BackgroundCorrection}b shows that all three curves overlap. The constant-current STM topographs were processed with Gwyddion 2.56 with a plane-subtraction and 2D-FFT filtering for clarity purposes.

\subsection{Artificial atoms and Dimers}

\begin{figure}
    \centering
    \includegraphics{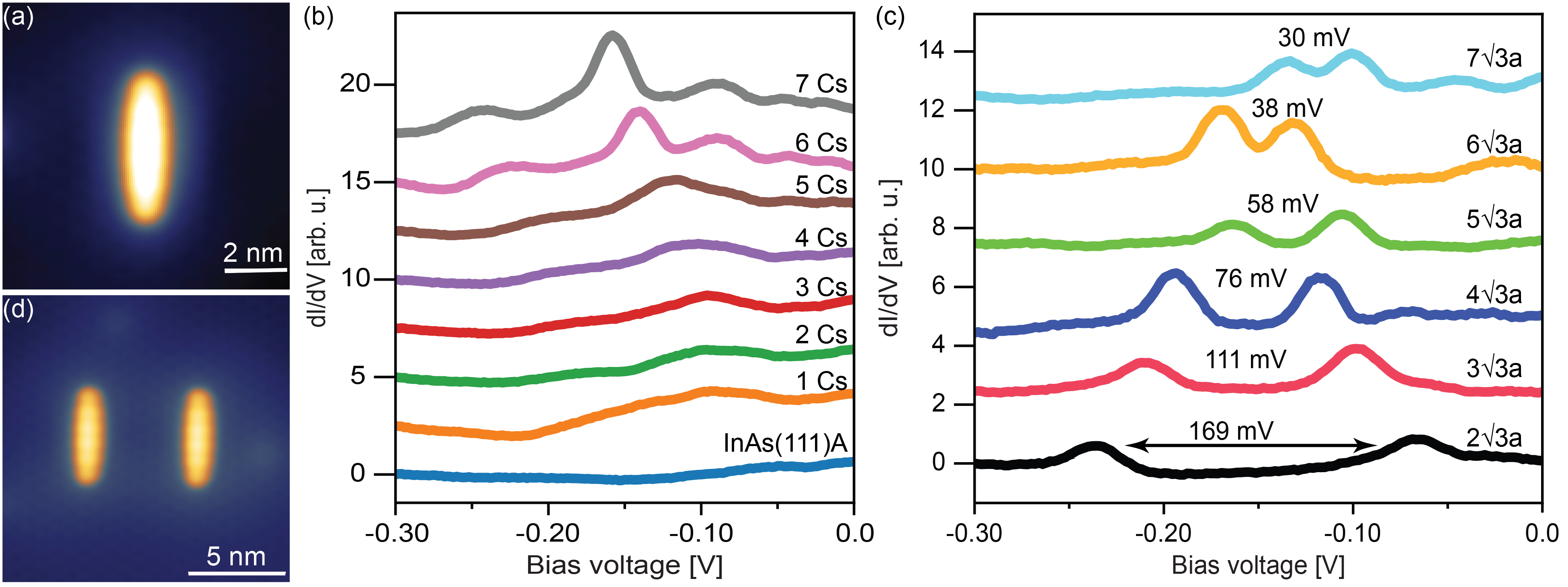}
    \caption{(a) STM topography image of 6 Cs atoms in a line, image taken at 0.1 V and 30 pA, scale bar is 2 nm. (b) Single $dI/dV$ spectra taken next to a line of 1 $-$ 7 Cs atoms, spectra separated in height for clarity. (c) Single $dI/dV$ spectra taken next to a dimer of artificial atoms with distances between the lines of $2a\sqrt{3}$ $-$ $7a\sqrt{3}$, spectra separated in height for clarity. (d) STM topography image of two lines of each 6 Cs atoms with a distance between the artificial atoms of $4a\sqrt{3}$ nm, image taken at 0.1 V and 30 pA, scale bar is 5 nm.}
    \label{LinesAndDimers}
\end{figure}

Electronic artificial lattices are prepared by confining the surface state electrons by single adatoms acting as repulsive or attractive scatterers. Controlled positioning of single atoms --- acting as scatterers --- can be done with atomic-scale precision using the tip of the STM. Positioning enough attractive scatterers in proximity will result in particle-in-a-box-like states; see Fig.~\ref{LinesAndDimers}b. In this case, from a line of 5 Cs onwards, a confined state is observed at around $-0.12$~mV. The confined state shifts down in energy with more Cs atoms due to the increased confinement. The confined state can be compared to a particle-in-a-box-like state. Its lowest state resembles the $s$-like wavefunction of hydrogen, which can be seen as an artificial atom used to construct designer materials. Recently, the energy resolution of the artificial electronic lattices has increased considerably by moving the platform from a metallic to a semiconductor surface; here, the confined states fall in the material's bandgap, such as InAs(111) or InSb(110). Figure~\ref{LinesAndDimers}c shows the coupling strength of two artificial atoms when placed parallel as in Fig.~\ref{LinesAndDimers}d. From the bond strength of the dimers, the hopping parameters for tight binding can be retrieved. This is not the exact hopping term since the overlap integral cannot be obtained. The bond strengths are comparable to the values of the dimers made with In on InAs(111)A by Fölsch \emph{et al.}~\cite{Folsch2014}. This implies that only the charge of the In/Cs plays a role in confining the surface state electrons, and the type of adatom itself plays no role.

\subsection{Additional 1D models}
\subsubsection{Rice-Mele}\label{exp_RM}
\begin{figure}[!b]
 \centering
 \includegraphics{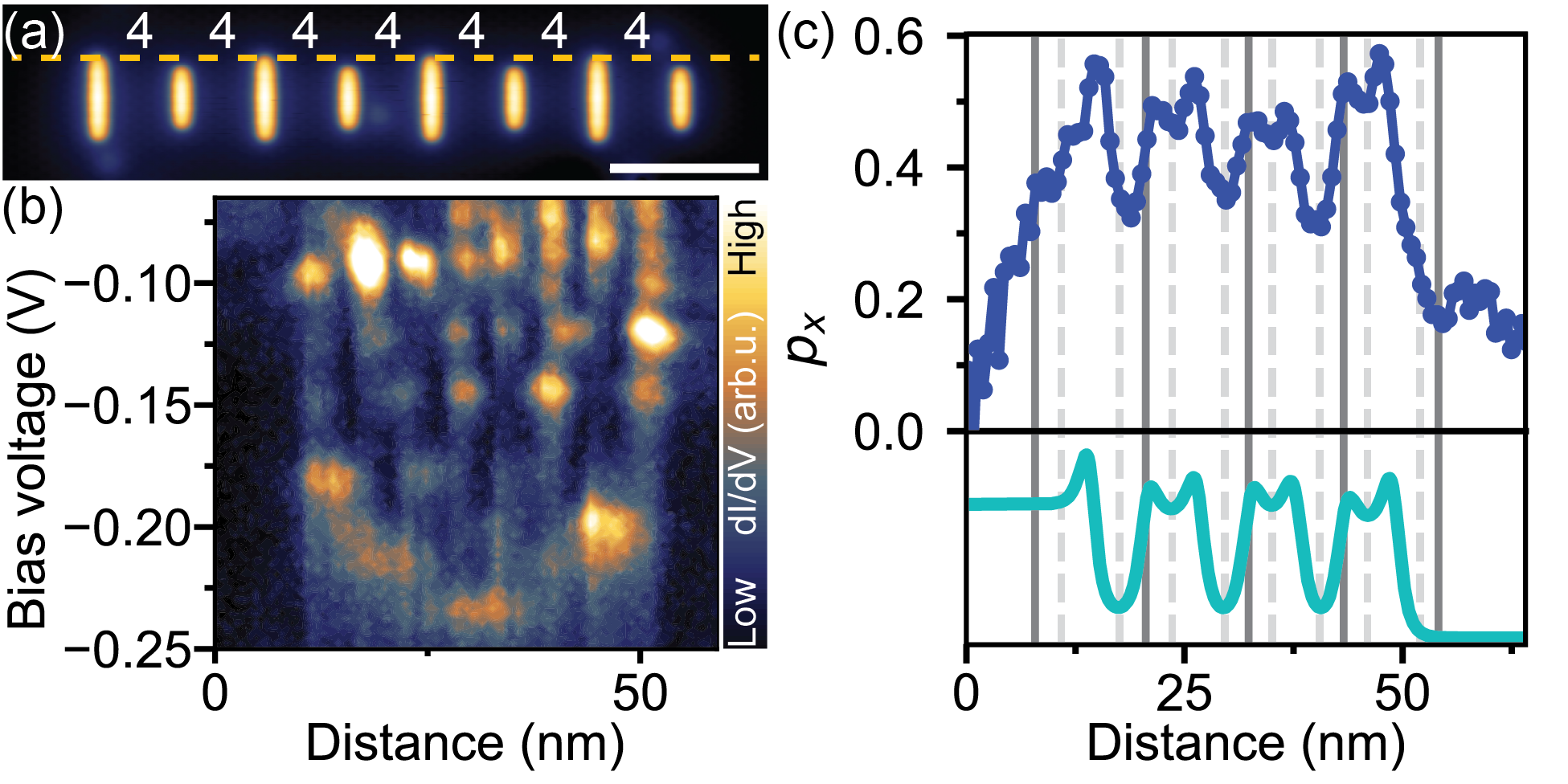}
 \caption{(a) STM topography image of the Rice-Mele chain with 4 unit cells. The on-site energy of the artificial atoms depends on the number of Cs atoms (7 and 5 for longer and shorter chains, respectively). The distance between the chains is $4a\sqrt{3}$ nm, with $a$ the lattice constant. The image was taken at $0.5$~V and $30$~pA, and the scale bar was $10$~nm. (b) Contour plot of $dI/dV$ spectra taken above the Rice-Mele chain in (a). (c) Charge density plot with the positions of the atoms (light grey) and the boundary of the unit cell (dark grey).}
 \label{RiceMele}
\end{figure}

The atomic chain we presented in the main text can be slightly altered to obtain the Rice-Mele chain~\cite{Rice_1982, Asboth2015, Allen_2022} by varying the on-site energy of the atomic chain. Figure~\ref{RiceMele}a shows a chain consisting of 8 artificial atoms (each made of 5 or 7 Cs atoms), with an equidistant spacing of $4a\sqrt{3}$, with $a = 0.857$~nm, the lattice constant of the InAs(111)A (2$\times$2) surface reconstruction. We control the on-site energy of the artificial atoms by the number of Cs atoms in the chain, here 5 and 7 Cs atoms for the shorter and longer lines. STS is performed above the chain (dashed orange line) with 108 points and is depicted as a contour plot in Fig.~\ref{RiceMele}b. A small bandgap of approximately $30$~mV is present around $-165$~meV; this corresponds to a ratio between the gap and the hopping parameter of $\Delta/t\sim0.39$. For the computation of the charge density/Wannier representation of the lowest energy band, the integration values were set to include all states with energies lower than the bandgap: we present the resulting charge density in Fig.~\ref{RiceMele}c (solid blue line). 
The system has no inversion symmetry~\cite{Allen_2022}; thus, the Wannier center can sit in any position within the unit cell. 
Furthermore, the same features are observed qualitatively for every unit cell. Tight binding calculations (cyan line) reproduce the experimental findings. 

\subsubsection{SSH 63 and 72}
\begin{figure}
    \centering
    \includegraphics{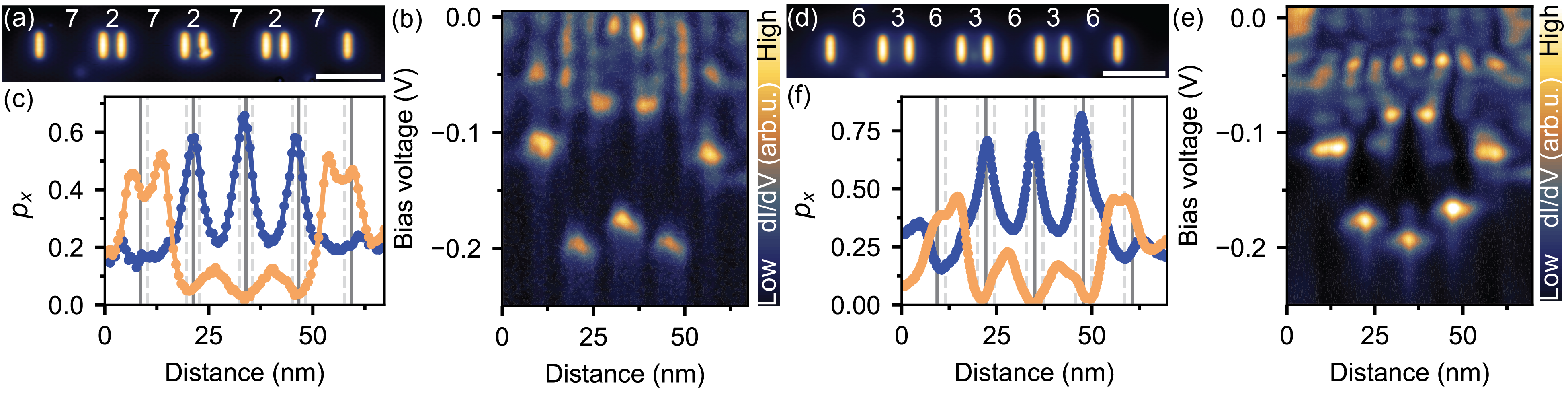}
    \caption{(a, d) STM topography image of two non-trivial SSH chains with a different hopping ratio of $t_1/t_2=0.18$ with distances of $7a\sqrt{3}$, $3a\sqrt{3}$ (a) and hopping ratio of $t_1/t_2=0.34$ with distances $6a\sqrt{3}$, $3a\sqrt{3}$ (d). Images taken at 0.5~V, 30~pA, scale bar is 10~nm. (b, e) Contour plot of $dI/dV$ spectra taken along a line above the two non-trivial SSH chains in (a) and (d), respectively. (c, f) Charge density plot with the positions of the atoms (light grey) and the boundary of the unit cell (dark grey). The orange line is the sole integration of the edge mode itself.}
    \label{SSH63_72}
\end{figure}

In this section, we discuss the role of the hopping ratio in defining the edge modes at the end of the SSH model. The relative difference between the two hopping terms $t_1$ and $t_2$ dictate the size of the energy gap between the two bands of the SSH model~\cite{Asboth2015}. The combination of the chiral symmetry and the gap size leads to a topological protection of the edge modes. An additional factor of protection comes from the system size (see Sec.~\ref{finite_size}), namely the spatial separation of the two modes. When two edge modes are in proximity, they can hybridize and are no longer localized at the end of a chain. This was shown in Ref.~\cite{Pham2022}, where the two edge modes were hybridized due to a too-large deviation from the dimerized hopping ratio.


We show that the position of the Wannier functions does not change with a different hopping ratio (in the non-trivial regime), realizing another SSH chain with a different hopping ratio. However, in this case, the edge modes start to hybridize. We start considering the case presented in the main text, with an SSH chain with the hopping ratio set to $t_1/t_2 = 0.18$ | Figs.~\ref{SSH63_72}ab. In the charge density plot (Fig.~\ref{SSH63_72}c), besides showing the Wannier states of the bonding band (blue), it now also displays the charge density of the edge state (orange).

We realized another SSH chain with a hopping ratio of $t_1/t_2 = 0.34$, which is less favorable because it results in a smaller gap. This chain was constructed by spacing the artificial atoms with distances of $6a\sqrt{3}$ and $3a\sqrt{3}$ apart | see Fig.~\ref{SSH63_72}d. The contour plot in Fig.~\ref{SSH63_72}e shows that edge modes are still present in the comparatively smaller bandgap. The end modes are present at $-115$~mV. Intensity at this energy is also seen on the 3rd and 6th atomic sites, suggesting that the edge modes are now hybridized. The edge's mode charge is not fully localized anymore due to the hybridization of the edge modes. Figure~\ref{SSH63_72}f shows that the edge mode  --- orange line ---  has increased intensity in the center of the chain as well in comparison with the SSH chain with the $0.18$ hopping ratio. Deviating from the fully dimerized hopping ratio leads to a charge density redistribution from the edge toward the center of the chain.
However, the Wannier center of the bonding band | blue line | is still present at the boundary of the unit cell. This has not changed since the hopping ratio still dictates the topological phase. However, a smaller gap results in a broader spatial extension of the edge modes, thus a greater hybridization to each other | see Sec.~\ref{finite_size}.

\section{Theoretical model}

\subsection{Hamiltonians of the 1D models}

In the main text and this supplemental material, we have considered various 1D chain models. In the following, we present the corresponding tight-binding and Bloch-like Hamiltonians. Specifically, we are going to consider (1) the atomic chain (AC), (2) the SSH model, (3) the Rice-Mele model (RM), and (4) the trimer chain. The corresponding tight-binding Hamiltonians read:
%
%
\begin{subequations}
\begin{align}
\mathcal{H}_\text{AC}^\text{tb}&=t\sum_i c_{i+1}^\dag c_i + h.c., \label{HanTBAC}\\
%
%
\mathcal{H}_\text{SSH}^\text{tb}&=\sum_i (t_1 b_i^\dag a_i + t_2 a_{i+1}^\dag b_i) + h.c.,\label{HanTBSSH}\\
%
%
\mathcal{H}_\text{RM}^\text{tb}&=\sum_i \left[t_1 b_i^\dag a_i + t_2 a_{i+1}^\dag b_i+\Delta(a_i^\dag a_i-b_i^\dag b_i)\right] + h.c.,\label{HanTBRM}\\
%
    \mathcal{H}_\text{Trimer}&= \sum_i (t_1 a_{i}^\dag b_{i}+ t_2 b_{i}^\dag c_{i}+ t_3 a_{n+1}^\dag c_{i}) + h.c.\,.\label{HanTBTrimer}
\end{align}
\end{subequations}
%
%
Assuming translational invariance, we can express these tight-binding Hamiltonians in a Bloch form as follows:
%
%
\begin{subequations}
\begin{align}
\mathcal{H}^\text{B}_\text{AC}&= \sum_{k\in\text{BZ}} c_k^\dag [2t \cos(ka)] c_k, \label{BAC} \\  
\mathcal{H}^\text{B}_\text{SSH}&= \sum_{k\in\text{BZ}} \begin{bmatrix}
a_k \\ 
b_k
\end{bmatrix}^\dag
\left[2\bar{t}
 \cos\left(\frac{ka}{2}\right)\sigma_x+\delta \sin\left(\frac{ka}{2}\right)\sigma_y\right]
\begin{bmatrix}
a_k \\ 
b_k 
\end{bmatrix}, \label{BSSH} \\
\mathcal{H}^\text{B}_\text{RM}&= \sum_{k\in\text{BZ}} \begin{bmatrix}
a_k \\ 
b_k
\end{bmatrix}^\dag
\left[2\bar{t}
 \cos\left(\frac{ka}{2}\right)\sigma_x+\delta \sin\left(\frac{ka}{2}\right)\sigma_y+\Delta\sigma_z\right]
\begin{bmatrix}
a_k \\ 
b_k 
\end{bmatrix}, \label{BRM}\\
\mathcal{H}^\text{B}_\text{Trimer}&=\sum_{k\in\text{BZ}} \begin{bmatrix}
a_k \\ 
b_k \\
c_k
\end{bmatrix}^\dag
\{ t_1 \lambda_1+t_2\lambda_2+t_3[\cos(ka) \lambda_4+\sin(ka) \lambda_5]\}\begin{bmatrix}
a_k \\ 
b_k \\
c_k
\end{bmatrix} \label{BTrimer},
\end{align}
\end{subequations}
%
%
where $a$ is the 1D lattice periodicity, $k\in[-\pi/a,\pi/a)$ the momentum defined in the first Brillouin zone (BZ), $\bar{t}=(t_1+t_2)/2$, $\delta=t_1-t_2$ is the dimerization parameter, $\sigma_i$ are the Pauli matrices and $\lambda_i$ are the following Gell-Mann matrices:
%
%
\begin{align}
    \lambda_1=\begin{pmatrix}
        0 & 1 & 0 \\
        1 & 0 & 0 \\
        0 & 0 & 0
    \end{pmatrix},
    ~~~
    \lambda_2=\begin{pmatrix}
        0 & 0 & 0 \\
        0 & 0 & 1 \\
        0 & 1 & 0
    \end{pmatrix},
    ~~~
    \lambda_4=\begin{pmatrix}
        0 & 0 & 1 \\
        0 & 0 & 0 \\
        1 & 0 & 0
    \end{pmatrix},
    ~~~
    \lambda_5=\begin{pmatrix}
        0 & 0 & -\text{i} \\
        0 & 0 & 0 \\
        \text{i} & 0 & 0
    \end{pmatrix}.
\end{align}
%
%
We note on passing that the RM differs from the SSH model by having a staggered onsite potential proportional to $\Delta$. Additionally, inversion symmetry is present for the SSH model, whereas it is lost in the RM model~\cite{Allen_2022}. In real space, inversion symmetry is encoded in the $\sigma_x$ Pauli matrix that exchanges the two orbitals of the SSH or RM models.

The energy spectrum of the AC is composed of a single band described by $\epsilon_\text{AC}=2t\cos(\kappa)$. The energy spectra of the SSH and the RM models are composed of two bands described by the following:
%
%
\begin{align}
    \epsilon_\text{SSH}&=\pm\sqrt{2\bar{t}^2\cos\left(\frac{\kappa}{2}\right)^2+\delta^2 \sin\left(\frac{\kappa}{2}\right)^2}, \label{EnSSH} \\
    \epsilon_\text{RM}&=\pm\sqrt{2\bar{t}^2\cos\left(\frac{\kappa}{2}\right)^2+\delta^2 \sin\left(\frac{\kappa}{2}\right)^2+\Delta^2} \label{EnRM}.
\end{align}
%
%
In the case of the SSH model, the energy spectrum will contain a gap as long as $\delta\neq0$, i.e., $t_1\neq t_2$. On the other side, for the RM model, a gap can also be obtained for finite values of $\Delta$. The difference is that the gap open by the $\delta$ parameter can induce a topological phase transition for $\delta<0$, whereas the one open by $\Delta$ \emph{alone} is always trivial~\cite{Asboth2015, Allen_2022}. The energy spectrum of the trimer chain is analyzed in detail in Sec.~\ref{Trimer_general}.

\subsection{The Bloch Hamiltonian and the Wannier center for the bulk the Rice-Mele and SSH models.}

In this section, we will show how to obtain the Bloch and the Wannier wavefunctions analytically for the SSH and RM models, we will follow the construction of the these functions following Ref.~\cite{Vanderbilt2018}. Please note that the RM one obtains the SSH model by setting the staggered onsite energy to zero $\Delta=0$.
We consider an RM chain with the A and B lattice sites centered at $\pm \frac{a}{4}$ so that the relative distance is $r_\text{B}-r_\text{A}=\frac{a}{2}$, where $a$ is the lattice periodicity. We can introduce an intracell Hamiltonian that reads:
%
%
\begin{equation}\label{H_intra}
H(\bm{0})=\begin{pmatrix}
\Delta & t_1 \\
t_1& -\Delta
\end{pmatrix},
\end{equation}
%
%
and two intercell Hamiltonians that read:
%
%
\begin{align}\label{H_inter}
H(\bm{a})=\begin{pmatrix}
0 & 0 \\
t_2 & 0
\end{pmatrix}, \hspace{1cm}
H(-\bm{a})=\begin{pmatrix}
0 & t_2 \\
0& 0
\end{pmatrix}.
\end{align}
%
%
The tight-binding Hamiltonian matrix for the SSH chain reads $H_\text{RM}=H(\bm{0})+H(\bm{a})+H(-\bm{a})$ and is equivalent to Eq.~\eqref{HanTBRM}.

We can construct the Bloch-like basis functions as follows:
%
%
\begin{align}\label{bloch_like}
|\chi_\text{A}^k\rangle=\sum_{m\in\mathbb{Z}}\text{e}^{\text{i} k(ma-\frac{a}{4})}|m\text{A}\rangle, ~~~~~~
|\chi_\text{B}^k\rangle=\sum_{m\in\mathbb{Z}}\text{e}^{\text{i} k(ma+\frac{a}{4})}|m\text{B}\rangle.
\end{align}
%
%
Here, the ket states $|m\text{A}\rangle$ and $|m\text{B}\rangle$ represent the tight-binding orbitals that can be expressed via an $s$-like orbital | later, we will represent them in terms of Gaussian functions. A generic Bloch state is expressed as a linear combination of the functions in Eq.~\eqref{bloch_like}:
%
%
\begin{equation}\label{Bloch}
|\Psi_{nk}\rangle=c_\text{A}^{nk}|\chi_\text{A}^k\rangle+c_\text{B}^{nk}|\chi_\text{B}^k\rangle.
\end{equation}
%
%
We will now evaluate the Hamiltonian $H_\text{RM}$ over the state~\eqref{Bloch}:
%
%
\begin{subequations}\label{H_SSH}
\begin{align}
H_{\alpha\alpha}^k&= \langle\chi_\alpha^k|H_\text{RM}|\chi_\alpha^k\rangle=\Delta f(\alpha)\\
H_\text{AB}^k&=\langle \chi_\text{A}^k|H_\text{RM}|\chi_\text{B}^k\rangle  = \text{e}^{\text{i} k \frac{a}{2}}\left( t_1 +t_2 \text{e}^{-\text{i} k a} \right), \\
H_\text{BA}^k&=\langle \chi_\text{B}^k|H_\text{RM}|\chi_\text{A}^k\rangle = \text{e}^{\text{i} k \frac{a}{2}}\left( t_1 +t_2 \text{e}^{\text{i} k a} \right),
\end{align}
\end{subequations}
%
%
with $f(\alpha)=\pm 1$ and $\alpha\in\{\text{A,B}\}$.
Using the results in Eqs.~\eqref{H_SSH}, we can write down the eigenvalue equation $\bm{H}_k\cdot \bm{c}_{nk}=E_{nk}\bm{c}_{ck}$  as
%
%
\begin{align}\label{}
\begin{pmatrix}
\Delta-E_{nk} & t_1+t_2 \text{e}^{-\text{i} k a} \\
t_1+t_2 \text{e}^{\text{i} ka} & -(\Delta+E_{nk})
\end{pmatrix}
\begin{pmatrix}
c_\text{A}^{nk} \\ c_\text{B}^{nk}
\end{pmatrix}=0.
\end{align}
%
%
The matrix associated with $H_k$ is equivalent to the Bloch Hamiltonian in Eq.~\eqref{BRM}.
This is equivalent to the following system of equations:
%
%
\begin{align}\label{Bloch_equations}
\begin{cases}
(t_1+t_2 \text{e}^{-\text{i} k a}) c_\text{B}^{nk}=(E_{nk}-\Delta) c_\text{a}^{nk} \\
(t_1+t_2 \text{e}^{\text{i} k a}) c_\text{A}^{nk}=(E_{nk}+\Delta) c_\text{B}^{nk}
\end{cases},
\end{align}
%
%
after substituting the first into the second, we obtain $E_{nk}=\pm \sqrt{|t_1+t_2 \text{e}^{\text{i} k a}|^2+\Delta^2}$, this expression is equivalent to to the energy spectrum in Eq.~\eqref{EnRM}, and
%
%
\begin{subequations}
\begin{align}
c_\text{A}^{nk}&=\pm\frac{1}{\sqrt{2}}\frac{t_1+t_2 \text{e}^{-\text{i} k a}}{\sqrt{|t_1+t_2 \text{e}^{\text{i} k a}|^2+\Delta^2}-\Delta}=\pm\frac{1}{\sqrt{2}} \text{e}^{\text{i} \theta(k)},\\
c_\text{B}^{nk}&=\frac{1}{\sqrt{2}},\\
\theta(k)&=-\text{i} \text{Log}\left[ \frac{t_1+t_2 \text{e}^{-\text{i} k a}}{\sqrt{|t_1+t_2 \text{e}^{\text{i} k a}|^2+\Delta^2}-\Delta}\right]\label{phase}.
\end{align}
\end{subequations}
%
%
The wave function phase in Eq.~\eqref{phase} contains all the information about the topology of the RM and SSH chains: for $\Delta=0$ (the SSH chain), it describes a unitary circle of the complex plane centered at $t_1/t_2$. 
This circle contains the origin for $t_1<t_2$, whereas for $t_1\ge t_2$ not. In the former case, the system is in a topological phase~\cite{Asboth2015}. When $\Delta\neq0$ (the RM chain), it is possible to open a gap that competes with the one due to the dimerization parameter $\delta=t_1-t_2$. The RM chain is in a topological phase as long as the unitary circle includes the origin~\cite{Asboth2015}. In the case of the RM presented in Sec.~\ref{exp_RM}, we have that $t_1=t_2$ and $\Delta\neq0$, this set of parameters places this RM chain in the trivial phase.

The normalized Bloch wave function will now read:
%
%
\begin{align}
|\Psi_{nk}\rangle=\frac{1}{\sqrt{2}}&\left[ \text{e}^{\text{i} \theta(k)} \sum_{m\in\mathbb{Z}} \text{e}^{\text{i} ka (m-\frac{1}{4})} |m\text{A}\rangle+\sum_{m\in\mathbb{Z}} \text{e}^{\text{i} ka (m+\frac{1}{4})} |m\text{B}\rangle \right].
\end{align}
%
%
Assuming that the states $|m,\text{A/B}\rangle$ can be represented in real space with Gaussian localized functions:
%
%
\begin{equation}
\langle \bm{x} |m,\text{A/B}\rangle=G_\sigma\left[x-\left(m a \mp \frac{a}{4}\right)\right],
\end{equation}
%
%
we now project this wave function in the real space:
%
%
\begin{align}\label{Bloch_real}
\langle \bm{x} |\Psi_{nk}\rangle&=\frac{1}{\sqrt{2}}\left\{ \text{e}^{\text{i} \theta(k)} \sum_{m\in\mathbb{Z}} \text{e}^{\text{i} ka (m-\frac{1}{4})} G_\sigma\left[x-\left(m a - \frac{a}{4}\right)\right]+ \sum_{m\in\mathbb{Z}} \text{e}^{\text{i} ka (m+\frac{1}{4})}G_\sigma\left[x-\left(m a + \frac{a}{4}\right)\right] \right\}.
\end{align}
%
%
This formal expression can be evaluated only by truncating the infinite sum. We can Fourier transform the wave function in~\eqref{Bloch_real} to obtain the Wannier wave function:
%
%
\begin{subequations}\label{Wannier}
	\begin{align}
		\langle \bm{x}| \mathcal{W}_{nR}\rangle&=\frac{a}{2\pi}\int_\text{BZ} dk\text{e}^{-\text{i} k R}\langle \bm{x} |\Psi_{nk}\rangle\\
		\langle \bm{x}| \mathcal{W}_{n\ell}\rangle& =  \frac{a}{2\pi}\int_\text{BZ} dk\text{e}^{-\text{i} k \ell a}\langle \bm{x} |\Psi_{nk}\rangle \\
		& = \frac{1}{\sqrt{2}}\sum_{m\in\mathbb{Z}} \left[ G_\sigma\left[x-\left(m a - \frac{a}{4}\right)\right]\mathcal{I}_1(m,\ell)			+ G_\sigma\left[x-\left(m a + \frac{a}{4}\right)\right] \mathcal{I}_2(m,\ell)\right], \nonumber
	\end{align}
\end{subequations}
%
%
in these expressions, we have assumed $R=\ell a$ and we have introduced the following quantities:
%
%
\begin{subequations}\label{integrals}
\begin{align}
\mathcal{I}_1(m,\ell)&=\frac{a}{2\pi}\int_\text{BZ} dk \text{e}^{\text{i} \left[\theta(k)+k(m-\ell)a-\frac{ka}{4}\right]}\\
\mathcal{I}_2(m,\ell)&= \frac{a}{2\pi}\int_\text{BZ} dk \text{e}^{\text{i} k\left[(m-\ell)a-\frac{a}{4}\right]} \nonumber\\
&= \frac{\sin\left[\pi a (m-\ell)-\frac{\pi a}{4}\right]}{\pi\left[(m-\ell)+\frac{1}{4}\right]}.
\end{align}
\end{subequations}
%
%
%
%
\begin{figure*}
	\centering
	\includegraphics[width=\columnwidth]{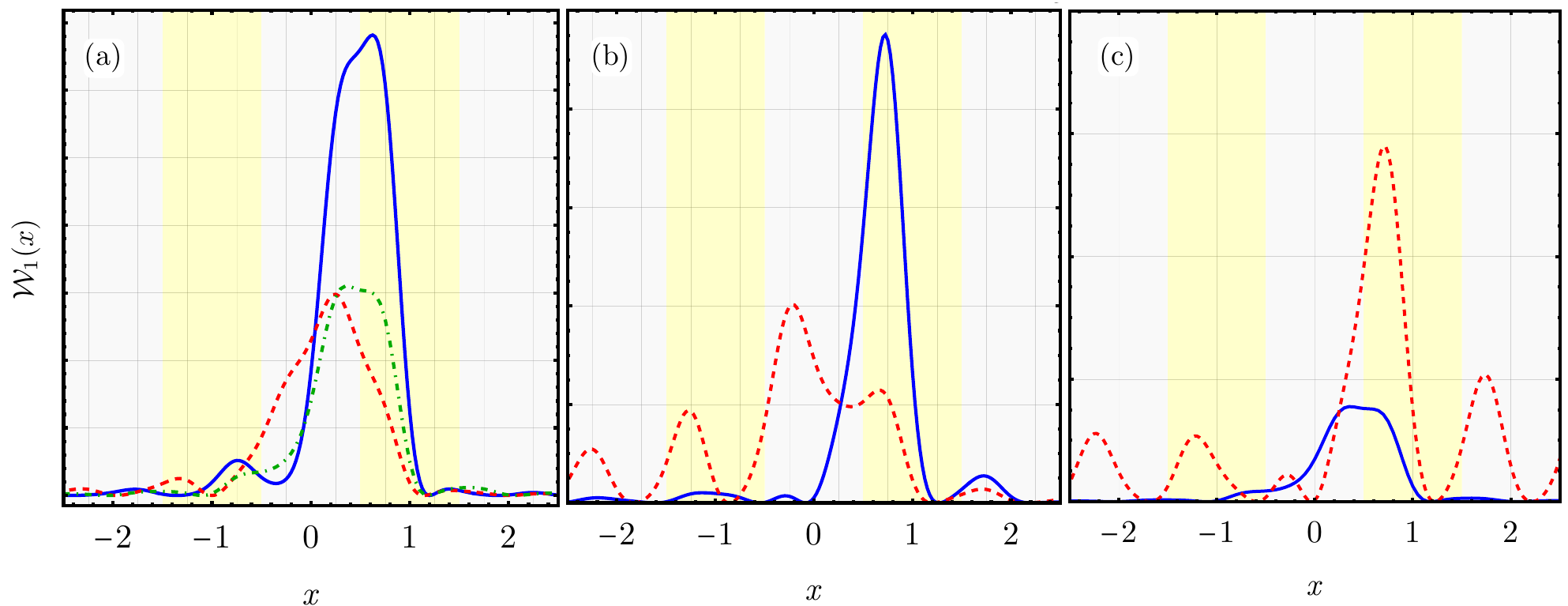}
	\caption{\label{Fig_5} The Wannier wave function for the RM and SSH models as in Eq.~\eqref{Wannier}, for $m_\text{max}=60$. (a) SSH case in the trivial (dashed red), non-trivial (blue solid) with $\delta=\pm0.6$ and \emph{metallic} $\delta=0$ (dot-dashed green) configurations. (b) RM case with $\delta=\pm0.6$ (solid blue and red dashed) and $\Delta=1$. (c) Comparison between the metallic case ($\delta=0$ of the SSH (solid blue) and RM model (dashed red) with $\Delta=0.067$. In all panels, the vertical grid lines represent the position of the lattice sites, whereas the striped regions (in yellow and gray) represent the various unit cells.}
\end{figure*}
%
%

\noindent The first of these two expressions cannot be evaluated analytically since it contains the transcendental function $\theta(k)$. When evaluating numerically Eq.~\eqref{Wannier}, we need to fix a threshold in the summation over $m$. We can use these results for evaluating the local density of states in terms of Bloch or Wannier function~\cite{Vanderbilt2018}:
%
%
\begin{align}
    \rho_n(\bm{r})=\frac{a}{2\pi}\int_\text{BZ}|dk\Psi_{nk}(\bm{r})|^2=\sum_\ell |\mathcal{W}_{n\ell}(\bm{r})|^2\,.
\end{align}
%
%
It can be proven~\cite{Vanderbilt2018} that the expectation value of the position operator over the Wannier function is given by the integral of the one-dimensional Berry connection or Zak phase:
%
%
\begin{align}
    \overline{\bm{r}}_{n}=\langle \mathcal{W}_{n0}|\overline{\bm{r}}|\mathcal{W}_{n0}\rangle= \frac{a}{2\pi}\int_\text{BZ} dk  \langle u_{nk}|\text{i}\nabla_k |u_{nk}\rangle\,,    
\end{align}
%
%
where $u_{nk}$ is the periodic part of the Bloch wave functions, the quantity $\overline{\bm{r}}_{n}$ is analogous to the center of mass of an object but defined based on wave function weight rather than mass.

In the various panels of Fig.~\ref{Fig_5}, we show the Wannier wave function of Eq.~\eqref{Wannier} at $\ell=0$ for the RM and the SSH model in the various topological phases by varying the $\delta$ and $\Delta$ parameters characterizing the topological and trivial energy gap. In Fig.~\ref{Fig_5}(a), we focus on the topological and trivial SSH model ($\Delta=0$) with $\delta=\mp0.6$ and the metallic case with $\delta=0$. We note that the Wannier function is peaked at the boundary of the unit cell in the topological phase. In contrast, the peak is inside the unit cell in the trivial phase. In the metallic case $\delta=0$, the peak is at the boundary since the model we consider corresponds to an infinite chain with a forced \emph{extension} of the unit cell. For the case of the RM model, in Fig.~\ref{Fig_5}(b) with $\delta=\pm0.6$ and $\Delta/t_2=1$, the center of the peak Wannier function inside the unit cell. Finally, in the \emph{metallic} case of the RM mode, the peak is away from the boundary of the unit cell |  Fig.~\ref{Fig_5}(c).

\subsection{Trimer chain: tight binding model}\label{Trimer_general}
%
%
\begin{figure}[!h]
    \begin{minipage}{0.45\textwidth}
    \begin{flushleft}
        \includegraphics[width=0.9\textwidth]{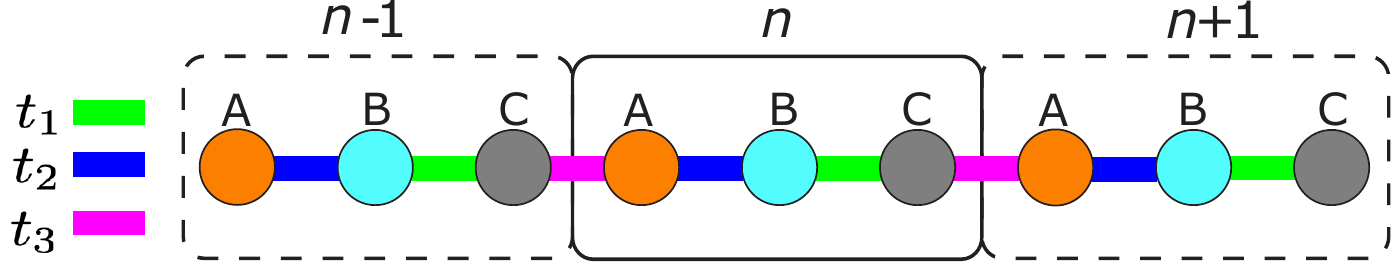}          
    \end{flushleft}
    \end{minipage}
    \begin{minipage}{0.45\textwidth}
    \caption{Sketch of the trimerized linear chain. The system contains three distinct lattice sites A, B and C. The hopping amplitudes between the three sites are characterized by the hopping strength $t_1$, $t_2$, and $t_3$. The period of the lattice is $a$.}
    \label{fig_1}
    \end{minipage}
\end{figure}
%
%
In this section, we will summarize the trimer chain's theoretical properties. We will follow the model presented in Ref.~\cite{MartinezAlvarez2019}; this was initially proposed in Ref.~\cite{Su1981}. The chain of periodicity $a$ contains three distinct lattice sites in the unit cell that we denote as A, B, and C. There are two intracell hopping parameters, $t_1$ and $t_2$, and an intercell one $t_3$. The tight-binding Hamiltonian in the nearest-neighbors approximation reads:
%
%
\begin{equation}\label{Ham_real_space}
    \mathcal{H}= \sum_n (t_1 a_{n}^\dag b_{n}+ t_2 b_{n}^\dag c_{n}+ t_3 a_{n+1}^\dag c_{n}) + h.c.
\end{equation}
%
%
where $\alpha_{n}^\dag$ ($\alpha_{n}$) denotes the creation (annihilation) operator
at the site $\alpha\in\{\text{a,b,c}\}$ of the $n$th unit cell. The lattice is presented schematically in Fig.~\ref{fig_1}. We observe that if the hopping parameters $t_1$ and $t_2$ are identical, the system presents inversion symmetry (IS), but in general, it does not have this symmetry~\cite{MartinezAlvarez2019}. We will present the modification to the spectral properties induced by the IS later in this section.

We can analyze the trimer chain under translational invariance condition; in this case, we can express the Hamiltonian~\eqref{Ham_real_space} in a Bloch form that reads:
%
%
\begin{align}\label{Ham_bloch_space}
    h(k)=\begin{pmatrix}
        0 & t_1  & t_3 \text{e}^{\text{i} k a} \\
        t_1 & 0 & t_2 \\
        t_3 \text{e}^{-\text{i} k a} & t_2 & 0
    \end{pmatrix},
\end{align}
%
%
where $k$ in the one-dimensional momentum defined as $k a\in[-\pi,\pi)$. This Bloch Hamiltonian is expressed in the following base: $\Psi=\{\psi_\text{A},\psi_\text{B},\psi_\text{C}\}$. We can obtain an analytical expression for the trimer chain spectrum. Defining $\Vec{t}$ and $f(\Vec{t},k)$ respectively as:
\begin{subequations}
\begin{equation}
    \Vec{t}=(t_1,t_2,t_3),
    \end{equation}
    \begin{equation}
    f(\Vec{t},k)=\left(\sqrt{2916 t_1^2 t_2^2 t_3^2 \cos ^2(k)-108 |\Vec{t}|^6}+54 t_1
   t_2 t_3 \cos (k)\right)^{1/3},
\end{equation}
    \end{subequations}
the analytical expression for the three bands ordered from lowest to highest in energy are:
    \begin{subequations}
     \begin{equation}
       E_{1k}= -\frac{\left(1+i \sqrt{3}\right) |\Vec{t}|^2}{\sqrt[3]{2}
   f(\Vec{t},k)}-\frac{\left(1-i \sqrt{3}\right) f(\Vec{t},k)}{6 \sqrt[3]{2}},
    \end{equation}
    \begin{equation}
       E_{2k}= -\frac{\left(1-i\sqrt{3}\right) |\Vec{t}|^2}{\sqrt[3]{2} f(\Vec{t},k)}-\frac{\left(1+i \sqrt{3}\right)
   f(\Vec{t},k)}{6 \sqrt[3]{2}},
    \end{equation}
    \begin{equation}
  E_{3k}= \frac{\sqrt[3]{2}
   |\Vec{t}|^2}{f(\Vec{t},k)}+\frac{f(\Vec{t},k)}{3
   \sqrt[3]{2}}
    \end{equation}
\end{subequations}

The generalized chiral symmetry dictates that the sum of the three eigenvalues is always zero | see Sec.~\ref{GCS} for additional details. 

\subsection{Generalized Chiral Symmetry}\label{GCS}
We can introduce a generalized chiral symmetry (GCS) similar to the one fulfilled by the kagome lattice~\cite{Ni_2018,Kempkes_2019,Herrera_2022}. This symmetry operator act in real space is permuting the position of the lattice sites. Specifically, we have:
%
%
\begin{align}\label{genchiralsymrealspace}
\Gamma_3^\text{rs}=\begin{pmatrix}
 0 & 1 & 0 \\
 0 & 0& 1 \\
 1 & 0 & 0
\end{pmatrix},
\end{align}
%
%
this operator exchanges the site A with B, B with C, and C with A.
Usually, for a generic chiral symmetry $\gamma$, we are going to require that $\{\mathcal{H},\gamma\}=0$ and $\gamma^2=\mathbb{I}_n$. However, for the case of a generalized chiral symmetry and in the absence of energy symmetry~\cite{Tassi_2024}, we are going to requires that the generalized chiral symmetry $\Gamma_3$ satisfies the following conditions:
%
%
\begin{subequations}\label{genchiralsym}
\begin{align}
\Gamma_3^{-1} \mathcal{H}_1 \Gamma_3& = \mathcal{H}_2,\label{genchiralsym_a} \\
\Gamma_3^{-1} \mathcal{H}_2 \Gamma_3 &=\mathcal{H}_3, \label{genchiralsym_b} \\
 \mathcal{H}_1+ \mathcal{H}_2+ \mathcal{H}_3&=0.\label{genchiralsym_c}
\end{align}
\end{subequations}
%
%
In the previous set of equations, we have assumed that $\mathcal{H}_1=h(k)$ | see Eq.~\eqref{Ham_bloch_space}.
When combining the Eq.~\eqref{genchiralsym_c} with~\eqref{genchiralsym_a} and~\eqref{genchiralsym_b}, it follows that $\Gamma_3^{-1} \mathcal{H}_3 \Gamma_3 =\mathcal{H}_1$. Following this reasoning, the generalized chiral symmetry introduced in Eqs.~\eqref{genchiralsym} shares similar properties to the chiral symmetry of the SSH model~\cite{Asboth2015}. However, in this case $[\mathcal{H}_1, \Gamma_3^3]=0$, which implies $\Gamma_3^3 = \mathbb{I}_3$ and the eigenvalues are given by $1, \text{exp}[\pm 2 \pi \text{i}/3]$.
Therefore, we can write
%
%
\begin{align}
\Gamma_3 = 
\begin{pmatrix}
 1 & 0 & 0 \\
 0 & \text{e}^{ 2\pi \text{i} /3} & 0 \\
 0 & 0& \text{e}^{- 2 \pi \text{i} /3}
\end{pmatrix}.
\end{align}
%
%
Furthermore, we now have three eigenvalues to consider ($\mathcal{H}_1$, $\mathcal{H}_2$, and $\mathcal{H}_3$ each have the same eigenvalues $\epsilon_1$, $\epsilon_2$ and $\epsilon_3$, since the Hamiltonians differ by a unitary transformation). By taking the trace of Eq.~\eqref{genchiralsym_c}, we find
%
%
\begin{equation}\label{trace_GCS}
    \text{Tr}[\mathcal{H}_1+\mathcal{H}_2+\mathcal{H}_3]=3\text{Tr}[\mathcal{H}_1]=0, 
\end{equation}
where we used the first Eqs.~\eqref{genchiralsym_a} and~\eqref{genchiralsym_b}  in addition to the cyclic property of the trace. This means that the sum of the three eigenvalues vanishes, $\epsilon_1 + \epsilon_2 +\epsilon_3 =0$.

\subsection{The Bloch and Wannier wavefunctions for the trimer chain model}

In the following, we will apply the same method to the trimer chain. As for the case of the RM and SSH ones, we start by defining the matrix describing the intra- and the inter-cell hopping processes:
%
%
\begin{align}
H_\text{T}^\text{intra}(\bm{0}) = \begin{pmatrix}
0 & t_1& 0 \\
t_1 & 0& t_2 \\
0 & t_2 & 0
\end{pmatrix}, \hspace{1cm}
H_\text{T}^\text{inter}(\bm{a}) = \begin{pmatrix}
0 & 0& 0 \\
0 & 0& 0 \\
t_3 & 0 & 0
\end{pmatrix}, \hspace{1cm}
H_\text{T}^\text{inter}(-\bm{a}) = \begin{pmatrix}
0 & 0& t_3 \\
0 & 0& 0 \\
0 & 0 & 0
\end{pmatrix}.
\end{align}
%
%
We now have three different Bloch states starting from localized orbitals on the three different lattice sites:
%
%
\begin{align}
|\chi_\text{A}^k\rangle  = \sum_{m\in\mathbb{Z}}\text{e}^{\text{i} k (m a-\frac{a}{3})} |m, \text{A}\rangle, ~~~
|\chi_\text{B}^k\rangle  = \sum_{m\in\mathbb{Z}}\text{e}^{\text{i} k m a} |m, \text{B}\rangle, ~~~
|\chi_\text{C}^k\rangle  = \sum_{m\in\mathbb{Z}}\text{e}^{\text{i} k (m a+\frac{a}{3})} |m, \text{C}\rangle.
\end{align}
%
%
We can make a generic Bloch state as a linear combination of the previous three states:
%
%
\begin{equation}\label{genBlochTrimer}
|\Psi_{n,k}\rangle = \sum_{\ell\in\{\text{A,B,C}\}} \alpha_\ell^{nk} |\chi_\ell^k\rangle.
\end{equation}
%
%
We can use this state to evaluate the matrix elements of the full Hamiltonian $H_\text{Trimer}=H_\text{T}^\text{intra}(\bm{0})+H_\text{T}^\text{inter}(\bm{a})+H_\text{T}(-\bm{a})$:
%
%
\begin{subequations}
\begin{align}
H_{\alpha\alpha}^k&=\langle \chi_\alpha^k|H_\text{Trimer}|\chi_\alpha^k\rangle= 0, \\
H_\text{AB}^k&=\langle \chi_\text{A}^k|H_\text{Trimer}|\chi_\text{B}^k\rangle= t_1 \text{e}^{\text{i} k\frac{a}{3}}, \\
H_\text{BC}^k&=\langle \chi_\text{B}^k|H_\text{Trimer}|\chi_\text{C}^k\rangle= t_2 \text{e}^{\text{i} k\frac{a}{3}}, \\
H_\text{AC}^k&=\langle \chi_\text{A}^k|H_\text{Trimer}|\chi_\text{C}^k\rangle= t_3 \text{e}^{\text{i} k(a+\frac{2}{3}a)}. 
\end{align}
\end{subequations}
%
%
Assuming that the eigenstates can be obtained numerically, we can find an expression for Eq.~\eqref{genBlochTrimer}. As for the case of the RM/SSH chain, we substitute a Gaussian representation in real space to the three localized orbitals $|m,\alpha\rangle$:
%
%
\begin{subequations}
\begin{align}
\langle r |m\text{A}\rangle & =G_\sigma\left[x-\left(ma-\frac{a}{3}\right)\right], \\
\langle r |m\text{B}\rangle & =G_\sigma[x-ma], \\
\langle r |m\text{C}\rangle & =G_\sigma\left[x-\left(ma+\frac{a}{3}\right)\right],
\end{align}
\end{subequations}
%
%
where $\sigma$ is the broadening of the Gaussian function. We can now express a generic Bloch state in real space as:
%
%
\begin{align}\label{Bloch_gaussian}
\langle r | \Psi_{n,k} \rangle= \mathcal{N} \sum_{m\in\mathbb{Z}}&\left\{ \alpha_\text{A}^{nk}\text{e}^{\text{i} k (m a-\frac{a}{3})} G_\sigma[x-(m a-\frac{a}{3})] + \alpha_\text{B}^{nk}\text{e}^{\text{i} kma} G_\sigma[x-ma]
\alpha_\text{C}^{nk}\text{e}^{\text{i} k (m a+\frac{a}{3})} G_\sigma[x-(m a+\frac{a}{3})] \right\},
\end{align}
%
%
with $\mathcal{N}$ the normalization constant. We can Fourier transform this expression to obtain the corresponding Wannier function:
%
%
\begin{align}
\langle r | \mathcal{W}_{nR} \rangle & = \frac{a}{2\pi} \int_\text{BZ} dk \text{e}^{-\text{i} kR} \langle r | \Psi_{nk} \rangle \\
&\underset{R=\ell a}{=} \frac{a}{2\pi} \int_\text{BZ} dk \text{e}^{-\text{i} k\ell a} \langle r | \Psi_{nk} \rangle \nonumber\\
&= \frac{a}{2\pi} \int_\text{BZ} dk \text{e}^{-\text{i} kR} \mathcal{N} \sum_{m\in\mathbb{Z}}\left\{ \alpha_\text{A}^{nk}\text{e}^{\text{i} k (m a-\frac{a}{3})} G_\sigma[x-(m a-\frac{a}{3})] + \alpha_\text{B}^{nk}\text{e}^{\text{i} kma} G_\sigma[x-ma]\right. \nonumber\\
& \hspace{3cm}\left. + \alpha_\text{C}^{nk}\text{e}^{\text{i} k (m a+\frac{a}{3})} G_\sigma[x-(m a+\frac{a}{3})] \right\}\nonumber \\
& = \frac{a}{2\pi} \sum_{m\in\mathbb{Z}} \left\{ G_\sigma[x-(m a-\frac{a}{3})] \mathcal{I}_\text{A}^n(m,\ell) +  G_\sigma[x-ma]\mathcal{I}_\text{B}^n(m,\ell)
+ G_\sigma[x-(m a+\frac{a}{3})] \mathcal{I}_\text{C}^n(m,\ell)\right\},\label{Wannier_trimer}
\end{align}
%
%
where we have introduced the following three integrals in momentum space:
%
%
\begin{subequations}
\begin{align}
	\mathcal{I}^n_\text{A}(m,\ell) &= \int_\text{BZ} dk\, \alpha_\text{A}^{nk}\, \text{e}^{\text{i} k [(m-\ell)a-\frac{a}{3}]}, \\
	\mathcal{I}^n_\text{B}(m,\ell) &= \int_\text{BZ} dk\, \alpha_\text{B}^{nk}\, \text{e}^{\text{i} k (m-\ell)a}, \\
	\mathcal{I}^n_\text{C}(m,\ell) &= \int_\text{BZ} dk\, \alpha_\text{C}^{nk}\, \text{e}^{\text{i} k [(m-\ell)a+\frac{a}{3}]}. 
\end{align}
\end{subequations}
%
%
%
%
\begin{figure}[!t]
    \begin{minipage}[c]{0.65\textwidth}
            \begin{flushleft}
            \includegraphics[width=0.89\linewidth]{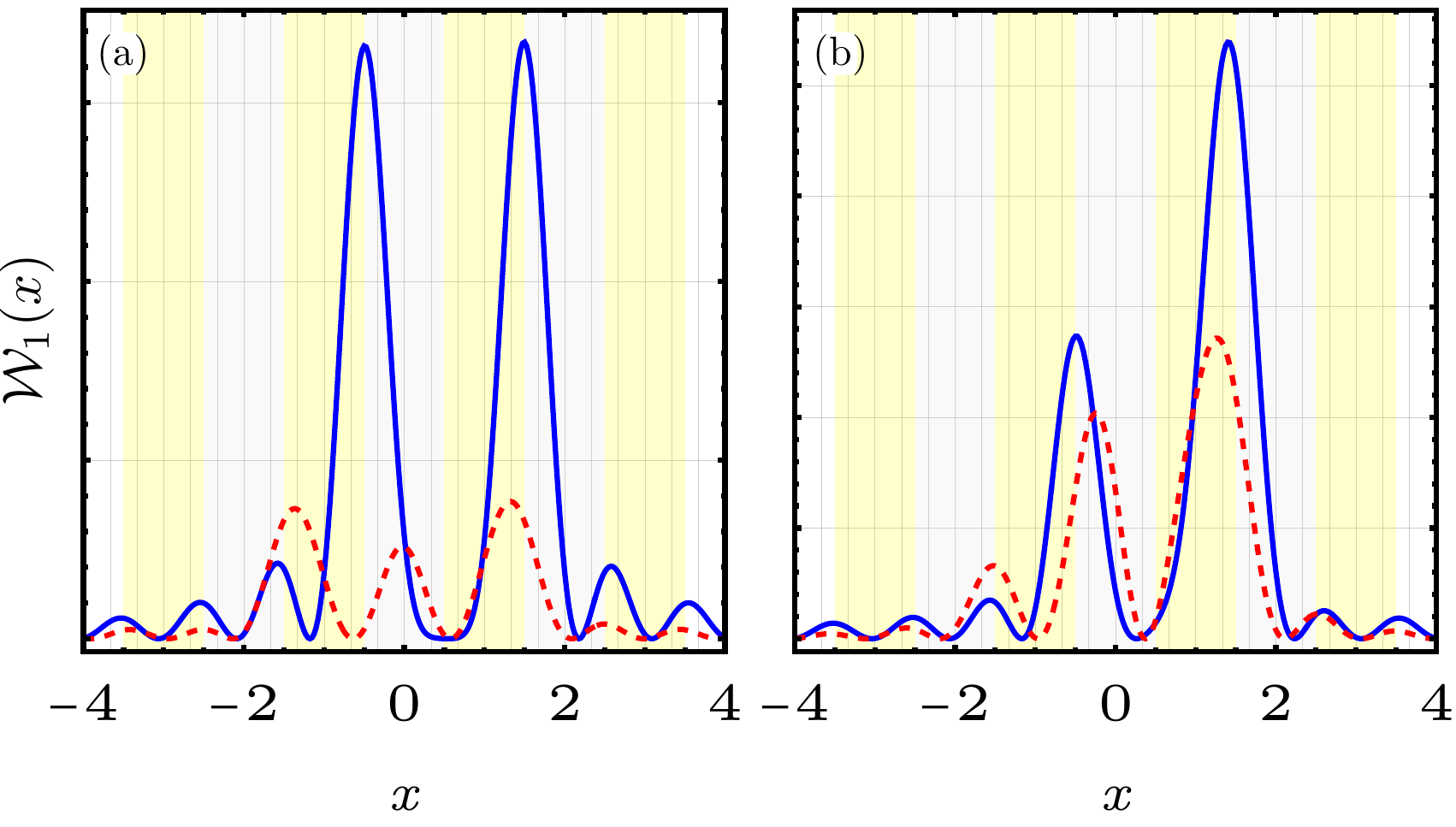}
            \end{flushleft}
    \end{minipage}
    \begin{minipage}[c]{0.3\textwidth}
    \caption{Wannier wave function for $\ell=0$ as in Eq.~\eqref{Wannier_trimer} for the trimer chain in the inversion symmetric (a) and non-inversion symmetric (b) configurations. The solid blue line in both panels corresponds to $\delta=-0.75$ (non-trivial phase) and the dashed red line to $\delta=0.75$ (trivial phase).  In all panels, the vertical grid lines represent the position of the three lattice sites, whereas the striped regions (in yellow and gray) represent the various unit cells.}
    \label{fig_WF_Trimer}
    \end{minipage}
\end{figure}
%
%
These are analogous to the integrals defined in Eqs.~\eqref{integrals}, with the difference that now we do not have an analytical and compact form for the coefficients $\alpha_i^{nk}$. In Fig.~\ref{fig_WF_Trimer}, we present the Wannier function with $\ell=0$ for the case of a trimer chain with and without inversion symmetry. In panel (a), we can observe that the \emph{center of mass} of the Wannier function is at the boundary of the unit cell in the non-trivial case, whereas it is in the center in the trivial case. In the non-inversion symmetric case in panel (b), the height of the peaks is not the same, and as a consequence, the center of mass of the Wannier function is no longer fixed at the boundary or the center of the unit cell. 

In the following sections, we present the corresponding spectral properties for the finite-size trimer chain in two possible configurations, one that is inversion symmetric and the other particular one that is not inversion symmetric (among all the possible choices).

\subsection{Trimer chain: spectral properties in the symmetric and non-symmetric configurations}

The trimer chain has many similarities with the SSH chain. There, when the intercell hopping is larger than the intracell hopping, two edge modes are pinned at zero energy. In the case of the trimer chain, since there is an extra atom in the unit cell compared to the SSH chain, there are more cases to study. Here, we cover two cases, one when inversion/mirror symmetry is present in the chain and the other not inversion or mirror symmetric. The schematic of the trimer chain is shown in Fig.~\ref{fig_1}, and the Hamiltonian is the one present in Eq.~\eqref{Ham_bloch_space}.

The band structure of the trimer chain is composed of three bands. Compared with a molecular diagram, this band structure can be seen as a bonding band at the lowest energy, a non-bonding band in the middle, and an antibonding band at the highest energy. As mentioned above, two cases of hopping parameters will be treated here. The first one where $t_1=t_2$, and the Hamiltonian is inversion symmetric, and the second one with $t_1  \neq t_2\neq t_3$.
%
%
\begin{figure}
    \centering
    \includegraphics[width=\linewidth]{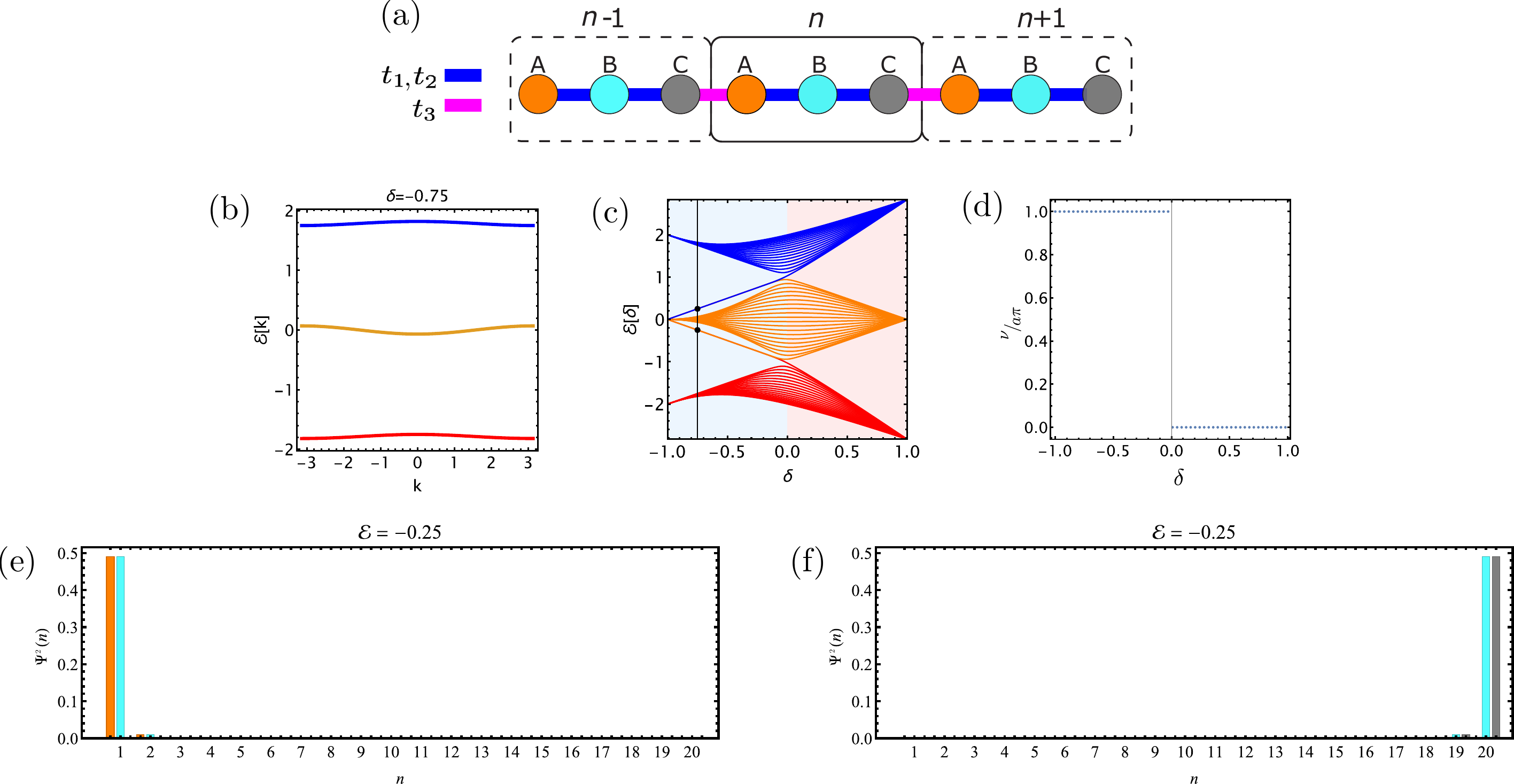}
    \caption{Spectral properties of the trimer chain tight-binding model for the inversion-symmetric configuration. Panel (a) displays a sketch of the chain, where lattice site B is the center of inversion. Panels (b) and (c) represent the trimer chain's bulk and finite-size spectra. Panel (d) Zak phase associated with the inversion-symmetric phase versus the dimerization parameter, displaying a step at $\delta=0$ (quantization). Panels (e) and (f) represent the two degenerate edge states appearing at $\delta=-0.75$ and $E=-0.25$ which are localized in opposite edges due to the inversion symmetry.}
    \label{fig_inv_sym}
\end{figure}
%
%
\subsubsection{Inversion-symmetric case}

For the inversion-symmetric case, we study the chain's properties with respect to a parameter $\delta$ that acts as the usual dimerization parameter studied in the RM/SSH chain. The expressions for the hopping terms under this ``dimerization'' are:
%
%
\begin{subequations}\label{hopping_IS}
    \begin{align}
        t_1=t_0(1+\delta), \\
        t_2=t_0(1+\delta), \\
        t_3=t_0(1-\delta),
    \end{align}
\end{subequations}
%
%
where $t_0$ is a reference value. Figure~\ref{fig_inv_sym} shows the spectral properties of the inversion-symmetric configuration for the trimer chain. Panel~\ref{fig_inv_sym}(a) displays a sketch of the chain where atom B is the center of inversion/mirror symmetries. Panel~\ref{fig_inv_sym}(b) shows the bulk energy spectrum of the chain for $\delta=-0.75$. Panel~\ref{fig_inv_sym}(c) shows the finite-size spectrum for a chain with 20 unit cells (60 sites in total). For negative $\delta$, edge modes are present in the first bandgap | between the bonding band and the non-bonding band | and in the second bandgap | between the non-bonding band and the antibonding band. The black line is set precisely at $\delta=-0.75$. Panels~\ref{fig_inv_sym}(d) and (e) represent the two states in the first gap at negative energy for $\delta=-0.75$. The states at $\delta=-0.75$ but in the second gap are localized on the same edge of the chain.
For the inversion symmetric case, the Zak phase is quantized | $0$ or $\pi$, see Fig.~\ref{fig_inv_sym}(b) | and when $t_3>t_1=t_2$ edge modes are present on both sides of the chain. The edge modes are similar to those found in the SSH model. However, the shape of edge modes has doubled since they are composed of two lattice sites compared to a single one of the RM/SSH model. Additionally, they are not pinned at zero energy, but their energy depends on the value of the $\delta$ parameter.

\subsubsection{Non inversion-symmetric case}
%
%
\begin{figure}
    \centering
    \includegraphics[width=\linewidth]{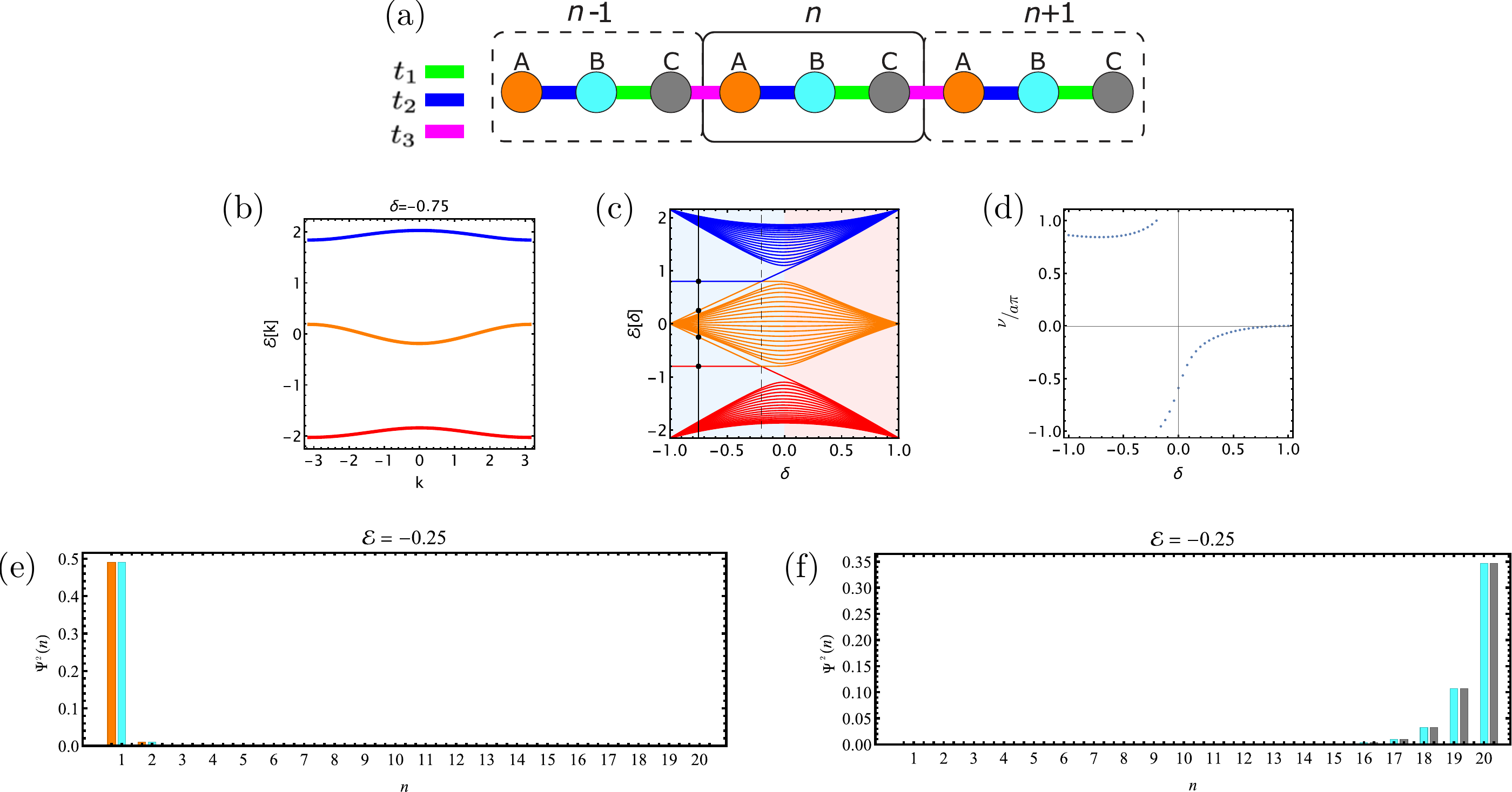}
    \caption{Spectral properties of the trimer chain tight-binding model for the chosen non-inversion-symmetric configuration. Panel (a) displays a sketch of the chain. Panels (b) and (c) represent the trimer chain's bulk and finite-size spectra. Panel (d) shows the Zak phase associated with the non-inversion symmetric phase, displaying no quantization. Panels (e) and (f) represent the two edge states at $\delta=-0.75$. They show different localization lengths.}
    \label{fig_non_inv_sym}
\end{figure}
%
%
For the non-inversion-symmetric case, we study the chain's properties with respect to a parameter $\delta$ that acts as the usual dimerization parameter studied in the SSH chain. The expressions for the hopping terms under this new ``dimerization'' are:
%
%
\begin{subequations}\label{hopping_NIS}
\begin{align}
        t_1=&0.8 t_0,\\
        t_2=&t_0(1+\delta),  \\
        t_3=&t_0(1-\delta),
    \end{align}
\end{subequations}
%
%
where $t_0$ is a reference value. Figure~\ref{fig_non_inv_sym} shows the spectral properties of the non-inversion-symmetric configuration for the trimer chain. Panel~\ref{fig_non_inv_sym}(a) displays a sketch of the chain where it is clear that there is no center of inversion/mirror symmetries. Panel~\ref{fig_non_inv_sym}(b) shows the bulk energy spectrum of the chain for $\delta=-0.75$. Panel~\ref{fig_non_inv_sym}(c) shows the finite-size spectrum for a chain with 20 unit cells (60 sites in total). For negative $\delta$, edge modes are present in the first and second gap. Remarkably, none of them are degenerate. Panel~\ref{fig_non_inv_sym}(d) shows the edge states depicted in panel~\ref{fig_non_inv_sym}(c) in red and blue, while panel~\ref{fig_non_inv_sym}(e) represents the edge states depicted in panel~\ref{fig_non_inv_sym}(c) in orange. These states have been obtained by using $\delta=-0.75$. In this configuration, the edge state that is more isolated in energy ($-.97$) appears to have a localization length shorted compared to the edge state with energy closer to the middle band ($-0.25$). The role of the edge states is exchanged in the upper gap, with the one at lower energy more extended compared to the successive.

A closer look to panel~\ref{fig_non_inv_sym}(c) reveals that at $\delta_\text{D}=(|t_1|-t_0)/t_0=-0.2$ the two edge modes are degenerate. This is represented by a dashed line in panel (c). This only occurs for $t_1<t_0$, since for $t_1=t_0$, $\delta_\text{D}=0$ but the band structure remains gapped, with no degeneracy point at $\delta=0$. The inversion symmetry is restored in the system for this specific value of $\delta$.

Finally, the Zak phase is thus not quantized for the non-inversion symmetric case~\cite{MartinezAlvarez2019} | see Fig.~\ref{fig_non_inv_sym}b.

\subsection{Finite size evolution}\label{finite_size}

In this section, we will characterize the properties of the SSH model's topological phase and the trimer chain's inversion-symmetric configuration. In the bulk case, we can describe the phase of these systems by evaluating the Zak phase~\cite{Asboth2015, MartinezAlvarez2019}; however, we cannot access the Bloch states for finite-size systems. In the case of the SSH model, we know that the energy of the edge states is identical and equal to zero. Similarly, for the case of the trimer chain with inversion symmetry, we know that there are two degenerate edge states in each gap in the bulk case.

In the following, we will use the difference in energy between the energy of these states to investigate the evolution with the system size $n$ and the degree of dimerization $\delta$. The results will coincide with the bulk case for $n\to\infty$. For both systems, the case $\delta=-1$ represents a \emph{sweet spot} in the parameter case since, for this value, the edge states are always degenerate. Based on this analysis, we will try to identify where the finite-size experimental results are compared to the ideal bulk case.
%
%
\begin{figure}
    \centering
    \includegraphics[width=\textwidth]{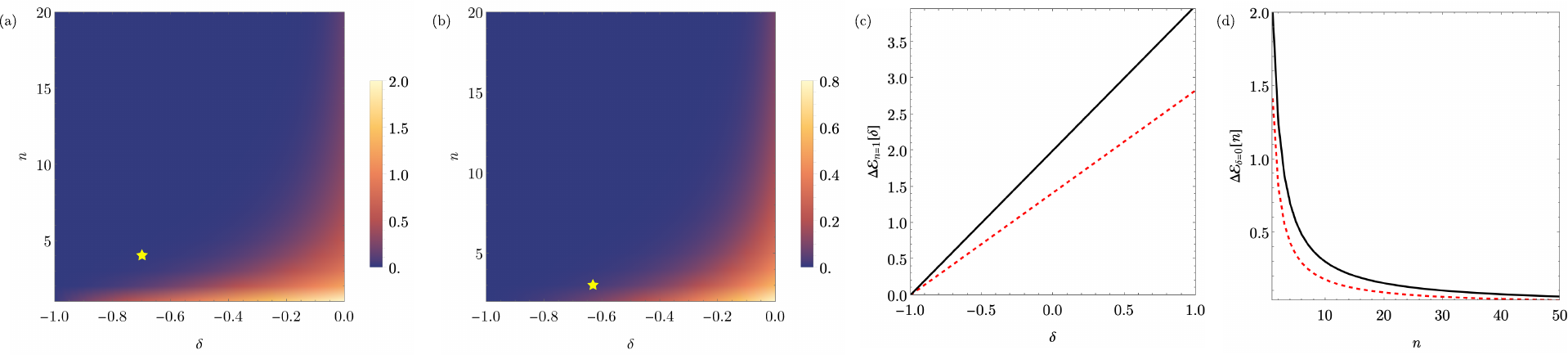}
    \caption{Study of the hybridization of the edge modes as a function of the system size $n$ and dimerization parameter: (a) case of the SSH model, (b) case of the trimer chain in inversion symmetric configuration. In these two panels, the yellow star shows the configurations for the non-trivial phase of the SSH chain (a) and the non-trivial inversion-symmetric trimer chain (b) that has been experimentally realized. Panel (c) is a cut of panels (a) and (b) along $n=1$ and (d) along $\delta=0$.}
    \label{fig_evolution}
\end{figure}
%
%
We present the SSH and trimer chain results in Fig.~\ref{fig_evolution}. In panel (a), we present the modulus of the energy difference of the two boundary modes as a function of the system size $n$ and the dimerization parameter $\delta\in[-1,0]$. We observe that the difference departs from zero for small chains and small dimerization. The case of the trimer chain is presented in panel (b). We note on passing that the results for the two gaps in the system are identical. As for the case of the SSH model, the energy difference differs from zero for small chains and small dimerization. In panel (c), we present a cut of both density plots when considering a chain of length $n=1$. For both chains, the energy difference grows linear with the dimerization parameter. We have included here the positive $\delta$ case corresponding to an opening of a gap. We note that in both chains, the energy difference is linear in the dimerization parameter. In panel (d), we present a cut for $\delta=0$ as a function of the chain length. This is a special point that characterizes the phase transition point. Here, the energy difference is zero in the bulk case. We observe that in the case of the trimer chain, the energy difference goes to zero faster than in the case of the SSH chain.

\section{Comparison experimental results and tight-binding calculations}
%
%
\begin{figure*}
    \centering
    \includegraphics{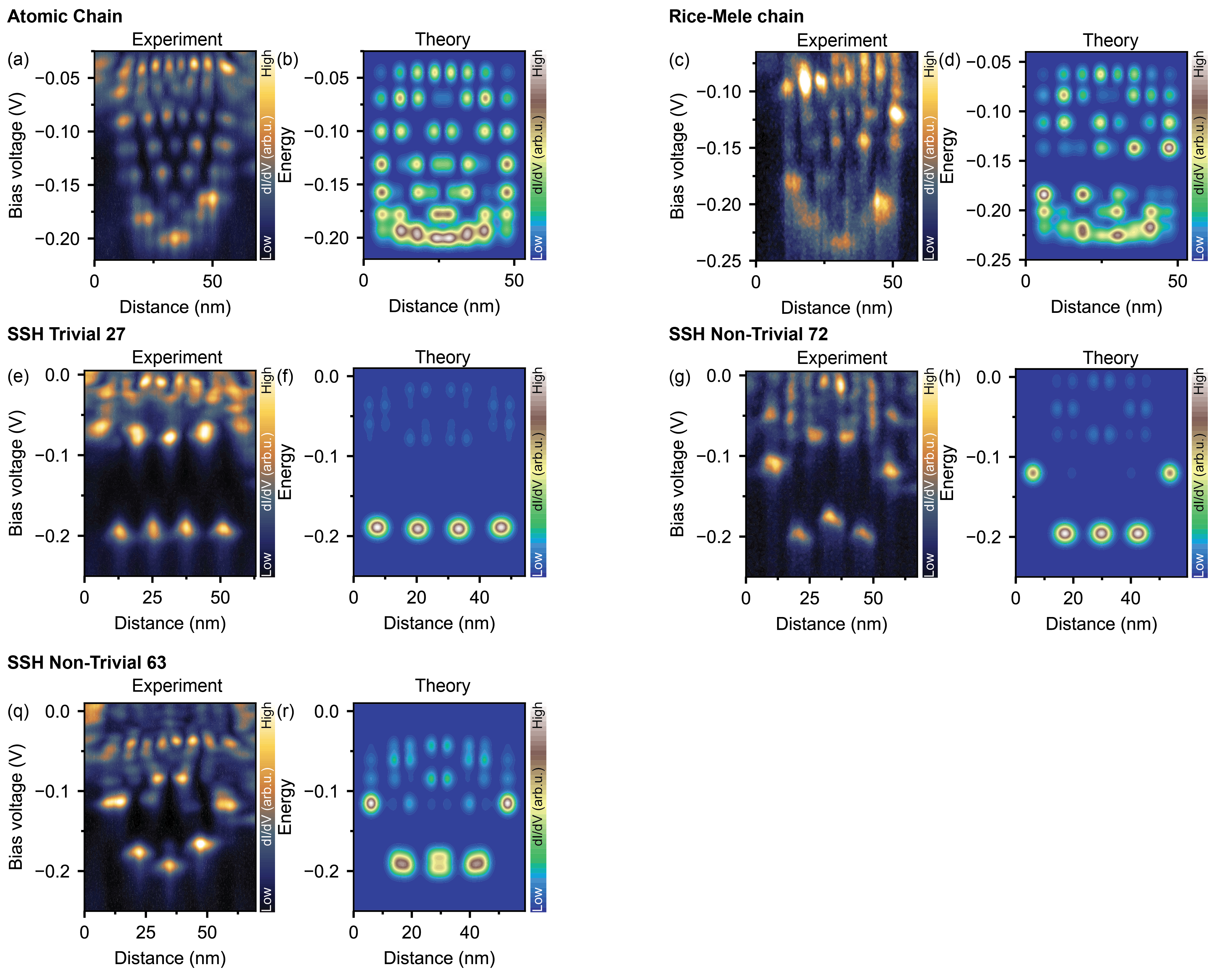}
    \caption{Comparison of the experimental line spectra with the tight-binding calculations for the atomic chain, the Rice-Mele chain, and the trivial and non-trivial SSH chains.}
    \label{ComparisonExpTheory}
\end{figure*}
\begin{figure*}
    \centering
    \includegraphics{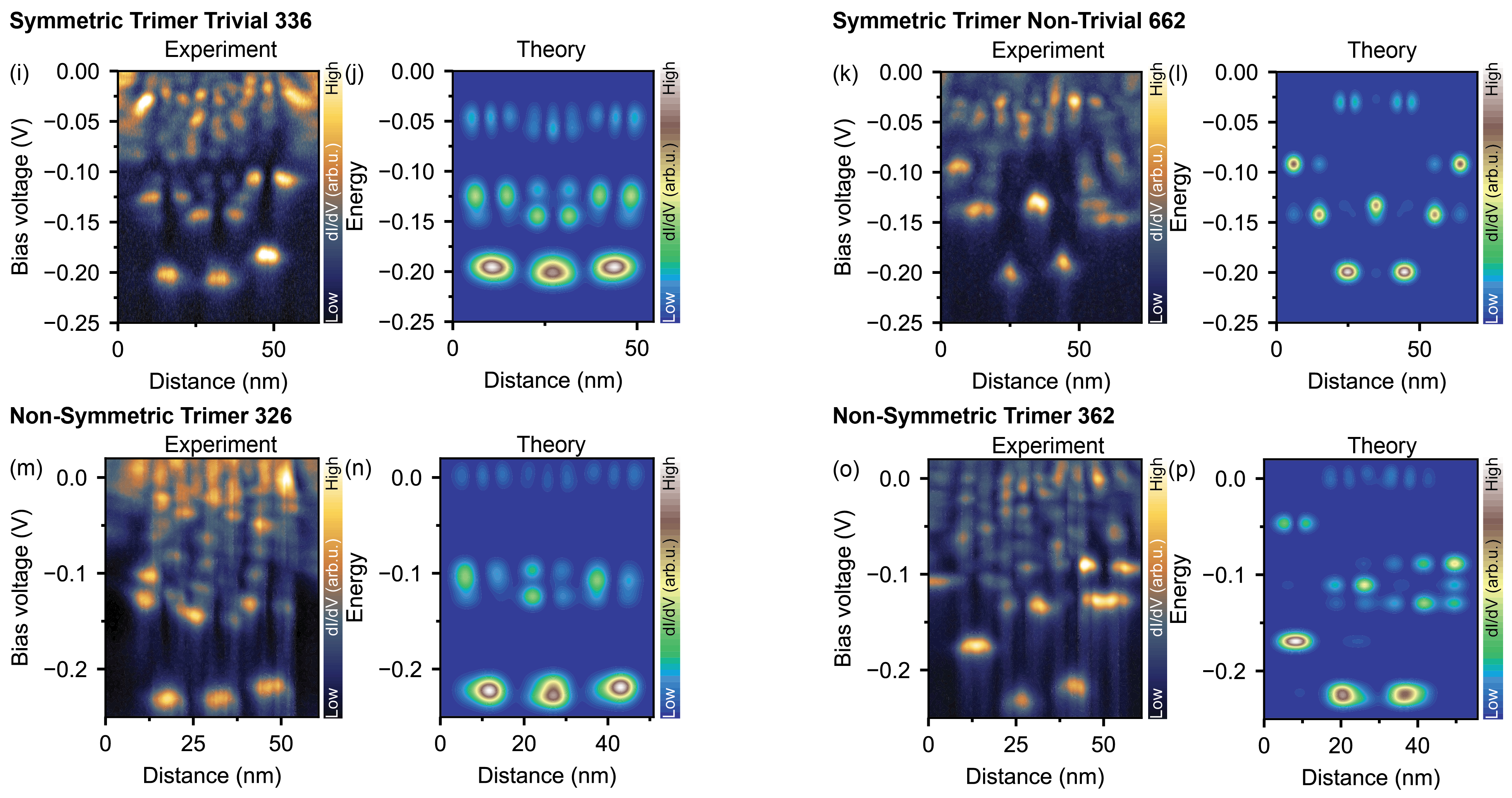}
    \caption{Comparison of the experimental line spectra with the tight-binding calculations for the symmetric/non-symmetric and trivial/non-trivial trimer chains.}
    \label{ComparisonExpTheory2}
\end{figure*}

%
%
In this section, we systematically compare the experimental and theoretical results obtained within the fine-size tight-binding model for the different 1D chains we have considered, see table \ref{tab: theoryparam} for all the parameters. In Fig.~\ref{ComparisonExpTheory2}, we compare the case of the Atomic chain [Panels (a) and (b)], the Rice-Mele model [Panels (c) and (d)], and the SSH model in the trivial and non-trivial configurations [Panels (e) to (r)].
We obtained a good agreement between the measurements and the theoretical model for all the cases shown in the various panels. We compared the trimer chain similarly in the two possible configurations with and without inversion symmetry in the trivial and non-trivial phases. The results are shown in Fig.~\ref{ComparisonExpTheory2}.  
%
%
\begin{table}[]
    \centering
    \caption{Theoretical parameters for the finite tight-binding calculations.}

    \begin{tabular}{ |c|c|c|c|c|c|c|c| } 
    
    \hline
     & $t_1$~[meV] & $t_2$~[meV] & $t_3$~[meV] & $s_1$~[meV] & $s_2$~[meV] & $s_3$~[meV] & on-site energy edges~[meV]  \\
    \hline
    Atomic Chain &  38 & & & 152 & & &\\
    Rice-Mele &  38 & 38& & 114 & 114& &$o_1$ = 0 $o_2$ = 38\\
    \hline
    SSH 27 Trivial &  50.7 & 9 & & 557.7 &  99 & & 5 \\
    SSH 72 Non-Trivial &  9 & 50.7 & & 99 &  557.7 & & 45 \\
    SSH 63 Non-Trivial &  18.6 & 54.4 & & 83.8 &  244.8 & & 35 \\
    \hline
    Symmetric Trimer 336 Trivial &  50 & 50 & 17.1 & 100 &  100 & 34.2 & 20 \\
    Symmetric Trimer 662 Non-Trivial & 17.1 & 17.1 & 76 & 51.3 &  51.3 & 228.1 & 35 \\
    \hline
    Non-symmetric Trimer 326 Trivial &  52.7 & 80.3 & 18.1 & 158.2 &  240.8 & 54.2 & 10 \\
    Non-symmetric Trimer 362 Non-Trivial &  52.7 & 18.1 & 80.3 & 158.1 &  54.2 & 240.8 & 10 \\
    \hline
    \end{tabular}
    \label{tab: theoryparam}
\end{table}
%
%

%
%
%
%

\subsection{Non-symmetric trimer chain}
\begin{figure*}
    \centering
    \includegraphics{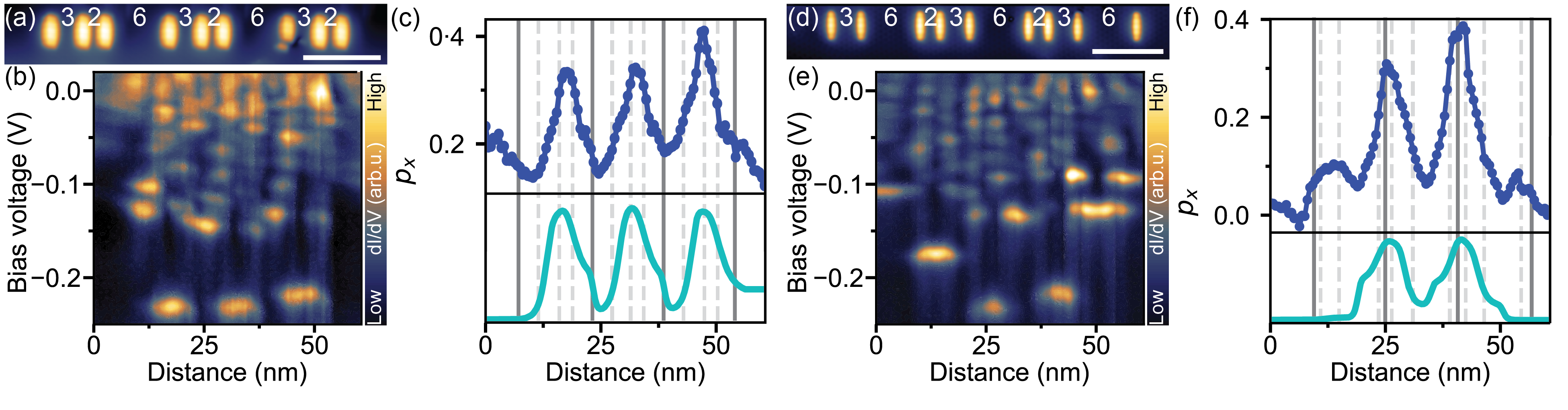}
    \caption{(a, d) STM topography image of the trivial (b) and non-trivial (e) non-symmetric trimer chain with 3 unit cells and a distance of $3a\sqrt{3}$, $2a\sqrt{3}$ and $7a\sqrt{3}$ nm between the artificial atoms and $t_2$ and $t_3$ swapped for the non-trivial chain. Images taken at 0.1 V, 180 pA (a) and 0.1 V, 310 pA (d), scale bar is 10 nm. (b, e) Contour plot of $dI/dV$ spectra taken along a line above the trivial and non-trivial non-symmetric trimer chains in (a) and (d), respectively. (c, f) Charge density plot with the positions of the atoms (light grey) and the boundary of the unit cell (dark grey).}
    \label{NonSymmTrimer}
\end{figure*}

The design of the non-symmetric non-trivial trimer chain is depicted in Fig.~\ref{NonSymmTrimer}(a); three unit cells were built, resulting in a chain length of 50 nm. The hopping parameters are chosen so the chiral edge mode is well centred in the first bandgap. A sizeable first bandgap is obtained by maximizing the ratio between $t_3$ and $t_2$. The ratio between $t_1$ and $t_2$ governs the separation in the right and left edge modes' energy; a larger ratio will center the left chiral edge mode more in the gap. The distances between the atomic sites were chosen to be $3a\sqrt{3}$, $6a\sqrt{3}$ and $2a\sqrt{3}$. The distances compare to hopping parameters of $t_1=55.5$~mV, $t_2 = 18$~mV, $t_3 = 84.5$~mV, this corresponds to a dimerization parameter $\delta\approx-0.78$ accordingly to Eq.~\eqref{hopping_NIS}. The LDOS along the chain | Fig.~\ref{NonSymmTrimer}(b) | shows the bonding state at $-232$~mV and the non-bonding state at $-133$~mV. At $-174$~mV  | in the first band gap | the chiral edge mode is observed on the left side of the chain. The second bandgap is too small to be observed due to electronic asymmetries in the chain likely caused by the surroundings, such as the hole above the $8^{th}$ atomic site. Therefore, it is also harder to observe the second edge mode on the left side of the chain in the second band gap. The right edge mode is absent and hybridized with the centre band.
Figure~\ref{NonSymmTrimer}(e) shows the charge density of the lowest band, the Wannier centres are still located at the boundary of the unit cell, although the peak is relatively broad. The electron density of the edge mode is located between the first and second atomic sites, similar to the symmetric trimer case.
To show that the edge mode solely arises due to the hopping ratio, the trivial case of the trimer chain was also built | see Fig.~\ref{NonSymmTrimer}(d). The hopping terms $t_2$ and $t_3$ are swapped for the trivial case, resulting in $\delta\approx0.78$. The contour plot of the line spectrum in Fig.~\ref{NonSymmTrimer}(e) shows that no in-gap states are present in the first gap. Due to asymmetries in the chain, the second band gap is not visible. The charge density | Figure \ref{NonSymmTrimer}f | shows the Wannier representation for the finite trivial non-symmetric trimer chain. The Wannier centres are localized right from the centre of the unit cell. Due to the breaking of the inversion symmetry of the chain, the Wannier centre is shifted from the centre of the unit cell towards the boundary. 
\end{document}